\documentclass[twocolumn]{aastex62}
\usepackage{amsmath,amsfonts}
\usepackage[super]{nth}
\usepackage{makecell}
\usepackage{amsmath}
\DeclareMathOperator{\arcsinh}{arcsinh}

\newcommand{\Mstar}{\text{M}_{\star}}

\defcitealias{LopezSanjuan:2014uj}{LS15}	

\graphicspath{{./}{figures/}}

\received{\today}
\revised{--}
\accepted{--}
\submitjournal{ApJ}

\shorttitle{Observational constraints on the merger history of galaxies since $z\approx6$}
\shortauthors{Duncan et al.}
\begin{document}

\title{Observational constraints on the merger history of galaxies since $z\approx6$:\\ Probabilistic galaxy pair counts in the CANDELS fields}

\correspondingauthor{Kenneth Duncan}

\author[0000-0001-6889-8388]{Kenneth Duncan}
\affiliation{Leiden Observatory, Leiden University, NL-2300 RA Leiden, Netherlands}
\affiliation{University of Nottingham, School of Physics \& Astronomy, Nottingham NG7 2RD, United Kingdom}
\email{duncan@strw.leidenuniv.nl}

\author{Christopher J. Conselice}
\affiliation{University of Nottingham, School of Physics \& Astronomy, Nottingham NG7 2RD, United Kingdom}

\author{Carl Mundy}
\affiliation{University of Nottingham, School of Physics \& Astronomy, Nottingham NG7 2RD, United Kingdom}

% \author{Guillermo Barro}
% \affiliation{University of California at Berkeley, 501 Campbell Hall \#3411, Berkeley, CA 94720-3411, USA}

\author{Eric Bell}
\affiliation{The University of Michigan, 300E West Hall, Ann Arbor, MI 48109-1107, USA}

\author{Jennifer Donley}
\affiliation{Los Alamos National Laboratory, P.O. Box 1663, Los Alamos, NM 87545, USA}

% \author{Sandra Faber}
% \affiliation{University of California Observatories/Lick Observatory, University of California, Santa Cruz, CA 95064, USA}

% \author{Henry C. Ferguson}
% \affiliation{Space Telescope Science Institute, 3700 San Martin Dr., Baltimore, MD, 21218, USA}

\author{Audrey Galametz}
\affiliation{Department of Astronomy, University of Geneva Chemin d'\'{E}cogia 16, CH-1290 Versoix, Switzerland}

\author{Yicheng Guo}
\affiliation{Department of Physics and Astronomy, University of Missouri, Columbia, MO, 65211, USA}

\author{Norman A. Grogin}
\affiliation{Space Telescope Science Institute, 3700 San Martin Dr., Baltimore, MD, 21218, USA}

\author{Nimish Hathi}
\affiliation{Space Telescope Science Institute, 3700 San Martin Dr., Baltimore, MD, 21218, USA}

\author{Jeyhan Kartaltepe}
\affiliation{School of Physics and Astronomy, Rochester Institute of Technology, Rochester, NY 14623, USA}

\author{Dale Kocevski}
\affiliation{Department of Physics and Astronomy, Colby College, Waterville, ME 04961, USA}

\author{Anton M. Koekemoer}
\affiliation{Space Telescope Science Institute, 3700 San Martin Dr., Baltimore, MD, 21218, USA}

% \author{Hooshang Nayyeri}
% \affiliation{Department of Physics \& Astronomy, University of California, Irvine, CA 92697, USA}

\author{Pablo G. P\'{e}rez-Gonz\'{a}lez}
\affiliation{Departamento de Astrof\'{i}sica, Facultad de CC. F\'{i}ısicas, Universidad Complutense de Madrid, E-28040 Madrid, Spain}
\affiliation{Centro de Astrobiolog\'{\i}a (CAB/INTA), Ctra. de Torrej\'on a Ajalvir, km 4, E-28850 Torrej\'on de Ardoz, Madrid, Spain}

\author{Kameswara B. Mantha}
\affiliation{Department of Physics and Astronomy, University of Missouri-Kansas City, Kansas City, MO 64110, USA}

\author{Gregory F. Snyder}
\affiliation{Space Telescope Science Institute, 3700 San Martin Dr., Baltimore, MD, 21218, USA}

\author{Mauro Stefanon}
\affiliation{Leiden Observatory, Leiden University, NL-2300 RA Leiden, Netherlands}

%% Note that the \and command from previous versions of AASTeX is now
%% depreciated in this version as it is no longer necessary. AASTeX 
%% automatically takes care of all commas and "and"s between authors names.

%% AASTeX 6.2 has the new \collaboration and \nocollaboration commands to
%% provide the collaboration status of a group of authors. These commands 
%% can be used either before or after the list of corresponding authors. The
%% argument for \collaboration is the collaboration identifier. Authors are
%% encouraged to surround collaboration identifiers with ()s. The 
%% \nocollaboration command takes no argument and exists to indicate that
%% the nearby authors are not part of surrounding collaborations.

%% Mark off the abstract in the ``abstract'' environment. 
\begin{abstract}
Galaxy mergers are expected to have a significant role in the mass assembly of galaxies in the early Universe, but there are very few observational constraints on the merger history of galaxies at $z > 2$. 
We present the first study of galaxy major mergers (mass ratios $>$ 1:4) in mass-selected samples out to $z \approx 6$. Using all five fields of the HST/CANDELS survey and a probabilistic pair count methodology that incorporates the full photometric redshift posteriors and corrections for stellar mass completeness, we measure galaxy pair-counts for projected separations between 5 and 30 kpc in stellar mass selected samples at $9.7 < \log_{10}(\Mstar / \text{M}_{\odot}) < 10.3$ and $\log_{10}(\Mstar / \text{M}_{\odot}) > 10.3$. 
We find that the major merger pair fraction rises with redshift to $z\approx6$ proportional to $(1+z)^{m}$, with $m = 0.8\pm0.2$ ($m = 1.8\pm0.2$) for $\log_{10}(\Mstar / \text{M}_{\odot}) > 10.3$ ($9.7 < \log_{10}(\Mstar / \text{M}_{\odot}) < 10.3$). 
Investigating the pair fraction as a function of mass ratio between 1:20 and 1:1, we find no evidence for a strong evolution in the relative numbers of minor to major mergers out to $z < 3$.
Using evolving merger timescales we find that the merger rate per galaxy ($\mathcal{R}$) rises rapidly from $0.07\pm 0.01$ Gyr$^{-1}$ at $z < 1$ to $7.6\pm 2.7$ Gyr$^{-1}$ at $z = 6$ for galaxies at $\log_{10}(\Mstar / \text{M}_{\odot}) > 10.3$.
The corresponding co-moving major merger rate density remains roughly constant during this time, with rates of $\Gamma \approx 10^{-4}$ Gyr$^{-1}$ Mpc$^{-3}$. 
Based on the observed merger rates per galaxy, we infer specific mass accretion rates from major mergers that are comparable to the specific star-formation rates for the same mass galaxies at $z > 3$ - observational evidence that mergers are as important a mechanism for building up mass at high redshift as in-situ star-formation.
\end{abstract}
%% Keywords should appear after the \end{abstract} command. 
%% See the online documentation for the full list of available subject
%% keywords and the rules for their use.

%% From the front matter, we move on to the body of the paper.
%% Sections are demarcated by \section and \subsection, respectively.
%% Observe the use of the LaTeX \label
%% command after the \subsection to give a symbolic KEY to the
%% subsection for cross-referencing in a \ref command.
%% You can use LaTeX's \ref and \label commands to keep track of
%% cross-references to sections, equations, tables, and figures.
%% That way, if you change the order of any elements, LaTeX will
%% automatically renumber them.
%%
%% We recommend that authors also use the natbib \citep
%% and \citet commands to identify citations.  The citations are
%% tied to the reference list via symbolic KEYs. The KEY corresponds
%% to the KEY in the \bibitem in the reference list below. 

 \section{Introduction}

Galaxies grow their stellar mass in one of two distinct ways. 
They can grow by forming new stars from cold gas that is either accreted from their surroundings or already within the galaxy. 
Alternatively, they can also grow by merging with other galaxies in their local environment. 
Although observations suggest that both channels of growth play have played equal roles in the build-up of massive galaxies over the last eleven billion years \citep[e.g.,][]{Bell:2006ey,Bundy:2009jw,Bridge:2010ft,2010ApJ...719..844R,Ownsworth:2014gt,mundy2017}, there are few observational constraints on their relative roles in the early Universe.

On-going star-formation within a galaxy is to date by far the easiest, and most popular, of the two growth mechanisms to measure and track through cosmic time.
The numerous ways of observing star-formation: UV emission, optical emission lines, radio and far-infrared emissions, have allowed star-formation rates of individual galaxies to be estimated deep into the earliest epochs of galaxy formation \citep[see e.g.][for compilations of these measurements]{Hopkins:2006bq,Behroozi:2013fg,Madau:2014gt}. 
However, in contrast to measuring galaxy star-formation rates, measuring the merger rates of galaxies is a significantly more tricky task, yet at least as equally important for many reasons. 
Despite the difficulty in measuring merger rates, studying the merger history of galaxies is vital for understanding more than just the mass build-up of galaxies.
Mergers are thought to play a crucial role in structure evolution \citep{1972ApJ...178..623T,2002MNRAS.333..481B,Dekel:2009bn}, as well as the triggering of star-bursts and active galactic nuclei activity \citep{Silk:1998up,Hopkins:2008gr,2011MNRAS.418.2043E,Chiaberge:2015ip}.
Mergers are also correlated with super-massive black hole mergers, which may be the origin of a fraction of gravitational wave events that future missions such as LISA \citep{2017arXiv170200786A} will detect.

Two main avenues exist for studying the fraction of galaxies undergoing mergers at a given epoch (and hence the merger rate). 
The first method relies on counting the number of galaxies that exist in close pairs, for example \citet{Zepf:1989gv}, \citet{Burkey:1994hn}, \citet{Carlberg:1994ke}, \citet{Woods:1995il}, \citet{Yee:1995bc}, \citet{Neuschaefer:1997vr}, \citet{Patton:2000kt}, and \citet{LeFevre:2000iq} \citep[see also ][for recent examples]{2016ApJ...830...89M,mundy2017,2017A&A...608A...9V, 2018MNRAS.475.1549M}. 
This method assumes that galaxies in close proximity, a galaxy pair, are either in the process of merging or will do so within some characteristic timescale. 
The second method relies on observing the morphological disturbance that results from either ongoing or very recent merger activity \citep[e.g.,][]{Reshetnikov:2000un,Conselice:2003jz,Conselice:2008de,Lavery:2004dd, Lotz:2006is,Lotz:2008kr,Jogee:2009iz}. 
These two methods are complementary, in that they probe different aspects and timescales within the process of a galaxy merger. 
However, it is precisely these different merger timescales which represent one of the largest uncertainties in measuring the galaxy merger rate \citep[e.g.,][]{Kitzbichler:2008gi,Conselice:2009bi,Lotz:2010ie,Lotz:2010hf,Hopkins:2010ip}.

The major merger rates of galaxies have been well studied out to redshifts of $z \leq 2.5$ \citep{Conselice:2003jz,Bluck:2009in,LopezSanjuan:2010cz,Lotz:2011bu,Bluck:2012dh}, but fewer studies have extended the analysis beyond this. 
Taking into account systematic differences due to sample selection and methodology, there is a strong agreement that between $z = 0$ and $z\approx 2 - 3$ the merger fraction increases significantly \citep{Conselice:2003jz,Bluck:2009in,LopezSanjuan:2010cz,Bluck:2012dh,Ownsworth:2014gt}. \citet{2009MNRAS.397..208C} presented the first tentative measurements of the merger fractions at redshifts as high as $4 \leq z \leq 6$, making use of both pair-count and morphological estimates of the merger rate. 
For both estimates, the fraction of galaxies in mergers declines past $z\gtrsim4$, supporting the potential peak in the galaxy merger fraction at $1 \lesssim z \lesssim 2$ reported by \citeauthor{Conselice:2008de} (\citeyear{Conselice:2008de}; morphology) and \citeauthor{RyanJr:2008ka} (\citeyear{RyanJr:2008ka}; close pairs). 
However, as the analysis of \citet{2009MNRAS.397..208C} was limited to only optical photometry in the very small but deep Ultra Deep Field \citep{2006AJ....132.1729B}, the results were subject to uncertainties due to small sample sizes and limited photometric redshift and stellar mass estimates.

When studying galaxy close pair statistics, to satisfy the close pair criterion two galaxies must firstly be within some chosen radius (typically 20 to 50 kpc) in the plane of the sky and, in many studies,  within some small velocity offset along the redshift axis \citep[other studies, e.g.][deproject into 3D close pairs]{2010ApJ...719..844R}. The typical velocity offset required is $\Delta 500~\rm{km~s}^{-1}$, corresponding to a redshift offset of  $\delta z / (1+z) = 0.0017$. However, this clearly leads to difficulties when studying the close pair statistics within deep photometric surveys, as the scatter on even the best photometric redshift estimates is $\delta z / (1+z) \approx 0.01$ to $0.04$ \citep[e.g.][]{Molino:2014iz}. %Moreover, measuring systemic spectroscopic redshifts is increasingly difficult at high redshift due to the increased difficulty in observing multiple emission lines for systemic redshift estimates \citep{2015MNRAS.450.1846S}. The required redshift accuracy is often beyond spectroscopy even if such data was available for all galaxies in a survey.

To estimate the merger fractions of galaxies in wide-area photometric redshift surveys or at high-redshift, a methodology that allows us to overcomes the limitations of redshift accuracy in these surveys is required. The method used must correct or account for the pairs observed in the plane of the sky that are due to chance alignments along the line-of-sight. Various approaches have been used to overcome this limitation, including the use of de-projected two-point correlation functions \citep{Bell:2006ey, 2010ApJ...719..844R}, correcting for chance pairs by searching over random positions in the sky \citep{Kartaltepe:2007dv}, and integrating the mass or luminosity function around the target galaxy to estimate the number of expected random companions \citep{LeFevre:2000iq,Bluck:2009in,Bundy:2009jw}. The drawback of these methods is that they are unable to take into account the effects of the redshift uncertainty on the derived properties, such as rest-frame magnitude or stellar mass, potentially affecting their selection by mass or luminosity.

\citet[][LS15 hereafter]{LopezSanjuan:2014uj} present a new method for estimating reliable merger fractions through the photometric redshift probability distribution functions (posteriors) of galaxies. By making use of all available redshift information in a probabilistic manner, this method has been shown to produce accurate merger fractions in the absence of spectroscopic redshift measurements. In this paper we apply this PDF close pair technique presented in \citetalias{LopezSanjuan:2014uj}, and further developed by us in \citet{mundy2017} using deep ground based NIR surveys.

In this paper we apply this methodology, with some new changes, to all five of the fields in the CANDELS \citep{2011ApJS..197...35G,Koekemoer:2011br} photometric survey in order to extend measurements of the major merger fraction of mass-selected galaxies out to the highest redshifts currently possible, $z \sim 6$.  
This allows us to determine how mergers are driving the formation of galaxies through 12.8 Gyr of its history when the bulk of mass in galaxies was put into place \citep[e.g.][]{Madau:2014gt}.  
By doing this we are also able to test the role of minor mergers at lower redshifts, and how major mergers compare with star-formation for the build up of stellar mass in galaxies over the bulk of cosmic time.
Crucially, thanks to the availability of extensive narrow- and medium-band surveys in a subset of these fields, we are also able to directly explore the effects of redshift precision on our method and resulting merger constraints.
%In addition to the greatly improved number statistics available due to increased volume probed by the CANDELS survey (compared to the Hubble UDF alone), this study also benefits from the use of deep \emph{Spitzer} IRAC observations for improved photometric redshift and stellar mass estimation at high-redshift. It has been shown that sampling galaxies SEDs above the Lyman break is essential for reliably selecting galaxies at $z >5$ \citep{2011MNRAS.418.2074M} and estimating accurate photometric redshifts and stellar masses. 

%The aim of this paper is to measure the major merger rate at $z \geq 2$ and make the first robust estimates for mass selected samples at $z\sim4$, 5 and 6.  By applying this novel method to a subset of the available CANDELS data we hope to show its potential for helping understand the early assembly history of galaxies. If successful, this method will allow us for the first time to constrain the contribution of mergers to the progenitors of today's most massive galaxies before the peak in the cosmic star-formation history. In tandem with complimentary analysis at lower redshifts (Mundy et al. \emph{in prep.}), it will be possible to study the merger history of massive galaxies in a consistent manner throughout the bulk of cosmic history.

The structure of this paper is as follows: In Section~\ref{merger-sec:data} we briefly outline the photometric data and the derived key galaxy properties used in this analysis. In Section~\ref{merger-sec:method} we describe the probabilistic pair-count method of \citetalias{LopezSanjuan:2014uj} and \citet{mundy2017} as implemented in this work. In Section~\ref{merger-sec:results} we present our results, including comparison of our observations with the predictions of numerical models of galaxy evolution and comparable studies in the literature. In Section~\ref{merger-sec:discussion}, we discuss our results and their implications. Finally, Section~\ref{merger-sec:summary} presents our summary and conclusions for the results in this paper. 
Throughout this paper all quoted magnitudes are in the AB system \citep{1983ApJ...266..713O} and we assume a $\Lambda$-CDM cosmology ($H_{0} = 70$ kms$^{-1}$Mpc$^{-1}$, $\Omega_{m}=0.3$ and $\Omega_{\Lambda}=0.7$) throughout. Quoted observables are expressed as actual values assuming this cosmology unless explicitly stated otherwise. Note that luminosities and luminosity-based properties such as observed stellar masses scale as $h^{-2}$ while distances such as pair separation scale as $h^{-1}$.

\section{Data}\label{merger-sec:data}

The photometry used throughout this work is taken from the matched UV to mid-infrared multi-wavelength catalogs in the CANDELS field based on the CANDELS WFC3/IR observations combined with the existing public photometric data in each field. The published catalogs and the data reduction involved are each described in full in their respective catalog release papers: GOODS South \citep{Guo:2013ig}, GOODS North (Barro et al. \emph{in prep}), COSMOS \citep{2017ApJS..228....7N}, UDS \citep{Galametz:2013dd} and EGS  \citep{2017ApJS..229...32S}.
  
% \subsection{Imaging Data (Short version - would merge with above)}
% Given the wide variation in included optical and near-infrared filters between fields, we refer the interested reader to the corresponding CANDELS catalog release papers (listed above) for full details. In the remainder of this section we outline the key analysis steps common to all five catalogs and the details of how secondary data products such as photometric redshifts and stellar mass estimates were produced.

\subsection{Imaging Data}
\subsubsection{HST Near-infrared and Optical Imaging}
The near-infrared WFC3/IR data observations of the CANDELS survey \citep{2011ApJS..197...35G,Koekemoer:2011br} comprise of two tiers, a DEEP and a WIDE tier. 
In the CANDELS DEEP survey, the central portions of the GOODS North and South fields were observed in the WFC3 F105W ($Y_{105}$), F125W ($J_{125}$) and F160W ($H_{160}$) filters in five separate epochs. 
In fields flanking the DEEP region, GOODS North and South were also observed to shallower depth (two epochs) in the same filters as part of the CANDELS WIDE tier. 

Additionally, the northern-most third of GOODS South comprises WFC3 Early Release Science \citep[ERS, ][]{2011ApJS..193...27W} region and was observed in F098M ($Y_{98}$), $J_{125}$ and $H_{160}$. 
Within the GOODS South DEEP region also lies the Hubble Ultra Deep Field \citep[WFC3/IR HUDF:][see also \citeauthor{Bouwens:2010dk}~\citeyear{Bouwens:2010dk} and \citeauthor{2013ApJS..209....6I}~\citeyear{2013ApJS..209....6I}]{Ellis:2012iw, Koekemoer:2013db} with extremely deep observations also in $Y_{105}$, $J_{125}$ and $H_{160}$.

As part of the CANDELS WIDE survey, the COSMOS, UDS and EGS fields were observed in the WFC3 $J_{125}$ and $H_{160}$ filters to two epochs. Finally, in addition to the CANDELS observations, all five CANDELS fields have also been observed in the alternative J band filter, F140W ($JH_{140}$), as part of the 3D-HST survey \citep{Brammer:2012bu,Skelton:do}. The 3D-HST observations, processed in the same manner as the CANDELS observations, are included in the photometry catalogs used in this work.

For the GOODS North and South fields, the optical HST images from the Advanced Camera for Surveys (ACS) images are version v3.0 of the mosaiced images from the GOODS HST/ACS Treasury Program, combining the data of \citet{2004ApJ...600L..93G} with the subsequent observations obtained by \citet{2006AJ....132.1729B} where available and the parallel F606W and F814W CANDELS observations \citep{2011ApJS..193...27W,Koekemoer:2011br}. Altogether, each GOODS field was observed in the F435W ($B_{435}$), F606W ($V_{606}$), F775W ($i_{775}$), F814W ($I_{814}$) and F850LP ($z_{850}$) bands. 

For COSMOS, UDS and EGS, optical ACS imaging in $V_{606}$ and $I_{814}$ is provided by the CANDELS parallel observations in combination with available archival observations \citep[EGS:][]{Davis:2007ek}.
All WFC3 and ACS data were reduced and processed following the method outlined in \citet{Koekemoer:2011br}.

\subsubsection{Spitzer Observations}
Being extremely well-studied extragalactic fields, all of the five fields have deep  \emph{Spitzer}/IRAC \citep{Fazio:2004eb} observations at 3.6, 4.5, 5.8 and 8.0$\mu m$ taken during \emph{Spitzer's} cryogenic mission. For the GOODS North and South fields the cryogenic mission observations GOODS Spitzer Legacy project (PI: M. Dickinson). The wider COSMOS field was observed as part of the S-COSMOS survey \citep{Sanders:ed}. The UDS was surveyed as part of the \emph{Spitzer} UKIDSS Ultra Deep Survey (SpUDS; PI: Dunlop). And finally, part of the EGS was observed by \citet{Barmby:hi}, with subsequent observations extending the coverage (PID 41023, PI: Nandra).

In addition to the legacy cryogenic data, subsequent observations in both the 3.6 and 4.5$\mu$m have since been made during the \emph{Spitzer} Warm Mission as part of both the SEDS \citep{Ashby:2013cc} and S-CANDELS \citep{Ashby:fh} surveys, significantly increasing the depth of 3.6 and 4.5$\mu$m over the wider CANDELS area.

All of the IRAC data available within the CANDELS footprints were combined and reprocessed, first as part of the SEDS survey \citep{Ashby:2013cc} and later as part of S-CANDELS \citep{Ashby:fh}. Due to their earlier publication date, the IRAC data in the published GOODS South and UDS catalogs make use of the SEDS data, while the remaining fields (GOODS North, COSMOS and EGS) use the latest S-CANDELS mosaics. Full details of the IRAC data and its reduction can therefore be found in the respective SEDS or S-CANDELS survey papers.
   
\subsubsection{Ground-based observations}

Complementary to the space based imaging of HST and Spitzer, each CANDELS field has also been surveyed by a large number of ground-based telescope and surveys.
As these extensive ancillary ground-based observations vary from field to field, we do not present the full details for each field, instead we again refer the interested reader to the corresponding individual release papers for each field: GOODS South \citep{Guo:2013ig}, GOODS North (Barro et al. \emph{in prep.}), COSMOS \citep{2017ApJS..228....7N}, UDS \citep{Galametz:2013dd} and EGS \citep{2017ApJS..229...32S}.

In addition to the ground-based photometry outlined in the primary CANDELS release papers, in the GOODS North field we also include the medium-band imaging from the Survey for High-z Absorption Red and Dead Sources \citep[SHARDS; ][]{2013ApJ...762...46P}. 
SHARDS uses 25 medium-band filters between wavelengths of 500-900 nm over an area of 130 arcmin$^{2}$ in the GOODS-N region.  
This imaging was taken with the 10.4 m Gran Telescopio Canarias (GTC), and by itself gives effectively a spectral resolution of about R=50 down to limits of AB$\approx 26.5$ mag.  
One of the major goals of the SHARDS survey is to find emission and absorption line galaxies at redshifts up to $z \sim 5$.  
However, the fine wavelength sampling also makes it a powerful dataset for producing precise photo-$z$ estimates for all source types. 
Similarly, in the GOODS South field we also include the \emph{Subaru} medium band imaging presented in \citet{Cardamone:2010jf}.

\subsection{Source photometry and deconfusion}
All of the CANDELS survey catalogs have been produced using the same photometry method, full details which can be found in the respective catalog papers \citep[e.g. ][]{Guo:2013ig,Galametz:2013dd}. In summary, photometry for the HST bands was done using \textsc{SExtractor}'s \citep{Bertin:1996hf} dual image mode, using the WFC3 H band mosaic as the detection image in each field and the respective ACS/WFC3 mosaics as the measurement image after matching of the point-spread function (PSF, individual to each field). 

For all ground-based and \textit{Spitzer} IRAC bands, deconvolution and photometry was done using template fitting photometry (TFIT). We refer the reader to \citet{2007PASP..119.1325L}, \citet{2012ApJ...752...66L}, and the citations within for further details of the TFIT process and the improvements gained on multi-wavelength photometry.

As with the broad-band imaging, photometry for the medium-band imaging was performed using the same TFIT forced photometry procedure employed during the main catalog production \citep{Guo:2013ig} - with positions based on the corresponding WFC3 $H_{160}$ imaging \citep[][and J. Donley, priv. communication for GOODS North and South respectively]{2013ApJ...762...46P}.

\subsection{Image depths and detection completeness estimates}\label{merger-sec:completeness}
Due to the tiered observing strategy employed for the CANDELS survey and the limitations imposed on the tiling of individual exposures, the final $H_{160}$ science images used for the catalog source detections are somewhat in-homogeneous. Not only is there significant variation in image depth across the five CANDELS fields, but each field itself is inhomogeneous. 
To overcome these limitations whilst still making full use of the deepest available areas, we divide each of the CANDELS fields into sub-fields based on the local limiting magnitude (as determined from the RMS maps of the $H_{160}$ science images).

\begin{figure}
{\centering
\includegraphics[width=0.99\columnwidth]{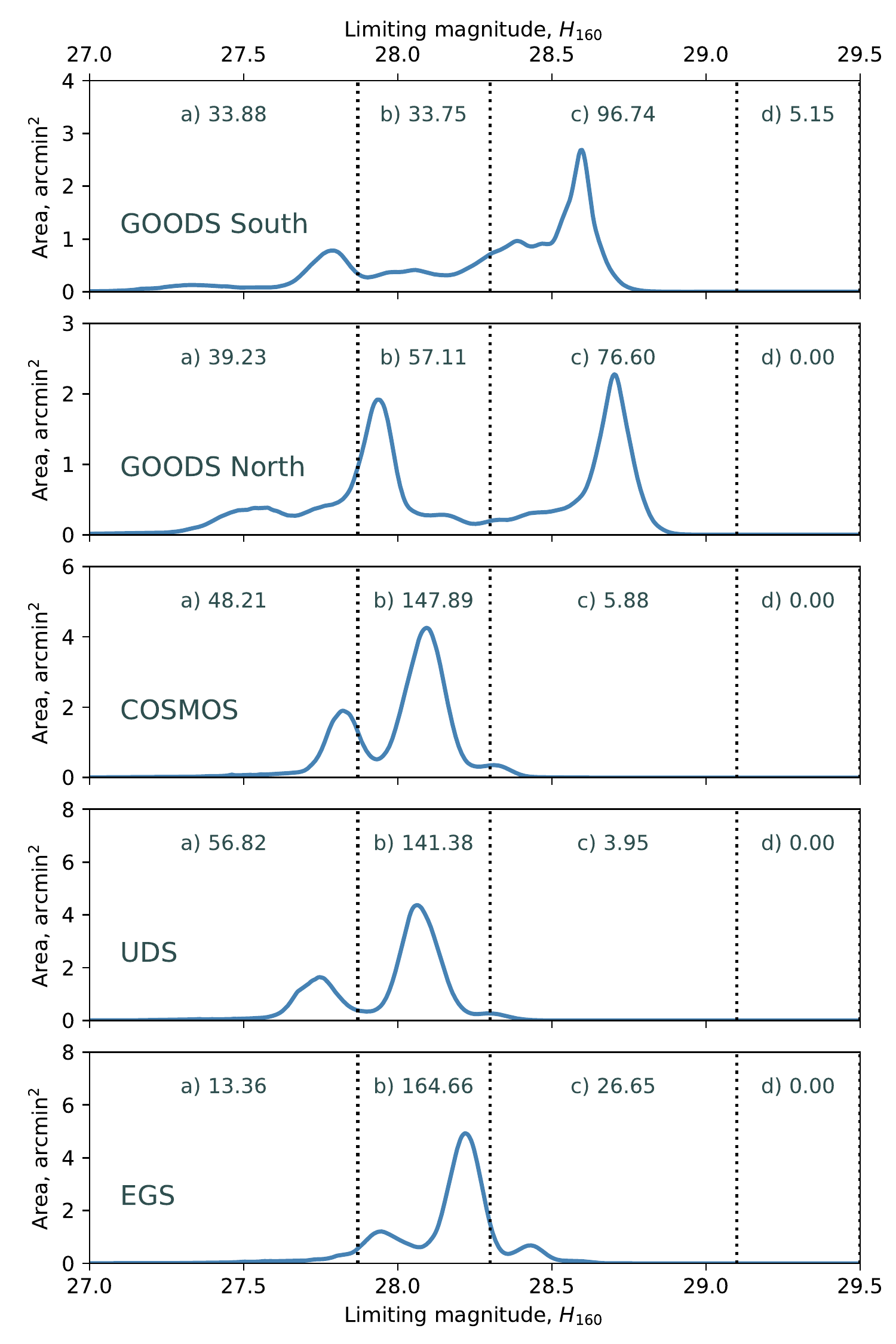}}
  \caption{Distribution of area with a given limiting magnitude (1$\sigma$ within an area of 1 arcsec$^2$) for each of the five CANDELS fields. The vertical dashed lines show the limiting magnitudes used to define the a) `Wide 1', b) `Wide 2', c) `Deep' and d) `Ultra-deep' sub-fields within each field. The corresponding total area covered (in arcmin$^2$) is also shown for each sub-field. Note that the range of limiting magnitudes shown excludes that reached by the HUDF, hence the area of GOODS South corresponding to the HUDF is not plotted. We refer the interested reader to the individual catalog release papers for an illustration of the spatial distribution of these depths (see Section.~\ref{merger-sec:data} for references).}
  \label{merger-fig:lim_depths}
 \end{figure}

Fig.~\ref{merger-fig:lim_depths} illustrates the distribution of area with a given limiting magnitudes (within an area of 1 arcsec$^2$ at 1$\sigma$; $H_{160}^{\text{lim}}$) for each of the five CANDELS fields. 
While the difference in depth between the WIDE and DEEP tiers of the survey are very clear, there is also noticeable variation in limiting magnitude between fields of with same number of HST observation epochs (COSMOS, UDS and EGS).  
The observed difference in field depth is primarily due to the different locations on the sky in which the CANDELS fields are located, the ability to schedule HST time to observe these fields, and how the orbits are divided into exposure times.  
Together these constraints determined the differences in the CANDELS tiling strategies and the resulting exposure times for each pointing \citep{Koekemoer:2011br,2011ApJS..197...35G}. 
As a result of this tiling and scheduling constraints, the EGS pointings are 10-15\% longer than in COSMOS and are as a result slightly deeper, with the UDS field in between these two.  

Additionally, the fields also have different background levels as they are in different portions of the sky, and these different background levels result in different effective depths being reached.  
This creates the variety of depths for the WIDE and DEEP epochs highlighted by Fig.~\ref{merger-fig:lim_depths}. 

Based on the distributions observed in Fig.~\ref{merger-fig:lim_depths}, we define four sets of sub-fields based on the following limiting magnitude ranges:  $H_{160}^{\text{lim}} < 27.87 ~\textup{mag}$ (Wide 1), $27.87 \leq H_{160}^{\text{lim}} < 28.3 ~\textup{mag}$  (Wide 2), $28.3 \leq H_{160}^{\text{lim}} < 29.1 ~\textup{mag}$ (Deep) and $H_{160}^{\text{lim}} \geq 29.1 ~\textup{mag}$ (Ultra-deep). The sub-sets of observed galaxies are then simply defined based on the measured $H_{160}^{\text{lim}}$ at the position of the galaxy.  

To ensure consistent estimates of the respective source detection limits, we performed new completeness simulations across all five CANDELS fields. 
These simulations include a realistic range of input (magnitude-dependent) morphologies based on the observed structural properties of galaxies in the CANDELS fields \citep{vanderWel:2012eu}.
Full details of how the completeness simulations were performed are outlined in Appendix~\ref{app:completeness}. 
In Fig.~\ref{merger-fig:gs_completeness}, we present an example plot illustrating the measured source recovery fraction as a function of magnitude for each of the sub-fields within the GOODS South field. 
Due to the effects of source confusion and chance alignment with brighter sources in the field, it can be seen that the catalogs are 100\% complete at only the very brightest magnitudes. 
For this field, the $80\%$ completeness limits range from $H_{160} = 25.29 \textup{mag}$ for the shallowest observations down to $H_{160} = 27.26 \textup{mag}$ mag for the Ultra-deep field. 
In Table~\ref{tab:completeness}, we present the measured completeness limits for image regions of different limiting magnitude for each CANDELS field. 
Figures illustrating the detection completeness for all fields are included for reference in Appendix~\ref{app:completeness}.

\begin{figure}
{\centering
\includegraphics[width=0.95\columnwidth]{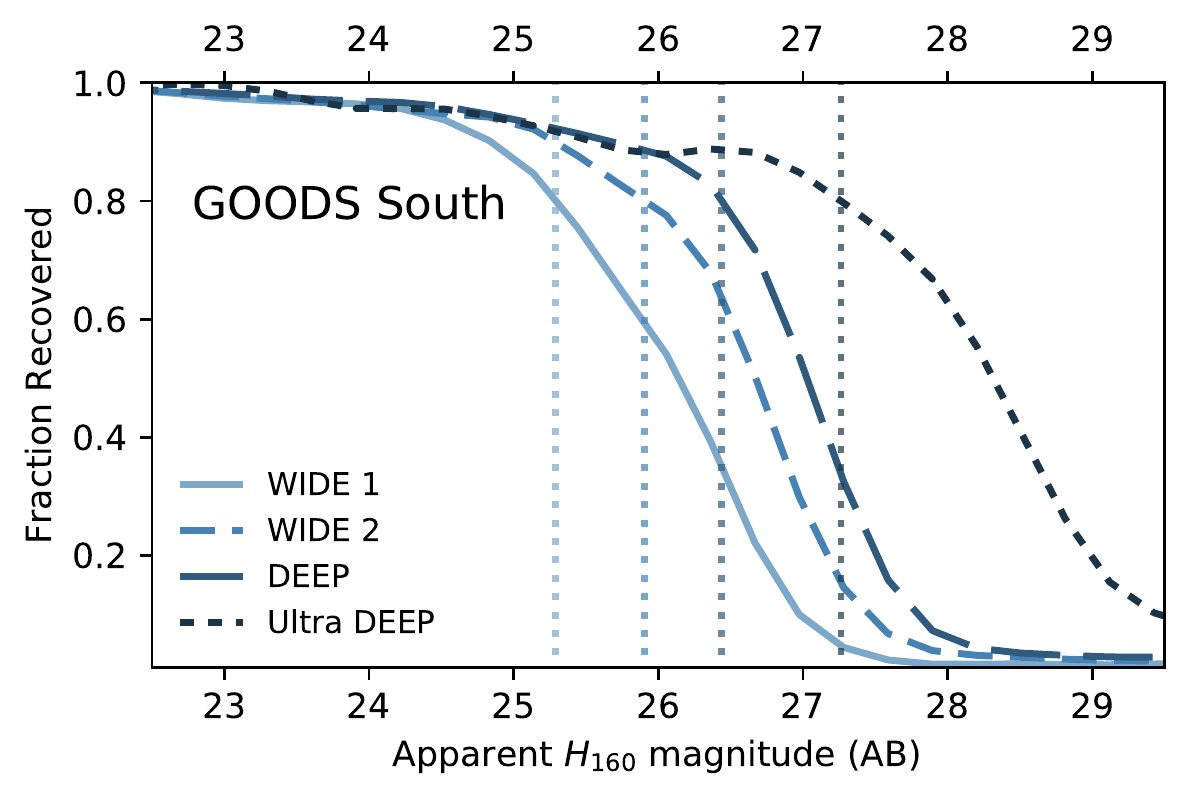}}
  \caption{Example detection completeness estimates, showing the fraction of recovered sources as a function of $H_{160}$ magnitude for the GOODS South field. The vertical dashed lines show the magnitude at which the recovery fraction equals 80\% for each sub-field.}
  \label{merger-fig:gs_completeness}
\end{figure}

\begin{deluxetable*}{lcccccccc}
\centering
\tablecaption{CANDELS Field Completeness Depths}
\label{tab:completeness}
\tablehead{ & \multicolumn{2}{c}{\textbf{Wide 1}} & \multicolumn{2}{c}{\textbf{Wide 2}} & \multicolumn{2}{c}{\textbf{Deep}} & \multicolumn{2}{c}{\textbf{Ultra Deep}} \\
            & Area\tablenotemark{a} & Depth & Area\tablenotemark{a}   & Depth  & Area\tablenotemark{a}  & Depth & Area\tablenotemark{a} & Depth}
\startdata
GOODS South & 33.88 & 25.29 & 33.75 & 25.91 & 96.74 & 26.44 & 5.15 & 27.26 \\
GOODS North & 39.23 & 25.28 & 57.11 & 25.77 & 76.6 & 26.56 & 0.0 & - \\
COSMOS & 48.21 & 25.35 & 147.89 & 25.74 & 5.88 & 26.23 & 0.0 & - \\
UDS & 56.82 & 25.46 & 141.38 & 25.95 & 3.95 & 26.28 & 0.0 & - \\
EGS & 13.36 & 25.43 & 164.66 & 26.06 & 26.65 & 26.29 & 0.0 & - \\
\hline
Total Area (Average) & 191.5 & 25.36 & 544.7 & 25.9 & 209.8 & 26.46 & 5.15 & 27.26 \\   
\enddata
\tablenotetext{a}{Area in arcmin$^2$}
\tablecomments{Summary of the estimated detection completeness levels in AB magnitudes for each of the five CANDELS fields and their corresponding sub-fields.}
\end{deluxetable*}

\subsection{Photometric redshifts}\label{merger-sec:photoz}
Photometric redshift (photo-$z$) estimates for all five fields are calculated following a variation of the method presented in \citet{Duncan:2017ul} and \citet{Duncan:2017wu}.
In summary, template-fitting estimates are calculated using the \textsc{eazy} photometric redshift code \citep{Brammer:2008gn} for three different template sets and incorporates zero-point offsets to the input fluxes and additional wavelength dependent errors \citep[we refer the reader to][for details]{Duncan:2017ul}.
Templates are fit to all available photometric bands in each field as outlined in Section~\ref{merger-sec:data}. 

Additional empirical estimates using a Gaussian process redshift code \citep[\textsc{GPz};][]{2016MNRAS.462..726A} are then estimated using a subset of the available photometric bands (further details discussed below).
Finally, after calibration of the individual redshift posteriors (Section~\ref{sec:pdf_calibration}), the four estimates are then combined in a robust statistical framework through a hierarchical Bayesian (HB) combination to produce a consensus redshift estimate.

For the GOODS North field, we also calculate an additional second set of photo-$z$ estimates incorporating the SHARDS medium-band photometry based using only template fitting.
The template fits for the GOODS North + SHARDS photometry are calculated using the default \textsc{eazy} template library.
To account for the spatial variation in filter wavelength intrinsic to the SHARDS photometry \citep[see][]{2013ApJ...762...46P}, the fitting for each source is done using its own unique set of filter response functions specific to the expected SHARDS filter central wavelengths at the source position.

\subsubsection{Luminosity priors in template fitting and HB combination}\label{merger-sec:priors}
When calculating the redshift posteriors for each template fit, we do not make use of a luminosity-dependent redshift prior as is commonly done to improve photometric redshift accuracy \citep{Brammer:2008gn,Dahlen:2013eu}, i.e. we assume a luminosity prior which is flat with redshift. 
Luminosity dependent priors such as the one implemented in \textsc{eazy} rely on mock galaxy lightcones which accurately reproduce the observed (apparent) luminosity function. 
Current semi-analytic models do agree well with observations at $z < 2$ \citep{Henriques:2012gsa}, but increasingly diverge at higher redshift \citep{Lu:2014kl} and may not represent an ideal prior.

Even in the case of an empirically calculated prior \citep[e.g.][]{Duncan:2017wu} that may not suffer from these limitations, the use of a prior which is dependent only on a galaxy's luminosity and not its color or wider SED properties could significantly bias the estimation of close pairs using redshift posteriors. 
As an example, we can imaging a hypothetical pair of galaxies at identical redshifts and with identical stellar population properties such that the only difference is the stellar mass of the galaxy (i.e. the star-formation histories differ only in normalization). 
If a luminosity-dependent prior is then applied, the posterior probability distribution for each galaxy will be modified differently for each galaxy and could erroneously decrease the integrated pair probability. 
%The effects of luminosity-based priors on redshift posteriors at $z < 3$ where SAMs are much better constrained are explored in further detail in \citet{mundy2017}.

\subsubsection{Gaussian process redshift estimates}
In addition to the primary template-based estimates outlined in the previous section, our consensus photo-$z$s also incorporate empirical photo-$z$ estimates based on the Gaussian process redshift code \textsc{GPz} \citep{2016MNRAS.455.2387A,2016MNRAS.462..726A}.
Our implementation of the \textsc{GPz} code in this work, includes magnitude and color-dependent weighting of the spectroscopic training sample, and follows the procedure outlined in \citet{Duncan:2017wu}, to which we refer the reader for additional details.
The spectroscopic training sample for each field was taken from a compilation of those available in the literature (Hathi, priv. communication), with additional spectroscopic quality cuts applied based on the quality flags provided by each survey.
To maximise the training sample available, we train \textsc{GPz} using only a subset of the available filters that are common to multiple fields: $V_{606}$, $I_{814}$, $J_{125}$, $H_{160}$ from HST (additionally $B_{435}$ for GOODS North and South) as well as the 3.6 and 4.5$\mu$m IRAC bands of \emph{Spitzer}.

In practice, the resulting \textsc{GPz} estimates have significantly higher scatter ($\sigma_{\textup{NMAD}} \approx 10\%$)\footnote{The normalized median absolute deviation is defined as $\sigma_{\text{NMAD}} = 1.48 \times \rm{median} \left (  \frac{\left | \Delta z \right |}{1+z_{\rm{spec}}} \right )$, see \citet{Dahlen:2013eu}.} and out-lier fraction ($\gtrsim 15\%$) than their corresponding template estimates.
Nevertheless, we include the \textsc{GPz} estimates within the Hierarchical Bayesian combination procedure as they can serve to break color degeneracies inherent within the template estimates in a more sophisticated manner than a simple luminosity prior (see Sec.~\ref{merger-sec:priors}).

\subsubsection{Calibrating redshift posteriors}\label{sec:pdf_calibration}
In \citet{Hildebrandt:2008jh}, \citet{Dahlen:2013eu} and more recently \citet{2016MNRAS.457.4005W} and \citet{Duncan:2017ul}, it is shown that the redshift probability density functions output by photometric redshift codes can often be an inaccurate representation of the true photometric redshift error. 
This inaccuracy can be due to under- or over-estimates of photometric errors, or a result of systematic effects such as the template choices. 
Whatever the cause, the effect can result in significantly over- or underestimated confidence intervals whilst still producing good agreement between the best-fit $z_{\rm{phot}}$ and the corresponding $z_{\rm{spec}}$. 
Although this systematic effect may be negated when measuring the bulk properties of larger galaxy samples, the method central to this paper relies on the direct comparison of individual redshift posteriors. 
It is therefore essential that the posterior distributions used in the analysis accurately represent the true uncertainties. 
Given this known systematic effect, we therefore endeavor to ensure the accuracy of our redshift posteriors before undertaking any analysis based on their posteriors. 

A key feature of the photo-$z$ method employed in this work is the calibration of the redshift posteriors for all estimates included in the Bayesian combination \citep{Duncan:2017ul,Duncan:2017wu}.
Crucially, this calibration is done as a function of apparent magnitude, rather than as a global correction, minimizing any systematic effects that could result from biases in the spectroscopic training sample.
An additional step in the calibration procedure introduced in this work is the correction of bias in the posteriors by shifting the posteriors until the Euclidean distance between the measured and optimum $\hat{F}(c)$ is minimized \citep{Gomes:2017ut}.
This additional correction is necessary due to the very high precision offered by the excellent photometry available in these fields (and the correspondingly low scatter in the resulting estimates) and prevents unnecessary inflation of the uncertainties to account for this bias during the subsequent calibration of the posterior widths.

\begin{figure*}
\centering
\includegraphics[width=0.8\textwidth]{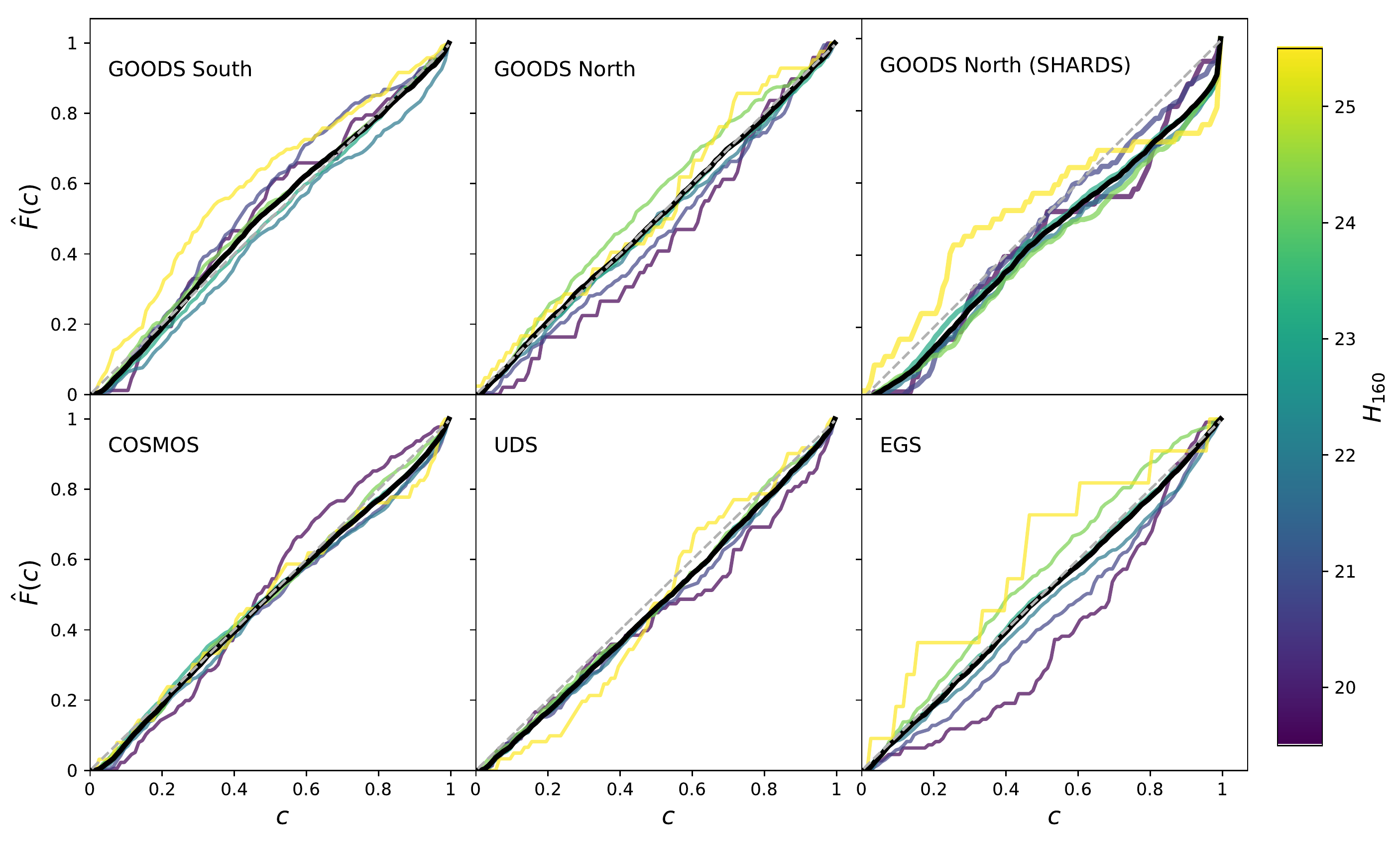}  
  \caption{Quantile-Quantile (Q-Q, or $\hat{F}(c)$, see text in Section~) plots for the final calibrated consensus redshift predictions for each of the CANDELS fields, plus the alternative GOODS North estimates incorporating the SHARDS medium band photometry. Colored lines represent the distributions in bins of apparent $H_{160}$ magnitude ($\pm0.5$ magnitudes), while the thick black line corresponds to the complete spectroscopic training sample. Lines that fall above the 1:1 relation illustrate under-confidence in the photo-$z$ uncertainties (uncertainties overestimated), while lines under illustrate over-confidence (uncertainties underestimated).}
  \label{merger-fig:pdf_calibration}
\end{figure*}

In Fig.~\ref{merger-fig:pdf_calibration} we present cumulative distribution, $\hat{F}(c)$, of threshold credible intervals, $c$, for our final consensus photo-$z$ estimate.
For a set of redshift posterior predictions which perfectly represent the redshift uncertainty, the expected distribution of threshold credible intervals should be constant between 0 and 1, and the cumulative distribution should therefore follow a straight 1:1 relation, i.e. a quantile-quantile plot.

If there is over-confidence in the photometric redshift errors, i.e the $P(z)$s are too sharp, the $\hat{F}(c)$ curves will fall below the ideal 1:1 relation.
Likewise, under-confidence results in curves above this line.
Remaining bias in the estimates can manifest as steeper or shallow gradients and offsets in the intercepts at $c = 0$ and $c = 1$.

From Fig.~\ref{merger-fig:pdf_calibration}, we can see that overall the accuracy of the photo-$z$ uncertainties is very high across a very broad range in apparent magnitude.
For the GOODS North + SHARDS estimates, there remains a small amount of over-confidence in the photo-$z$ uncertainties.
Additionally, for the EGS field there remains a magnitude dependent trend in the photo-$z$ posterior accuracy.
Uncertainties for bright sources are slightly under-estimated while those for faint sources are slightly over-estimated. 

\begin{figure}
{\centering
\includegraphics[width=1.02\columnwidth]{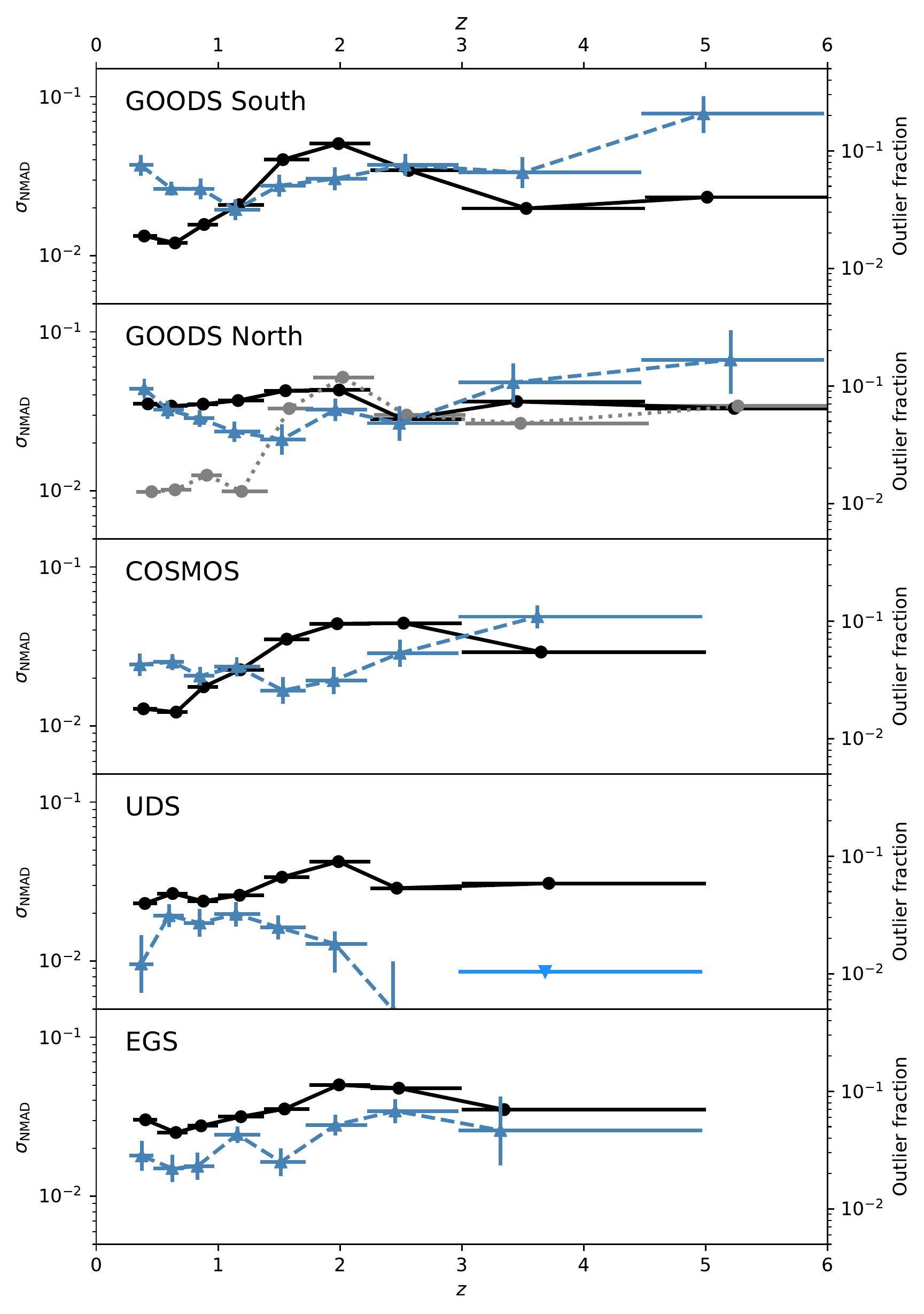}} 
  \caption{Robust scatter ($\sigma_{\textup{NMAD}}$: black circles, left-hand scale) and outlier fraction ($\frac{\left | \Delta z \right |}{1+z_{\rm{spec}}} > 0.15$, blue triangles, right-hand scale) for the galaxies in our samples with available high quality spectroscopic redshifts and with photometric redshift fits which pass our selection criteria. The position of the filled circle/triangle within each bin shows the average spectroscopic redshifts within that bin. Error-bars for the outlier fractions indicate the 1-$\sigma$ binomial uncertainties and lighter blue downward triangles indicate upper limits. For the GOODS North field, gray points (and dotted line) illustrate the scatter for the GOODS North + SHARDS redshift estimates.}
  \label{merger-fig:specz_comp1}
\end{figure}

\subsubsection{Photo-$z$ quality statistics}
In Fig.~\ref{merger-fig:specz_comp1} we illustrate the photometric redshift quality for each CANDELS field as a function of redshift. Following the same metrics as in \citet{Molino:2014iz} and \citetalias{LopezSanjuan:2014uj}, we find that the quality of our photometric redshifts is excellent given the high-redshifts being studied and the broadband nature of the photometry catalog. We find a normalized median absolute deviation of between $\sigma_{\text{NMAD}} \lesssim 1\%$ and $\sigma_{\text{NMAD}} \lesssim 5\%$, depending on redshift.

As with most spectroscopic redshift comparison samples, the typically bright nature of the galaxies with high quality spectroscopic redshift may present a biased representation of the quality of the photometric redshifts.  
We can see this effect in the comparison in Fig.~\ref{merger-fig:specz_comp1}, by comparing the different $\sigma_{\textup{NMAD}}$ values for the different fields.  
It may be initially surprising that we find poorer agreement between the photometric and spectroscopic redshifts (w.r.t outlier fraction) at $z>3$ for the GOODS North and South fields compared to EGS and UDS, given that these fields significantly deeper HST data available.  
In fact, it is the increased level of spectroscopic completeness at fainter magnitudes and higher redshifts that is the reason for the apparently poorer performance in GOODS fields, with spectroscopic redshifts for a greater number of sources for which photo-$z$ are more difficult to measure.

However, overall we are still getting good photometric redshifts for the fainter systems. 
The basis of our analysis is the full redshift posteriors for which we have high confidence in the accuracy and precision.

%TODO \emph{Discussion of including 3D-HST redshifts: increase in outlier fraction probably due to this...actually a fairer representation of photo-z accuracy but looks poor in comparison to previous sets. When matched with the same spectroscopic sample, the redshifts produced by \citet{Hsu:eu} show similar outlier fractions with only a slight improvement in quality compared to mine. Also compare to CANDELS team photo-z's, which exhibit only a slight improvement.} 

\subsection{Stellar mass estimates}\label{merger-sec:stellarmass}
The stellar mass as a function of redshift, $\textup{M}_{*} (z)$, for each galaxy is measured using a modified version of the SED code introduced in \citet{Duncan:2014gh}. Rather than estimating the best-fit mass (or mass likelihood distribution) for a fixed input photometric or spectroscopic redshift, we instead estimate the stellar mass at all redshifts in the photo-$z$ fitting range. Specifically, we calculate the least-squares weighted mean:
\begin{equation}
	\textup{M}_{*} (z) = \frac{\sum_{t}^{} w_{t}(z) \text{M}_{*,t}(z)} {\sum_{t} w_{t}(z)}
\end{equation}
where the sum is over all galaxy template types, $t$, with ages less than the age of the Universe at the redshift $z$, and $\textup{M}_{\star,t}(z)$ is the optimum stellar mass for each galaxy template (Equation~\ref{eq:temp_norm}). The weight, $w_{t}(z)$, is determined by
\begin{equation}
	w_{t}(z) = \exp(-\chi_{t}^{2}(z)/2),	
\end{equation}
where $\chi^{2}_{t}(z)$ is given by:
\begin{equation}\label{eqn:chi}
  \chi^{2}_{t}(z) = \sum_{j}^{N_{filters}} \frac{(\textup{M}_{\star,t}(z) F_{j,t}(z) - F_{j}^{obs})^2} {\sigma_{j}^{2}}.
\end{equation}
The sum is over $j$ broadband filters available for each galaxy, its observed photometric fluxes, $F_{j}^{\textup{obs}}$ and corresponding error, $\sigma_{j}$.
We note that due to computing limitations, we do not include the available medium-band photometry when estimating stellar masses.
The optimum scaling for each galaxy template type (normalized to 1 M$_{\odot}$), $\textup{M}_{\star,t}$, is calculated analytically by setting the differential of Equation~\ref{eqn:chi} equal to 0 and rearranging to give:
\begin{equation}\label{eq:temp_norm}
	\textup{M}_{\star,t}(z) = 
\frac{\sum_{j} \frac{F_{j,t}(z)F_{j}^{obs}}{\sigma^{2}_{j}} }  { \sum_{j} \frac{F_{j,t}(z)^{2}}{\sigma^{2}_{j}}  }.
\end{equation}
In this work we also incorporate a so-called ``template error function" to account for uncertainties caused by the limited template set and any potential systematic offsets as a function of wavelength. The template error function and method applied to our stellar mass fits is identical to that outlined in \citet{Brammer:2008gn} and included in the initial photometric redshift analysis outlined in Section~\ref{merger-sec:photoz}. Specifically, this means that the total error for any individual filter, $j$, is given by:
\begin{equation}
	\sigma_{j} = \sqrt{\sigma_{j,obs}^{2} + \left ( F_{j,obs}\sigma_{\textup{temp}}(\lambda_{j}) \right )^{2} }
\end{equation}
where $\sigma_{j,obs}$ is observed photometric flux error, $F_{j,obs}$ its corresponding flux and $\sigma_{\textup{temp}}(\lambda_{j})$ the template error function interpolated at the pivot wavelength for that filter, $\lambda_{j}$. 

We note that in addition to estimating the stellar mass, this method also provides a secondary measurement of the photometric redshift, whereby \(P(z) \propto \sum_{t} w_{t}(z) \). 
We use an independently estimated redshift posterior in the pair analysis in place of those generated by the marginalised redshift likelihoods from the stellar mass fits due to the higher precision and reliability offered by our hierarchical Bayesian consensus photo-$z$ estimates.

For the \citet{Bruzual:2003ckb} templates used in our stellar mass fitting we allow a wide range of plausible stellar population parameters and assume a \citet{Chabrier:2003ki} IMF. Model ages are allowed to vary from 10 Myr to the age of the Universe at a given redshift, metallicities of 0.02, 0.2 and 1 Z$_{\odot}$, and dust attenuation strength in the range $0 \le A_{V} \le 3$ assuming a \citet{2000ApJ...533..682C} attenuation curve. The assumed star-formation histories follow exponential $\tau$-models ($SFR \propto e^{-t/\tau}$), both decreasing and increasing (negative $\tau$), for characteristic timescales of $\left | \tau \right | = $ 0.25, 0.5, 1, 2.5, 5, 10, plus an additional short burst ($\tau = 0.05$) and continuous star-formation models ($\tau \gg1/H_{0}$). 

Nebular emission is included in the model SEDs assuming a relatively high escape fraction $f_{\text{esc}} = 0.2$ \citep{Yajima:2010fb,Fernandez:2011cw,Finkelstein:2012hr,Robertson:2013ji} and hence a relatively conservative estimate on the contribution of nebular emission. As in \citet{Duncan:2014gh}, we assume for the nebular emission that the gas-phase stellar metallicities are equivalent and that stellar and nebular emission are attenuated by dust equally.

To ensure that our stellar mass estimates do not suffer from significant systematic biases we compare our best-fitting stellar masses (assuming $z = z_{\text{peak}}$) with those obtained by averaging the results of several teams within the CANDELS collaboration \citep{Santini:2015hh}. 
Although there is some scatter between the two sets of mass estimates, we find that our best-fitting masses suffer from no significant bias relative to the median of the CANDELS estimates (see Fig.~\ref{merger-fig:mass_candels} in the Appendix). 
Some of the observed scatter can be attributed to the fact that the photometric redshift assumed for the two sets of mass estimates is not necessarily the same.
Overall, we are therefore confident that the stellar population modelling employed here is consistent with that of the wider literature.
We find no systematic error relative to other mass estimates which make use of stellar models and assume the same IMF.
However, standard caveats with regards to stellar masses estimated using stellar population models still apply (see discussion in \citealt{Santini:2015hh}).

\section{Close pair methodology}\label{merger-sec:method}
The primary goal of analysing the statistics of close pairs of galaxies is to estimate the fraction of galaxies which are in the process of merging. From numerical simulations such as \citet{Kitzbichler:2008gi}, it is well understood that the vast majority of galaxy dark matter halos within some given physical separation will eventually merge. 
For spectroscopic studies in the nearby Universe, a close pair is often defined by a projected separation, \(r_{\textup{p}} \), in the plane of the sky of \(r_{\textup{p}} < 20~\rm{to}~50~ h^{-1}\) kpc, and a separation in redshift or velocity space of \( \Delta v \leq 500 \textup{ km s}^{-1}\).

Armed with a measure of the statistics of galaxies that satisfy these criteria within a sample, we can then estimate the corresponding pair fraction, $f_{\textup{P}}$, defined as
\begin{equation}
	f_{\textup{P}} = \frac{N_{\textup{pairs}}} {N_{\textup{T}}},
\end{equation}
where \(N_{\textup{pairs}}\) and \(N_{\textup{T}}\) are the number of galaxy pairs and the total number of galaxies respectively within some target sample, e.g. a volume limited sample of mass selected galaxies. 
Note that $N_{\text{pairs}}$ is the number of galaxy pairs rather then number of galaxies \emph{in} pairs which is up to factor of two higher \citep{Patton:2000kt}, depending on the precise multiplicity of pairs and groups.

%A method of analysing pair-counts using photometric surveys is clearly required if we wish to study the merger rates of galaxies in the next generation of wide-area or high redshift surveys. 
In this work, we analyse the galaxy close pairs through the use of their photo-$z$ posteriors. 
The use of photo-$z$ posterior takes into account the uncertainty in galaxy redshifts in the pair selection, and the effect of the redshift uncertainty on the projected distance and derived galaxy properties. 
As presented in \citetalias{LopezSanjuan:2014uj} this method is able to directly account for random line-of-sight projections that are typically subtracted from pair-counts through Monte Carlo simulations.
In the following section we outline the method as applied in this work and how it differs to that presented in \citetalias{LopezSanjuan:2014uj} in the use of stellar mass instead of luminosity when defining the close pair selection criteria, as well as our use of flux-limited samples and the corresponding corrections.

\subsection{Sample cleaning}
Before defining a target-sample, we first clean the photometric catalogs for sources that have a high likelihood of being stars or image artefacts. 

A common method for identifying stars in imaging is though optical morphology of the sources in the high-resolution HST imaging.
The exclusion of objects with high \textsc{SExtractor} stellarity parameters (i.e. more point-like sources) could potentially bias the selection by erroneously excluding very compact neighbouring galaxies and AGN instead of stars. 
Therefore, when cleaning the full photometric catalog to produce a robust sample of galaxies, we define stars as sources that have a high \textsc{SExtractor} stellarity parameter ($> 0.9$)  in the $H_{160}$ imaging \emph{and} have an SED that is consistent with being a star. 

Using \textsc{eazy}, we fit the available optical to near-infrared photometry (with rest-frame wavelength $< 2.5\mu$m) for each field with the stellar library of \citet{Pickles:1998er} while fixing the redshift to zero. 
We then classify as a star any object which has $\chi^{2}_{\text{Star}}/N_{\textup{filt,S}} < \chi^{2}_{\text{Galaxy}}/N_{\textup{filt,G}}$, where $\chi^{2}_{\text{Galaxy}}$ and $\chi^{2}_{\text{Star}}$ are the best-fit $\chi^{2}$ obtained when fitting the galaxy templates used in Section~\ref{merger-sec:photoz} and stellar templates respectively, normalized by the corresponding number of filters used in the fitting ($N_{\textup{filt,G}}$, $N_{\textup{filt,S}}$). 
Based on the combined classification criteria, we exclude $\lesssim 0.4\%$ of objects per field. 
%Of this fraction, XX are excluded due to their stellarity, YY due to their SEDs and ZZ satisfy both criteria.
Thus, the fraction of sources excluded by this criterion is very small so should not present a significant bias in the following analysis. 

Additionally, to prevent erroneous SED fits (either photo-$z$ or stellar mass estimates) in sources with photometry contaminated by artefacts due to bright stars in the field (and their diffraction spikes) or edge effects, we also exclude sources which have flags in the photometry flag map \citep[see e.g.][]{Guo:2013ig,Galametz:2013dd}.
Based on inspection of the photo-$z$ quality for all of the sources identified in this initial cut we find the published catalog flags to be overly conservative, with the overall quality of the photo-$z$ for flagged sources comparable to those of un-flagged objects.
To exclude only objects for which the photometric artefacts will adversely affect the results in this work, we apply an additional selection criteria: excluding sources which are flagged and have $\chi^{2}_{\text{Galaxy}}/N_{\textup{filt,G}} > 4$, indicative of bad SED fits. 
Given these criteria, we exclude between 0.71\% and 3.3\% of sources in each field.

\subsection{Selecting initial potential close pairs}\label{merger-sec:initial}

%In the following paragraph we give a brief description of our methodology with the details presented in the subseqent sections. 
Once an initial sample has been selected based on redshift (see Section~\ref{merger-sec:photoz}), we then search for projected close pairs between the target and full galaxy samples. 
The initial search is for close pairs which have a projected separation less than the maximum angular separation across the full redshift range of interest (corresponding to the desired physical separation). 
Duplicates are then removed from the initial list of close pairs (with the primary galaxy determined as the galaxy with the highest stellar mass at its corresponding best-fit photo-$z$) to create the list of galaxy pairs for the posterior analysis. 
Because the posterior analysis makes use of all available information to determine the pair fractions, it is applied to all galaxies within the initial sample simultaneously, with the redshift and mass ranges of interest determined by the selection functions and integration limits outlined in the following sections.

\subsection{The pair probability function}\label{merger-sec:ppf}
\begin{figure}
{\centering
\includegraphics[width=\columnwidth]{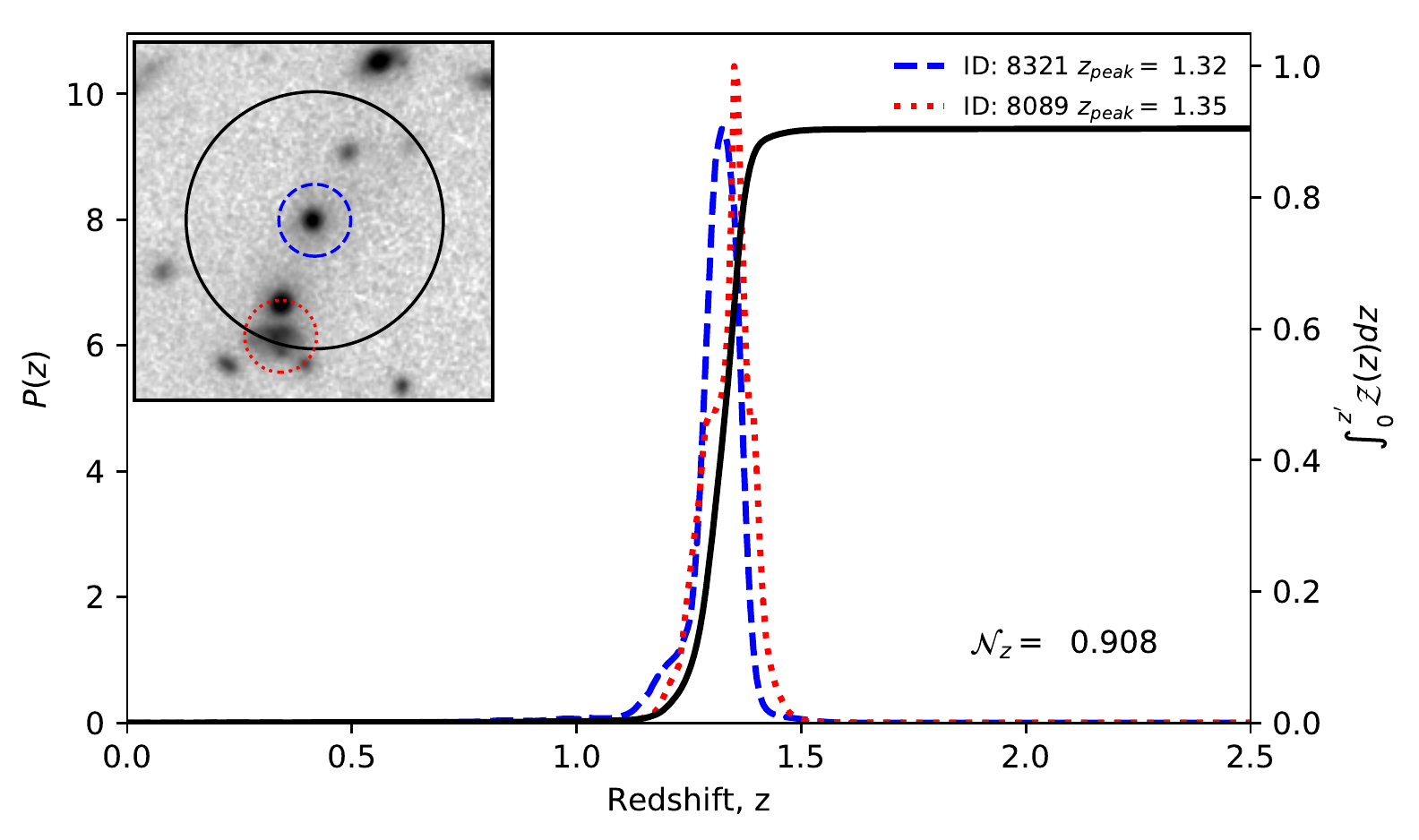}
\includegraphics[width=\columnwidth]{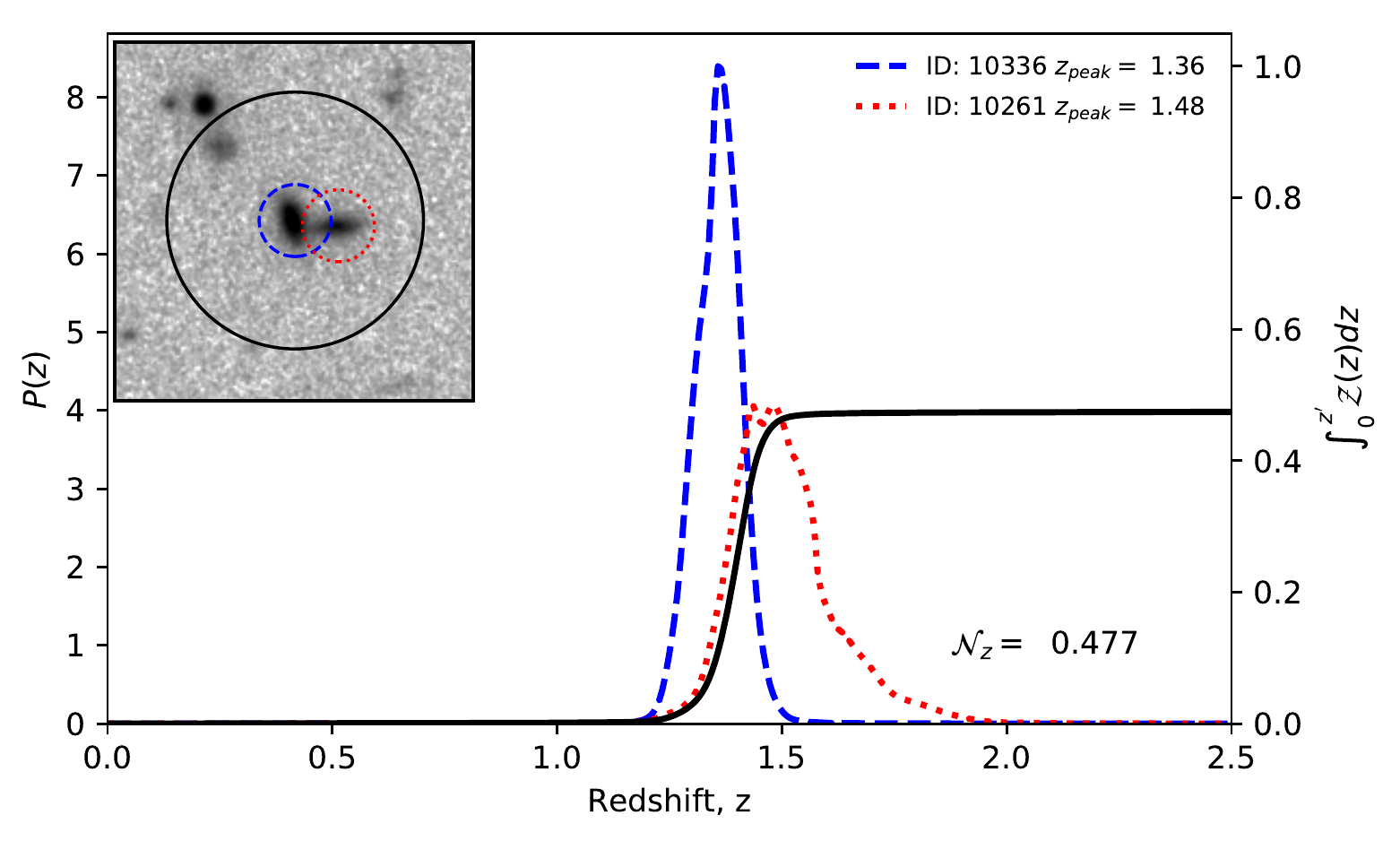}
\includegraphics[width=\columnwidth]{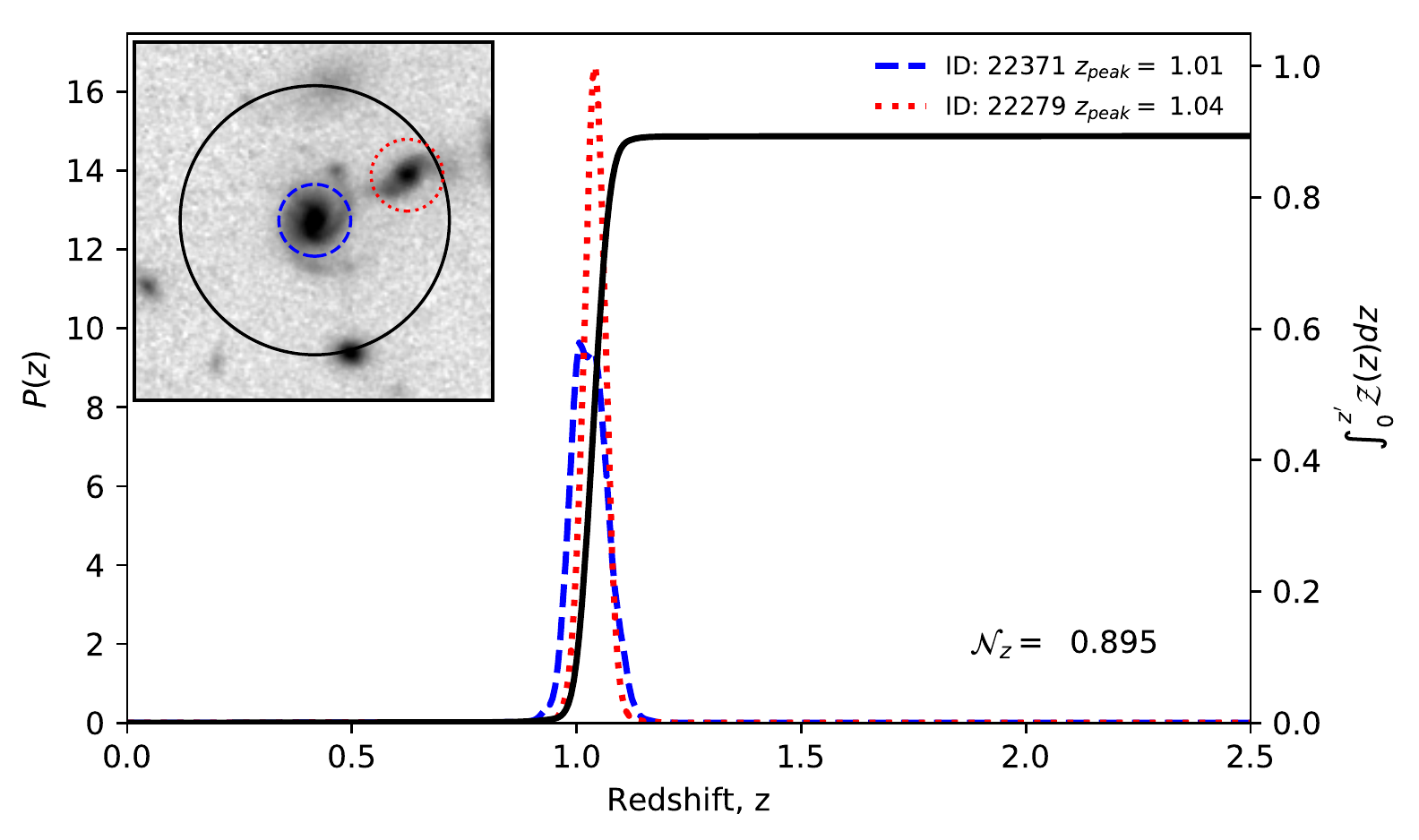}
}
\caption{Example redshift posteriors and integrated $\mathcal{Z}(z)$ for three projected pairs within the DEEP region of the GOODS South fields. In all panels, the blue dashed line corresponds to the redshift PDF for the primary galaxy, while the red dotted line is that of the projected companion. The solid black line shows the cumulative integrated $\mathcal{Z}(z)$ for the galaxy pair. Inset cutouts show the $H_{160}$ image centered on the primary galaxy (with $\arcsinh$ scaling), with the primary and secondary galaxies to match their corresponding $P(z)$. The black circle illustrates the maximum pair search radius at the peak of the primary galaxy $P(z)$.}
  \label{merger-fig:pairs_pz}
\end{figure}

For a given projected close pair of galaxies within the full galaxy sample, the combined redshift probability function, $\mathcal{Z}(z)$, is defined as
\begin{equation}\label{eq:Zz}
\mathcal{Z}(z) = \frac{2 \times P_{1}(z) \times P_{2}(z)}{P_{1}(z) + P_{2}(z)} =
\frac{P_{1}(z) \times P_{2}(z)}{N(z)}
\end{equation}
where $P_{1}(z)$ and $P_{2}(z)$ are the photo-$z$ posteriors for the primary and secondary galaxies in the projected pair.
The normalization, \(N(z) = (P_{1}(z) + P_{2}(z))/2\), is implicitly constructed such that \(\int_{0}^{\infty} N(z) dz = 1 \) and $\mathcal{Z}(z)$ therefore represents the number of fractional close pairs at redshift $z$ for the projected close pairs being studied. 
Following Equation~\ref{eq:Zz}, when either $P_{1}(z)$ or $P_{2}(z)$ is equal to zero, the combined probability $\mathcal{Z}(z)$ also goes to zero. 
This can be seen visually for the example galaxy pairs in Fig.~\ref{merger-fig:pairs_pz} (black line). 
The total number of fractional pairs for a given system is then given by
\begin{equation}
	\mathcal{N}_{z} = \int_{0}^{\infty} \mathcal{Z}(z) dz.
\end{equation}
and can range between 0 and 1. As each initial target galaxy can have more than one close companion, each potential galaxy pair is analysed separately and included in the total pair count. Note that because the initial list of projected pairs is cleaned for duplicates before analysing the redshift posteriors, if the two galaxies in a system (with redshift posteriors of $P_{1}(z)$ and $P_{2}(z)$) both satisfy the primary galaxy selection function, the number of pairs is not doubly counted. 

%TODO Change plots to: One figure showing P(z) for all three galaxies. Second figure showing Z(z) as a function of redshift for each of the pairs (with colors matching the P(z) lines above) - text then shows the integrated N(z).

In Fig.~\ref{merger-fig:pairs_pz} we show three examples of projected pairs within the DEEP region of CANDELS GOODS South that satisfy the selection criteria applied in this work (Section~\ref{merger-sec:results}). 
Two of the the pairs have a high probability of being a real pair within the redshift range of interest ($\mathcal{N}>0.8$) while the third pair (middle panel) has only a partial chance of being at the same redshift.

\subsubsection{Validating photometric line-of-sight probabilities with spectroscopic pairs}\label{merger-sec:specz_pairs}
Due to the relatively high spectroscopic completeness within the CANDELS GOODS-S field thanks to deep surveys such as the MUSE UDF and WIDE surveys \citep[][respectively]{Bacon:2015eh,2018arXiv181106549U}, precise spectroscopic redshifts are available for a number of close projected pairs within the field.
Calculating a mass-selected pair-fraction based on spectroscopic pairs is beyond the scope of this work due to the corrections required for the complicated spectroscopic selection functions.
However, the sample of available spectroscopic pairs does allow us to test the reliability of the photo-$z$ based line-of-sight pair probabilities ($\mathcal{N}_{z}$).

\begin{figure}
\centering
    \includegraphics[width=0.95\columnwidth]{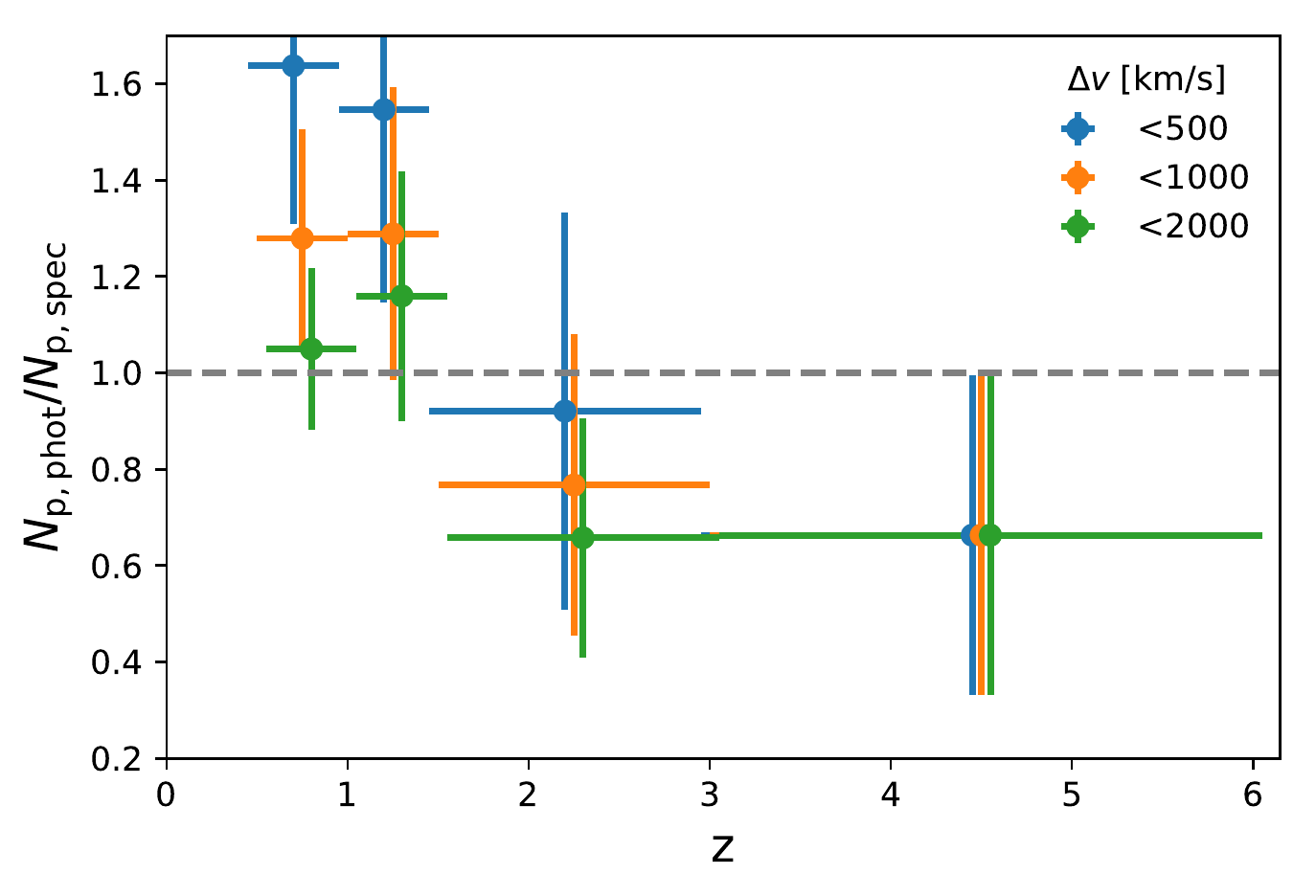}
  \caption{Ratio of total integrated photo-$z$ pairs ($\sum_{i}\mathcal{N}_{z,i}$) to total number of spectroscopic pairs as a function of velocity separation ($\Delta v$) and redshift for projected close pairs within the CANDELS GOODS-S spectroscopic sample. Datapoints for different velocity cuts are offset in redshift for clarity.}
  \label{merger-fig:spec_pairs}
\end{figure}

After applying a magnitude cut based on the GOODS South completeness limits and a stellar-mass cut on the primary galaxy of $9.7 < \log_{10}(\Mstar / \text{M}_{\odot})$, we find all potential pairs by searching for other galaxies with spectroscopic redshifts within 30 kpc of each primary galaxy. 
For each of these potential pairs, we then calculate the integrated number of photo-$z$ pairs, $\mathcal{N}_{z} = \int_{z_{\textup{min}}}^{z_{\textup{max}}} \mathcal{Z}(z) dz$, in four redshift bins from $z=0.5$ to $z=6$.
Figure~\ref{merger-fig:spec_pairs} shows how the number of integrated photo-$z$ pairs compares to the number of spectroscopic pairs after applying different cuts on velocity separation. 
We find that the integrated number of photo-$z$ pairs is comparable to the spectroscopic pair counts with velocity separations of up to $<2000~\textup{km~s}^{-1}$ at all redshifts.

At low redshift the photo-$z$ pair probabilities over-estimate the number of pairs at separations of $<500~\textup{km~s}^{-1}$, the typical definition used in spectroscopic pair fraction studies, by $\approx 50\%$.
However, above $z > 1.5$ we find that the photo-$z$ pairs are fully consistent with the spectroscopic definition within the uncertainties.
In Section~\ref{merger-sec:results} and ~\ref{merger-sec:discussion} we will discuss how the redshift dependence observed in Figure~\ref{merger-fig:spec_pairs} on our final results and the conclusions drawn.
The cause of the redshift dependency observed in Figure~\ref{merger-fig:spec_pairs} is not immediately clear.
Naively, we would expect the increased photo-$z$ scatter/outlier fraction at high redshift to result in the photo-$z$ measurements probing broader velocity offsets.
For now, we note that the photo-$z$ pair probabilities are able to effectively probe velocity separations that are a factor of 
$\approx 3-12\times$ smaller than the scatter within photo-$z$s themselves ($\Delta v = 500~\textup{km~s}^{-1} \approx 0.0017\times(1+z)$) - illustrating the power of the statistical pair count approach.

\subsubsection{Incorporating physical separation and stellar mass criteria}
The combined redshift probability function defined in Equation~\ref{eq:Zz} ($\mathcal{Z}(z)$) takes into account only the line-of-sight information for the potential galaxy pair, therefore two additional redshift dependent masks are required to enforce the remaining desired pair selection criteria.
These masks are binary masks, equal to one at a given redshift if the selection criteria are satisfied and zero otherwise. 
As above, we follow the notation outlined in \citetalias{LopezSanjuan:2014uj} and define the angular separation mask, $\mathcal{M}^{\theta}(z)$, as
\begin{equation}
\mathcal{M}^{\theta}(z) =  
\begin{cases}
 1, & \text{ if } \theta_{\textup{min}}(z) \leq \theta \leq \theta_{\textup{max}}(z)\\ 
 0, & \text{ otherwise, }
\end{cases}, 
\end{equation}
where the angular separation between the galaxies in a pair as a function of redshift is denoted $\theta (z)$. The angular separation is a function of the projected distance $r_{p}$ and the angular diameter distance, $d_{A}(z)$, for a given redshift and cosmology, i.e. $\theta_{\textup{max}}(z) = r^{\textup{max}}_{p} / d_{A}(z)$ and $\theta_{\textup{min}}(z) = r^{\textup{min}}_{p} / d_{A}(z)$.

The pair selection mask, denoted as $\mathcal{M}^{\rm{pair}}(z)$, is where our method differs to that outlined by \citetalias{LopezSanjuan:2014uj}. 
Rather than selecting galaxy pairs based on the luminosity ratio, we instead select based on the estimated stellar mass ratio. 
We define our pair-selection mask as
\begin{equation}\label{eq:sel_mask}
\mathcal{M}^{\rm{pair}}(z) = 
\begin{cases}
1, & \text{ if } \Mstar^{\rm{lim},1}(z) \leq \rm{M}_{\star,1}(z) \leq \rm{M}_{\star,max}  \\ 
	& \text{ and }~ \Mstar^{\rm{lim},2}(z) \leq \rm{M}_{\star,2}(z) \\
0, & \text{ otherwise. } 
\end{cases}
\end{equation}
where $\rm{M}_{\star,1}(z)$ and $\rm{M}_{\star,2}(z)$ are the stellar mass as a function of redshift, details of how $\rm{M}_{\star}(z)$ is calculated for each galaxies are discussed in Section~\ref{merger-sec:stellarmass}. The flux-limited mass cuts, $\Mstar^{\rm{lim},1}(z)$ and $\Mstar^{\rm{lim},2}(z)$, are given by
\begin{equation}\label{eq:fluxlim_pri}
\Mstar^{\rm{lim},1}(z)= max \{ \Mstar^{\rm{min}}, \Mstar^{\rm{flux}}(z) \}
\end{equation}
and
\begin{equation}\label{eq:fluxlim_sec}
\Mstar^{\rm{lim},2}(z) = max \{ \mu \Mstar^{1}(z), \Mstar^{\rm{flux}}(z) \}
\end{equation}
respectively, where $\Mstar^{\rm{flux}}(z)$ is the redshift-dependent mass completeness limit outlined in Section~\ref{merger-sec:weights_flux} and $\Mstar^{min}$ and $\Mstar^{max}$ are the lower and upper ranges of our target sample of interest. 
The mass ratio $\mu$ is typically defined as $\mu > 1/4$ for major mergers and $1/10 < \mu < 1/4$ for minor mergers. 
Throughout this work we set $\mu = 1/4$ by default, unless otherwise stated. 

The pair selection mask ensures the following criteria are met at each redshift: firstly, it ensures the primary galaxy is within the mass range of interest. 
Secondly, that the mass ratio between the primary and secondary galaxy is within the desired range (e.g. for selecting major or minor mergers). 
Finally, that both the primary and secondary galaxy are above the mass completeness limit at the corresponding redshift. 
We note that the first criteria of Equation~\ref{eq:sel_mask} also constitutes the selection function for the primary sample, given by
\begin{equation}\label{eq:pri_sel}
S(z) = 
\begin{cases}
1, & \text{ if } \Mstar^{\rm{lim},1}(z) \leq \rm{M}_{\star,1}(z) \leq \rm{M}_{\star,max}  \\ 
0, & \text{ otherwise. } 
\end{cases}
\end{equation}

\begin{figure}
{\centering
\includegraphics[width=\columnwidth]{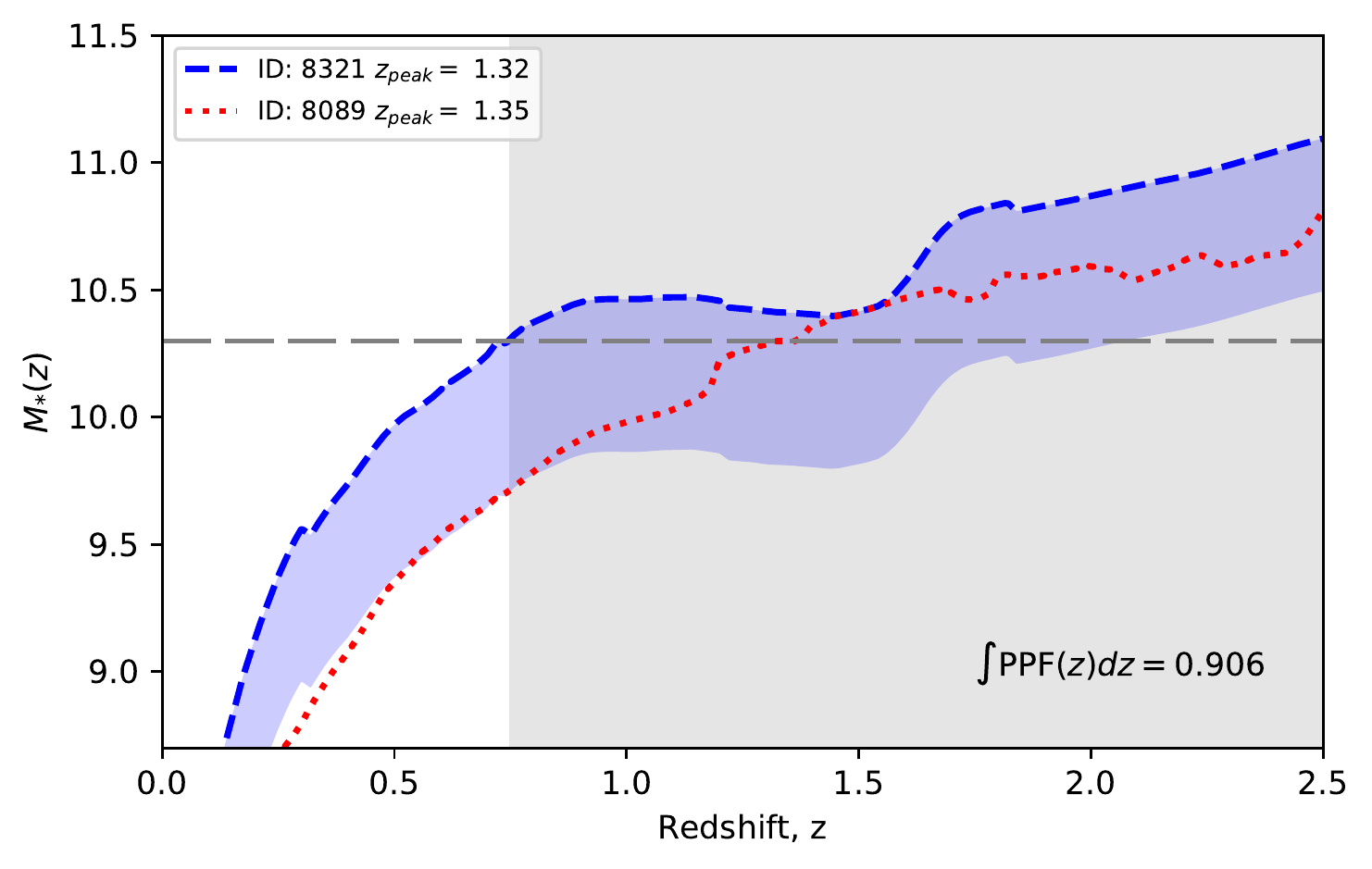}
\includegraphics[width=\columnwidth]{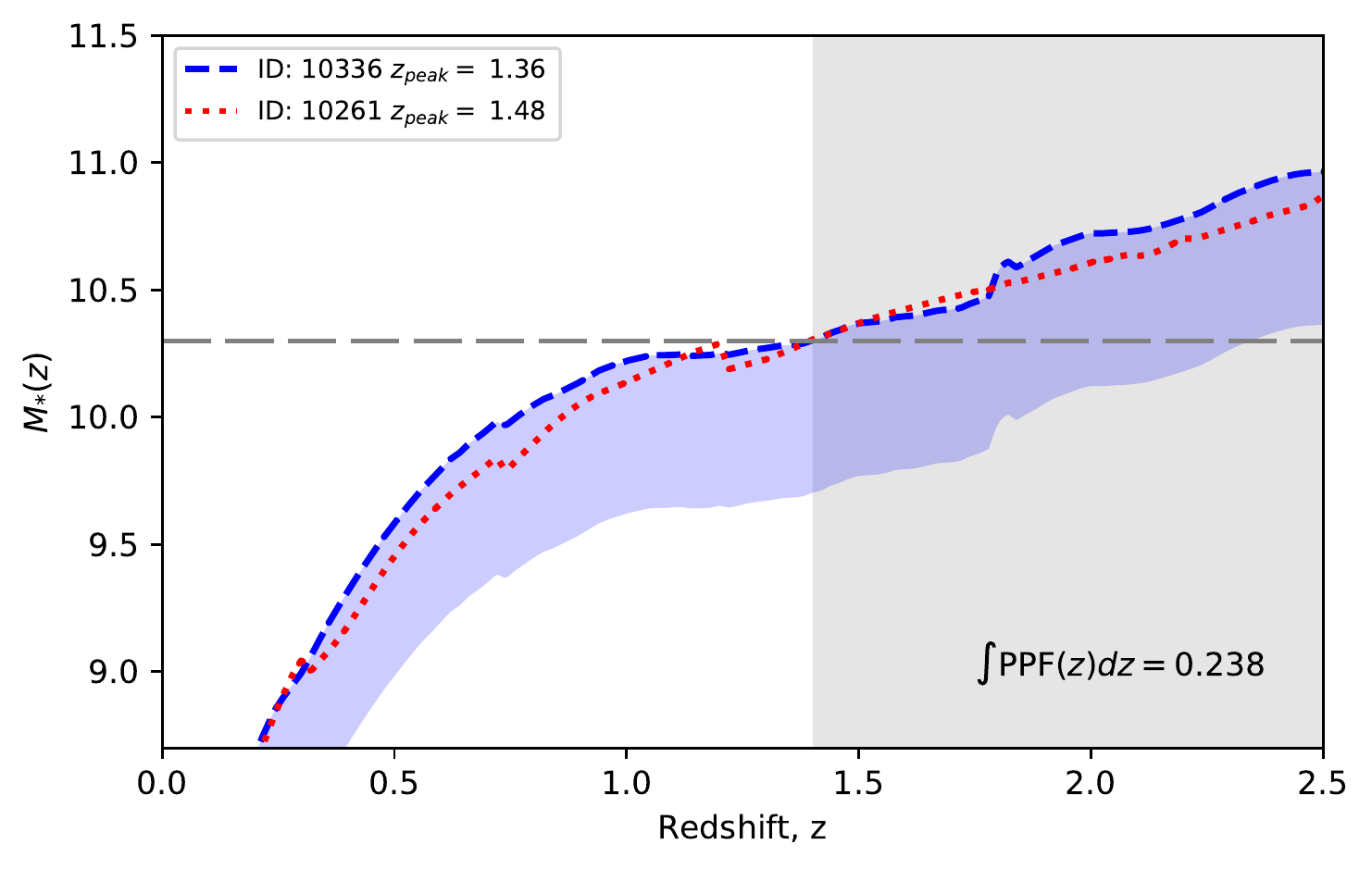}
\includegraphics[width=\columnwidth]{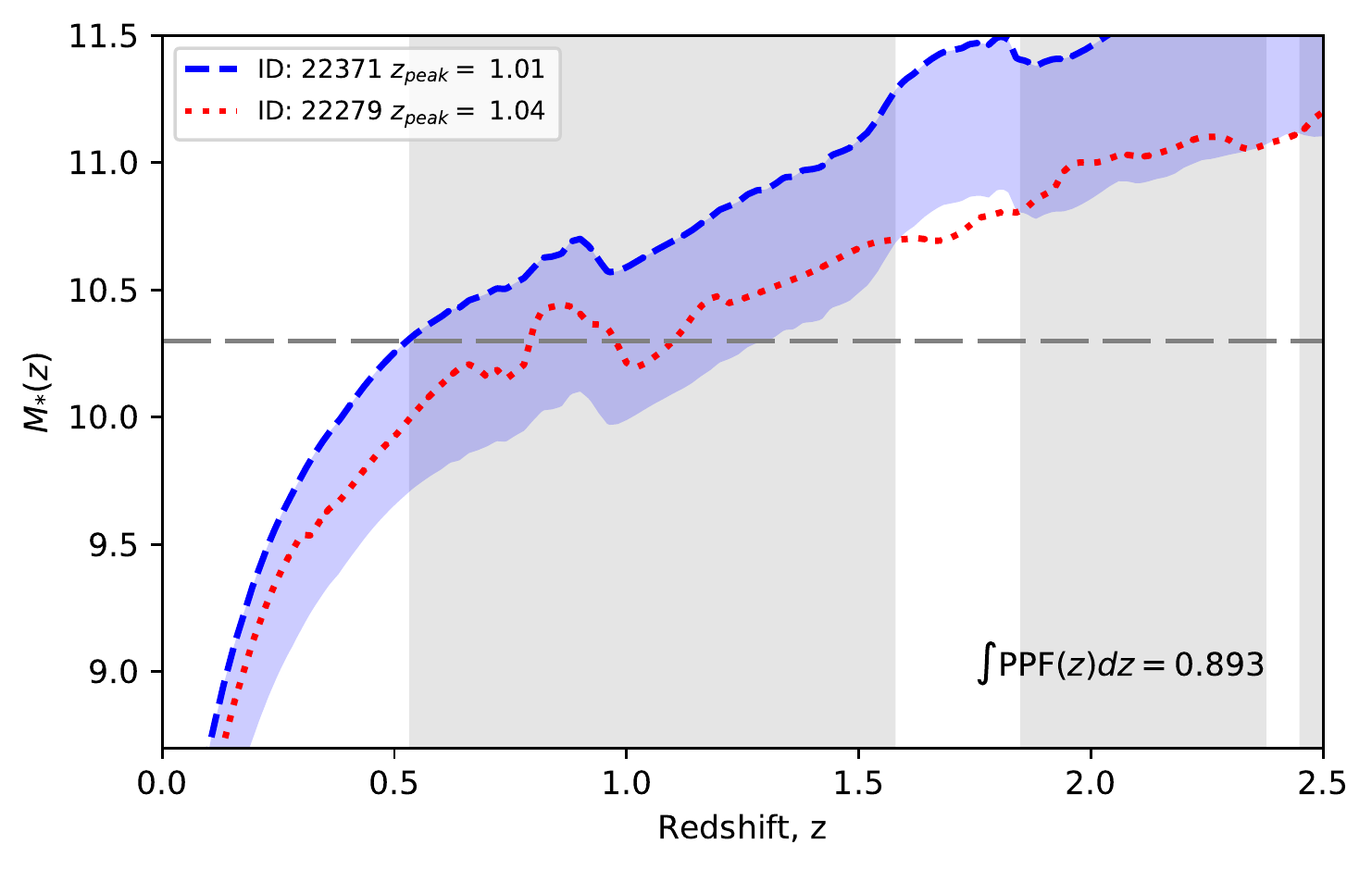}}
\caption{Redshift-dependent stellar mass estimations for the example close pairs shown in Fig.~\ref{merger-fig:pairs_pz}. In all panels the blue dashed line corresponds to the stellar mass for the primary galaxy, while the red dotted line is that of the projected companion. The blue shaded regions illustrate the range of secondary galaxy masses that satisfy the selected merger ratio criteria - $\mu > 1/4$. Dashed gray lines indicate the stellar mass selections applied in this study.}
  \label{merger-fig:pairs_mass}
\end{figure}

With these three properties in hand for each potential companion galaxy around our primary target, the pair-probability function, $\textup{PPF}(z)$, is then given by
\begin{equation}\label{eq:PPF}
\textup{PPF}(z) = \mathcal{Z}(z) \times \mathcal{M}^{\theta}(z) \times \mathcal{M}^{\text{pair}}(z).
\end{equation}

In Fig.~\ref{merger-fig:pairs_mass}, we show the estimated stellar mass as a function of redshift for the three example projected pairs shown in Fig.~\ref{merger-fig:pairs_pz}. 
Additionally, the redshift ranges where all three additional pair selection criteria are shown by the gray shaded region.
For the first and third galaxy pairs with high probability of being a pair along the line-of-sight, the separation criteria and mass selection criteria are also satisfied at the relevant redshift. 
In contrast, the second potential pair (with $\mathcal{N}_{z} = 0.477$) does not satisfy the stellar mass criteria at all redshifts of interest and therefore has a significantly reduced final pair-probability of $\int_{0}^{\infty} \textup{PPF}(z) dz = 0.238$.

In Section~\ref{merger-sec:pair_frac} we outline how these individual pair-probability functions are combined to determine the overall pair-fraction, but first we outline the steps taken to correct for selection effects within the data.

\subsection{Correction for selection effects}
As defined by \citetalias{LopezSanjuan:2014uj}, the pair-probability function in Equation~\ref{eq:PPF} is affected by two selection effects. Firstly, the incompleteness in search area around galaxies that are near the image boundaries or near areas affected by bright stars (Section~\ref{merger-sec:weights_area}). And secondly, the selection in photometric redshift quality (Section~\ref{merger-sec:weights_osr}). In addition, because in this work we use a flux-limited sample rather than one that is volume limited (as used by \citetalias{LopezSanjuan:2014uj}), we must also include a further correction to account for this fact.

\subsubsection{The redshift-dependent mass completeness limit}\label{merger-sec:weights_flux}
Since the photometric survey we are using includes regions of different depth and high-redshift galaxies are by their very nature quite faint, restricting our analysis to a volume-limited sample would necessitate excluding the vast majority of the available data. As such, we choose to use a redshift-dependent mass completeness limit determined by the flux limit determined by the survey. 

Due to the limited number of galaxy sources available, determining the strict mass completeness continuously as a function of redshift entirely empirically \citep{Pozzetti:2010gw} is not possible. Instead, we make use of a method based on that of \citet{Pozzetti:2010gw}, using the available observed stellar mass estimates to fit a functional form for the evolving 95\% stellar mass-to-light limit. 

Following \citet{Pozzetti:2010gw}, the binned empirical mass limit is determined by selecting galaxies which are within a given redshift bin, then scaling the masses of the faintest 20\% such that their apparent magnitude is equal to the flux limit. 
The mass completeness limit for a given redshift bin is then defined as the mass corresponding to the 95th percentile of the scaled mass range. 
To accurately cover the full redshift range of interest, we apply this method to two separate sets of stellar mass measurements.
Firstly at $z\leq 4$ we use the best-fitting stellar masses estimated for each of the CANDELS photometry catalogs used in this work. 
Secondly, at $z \geq 3.5$ we make use of the full set of high-redshift Monte Carlo samples of \citet{Duncan:2014gh} to provide improved statistics and incorporate the significant effects of redshift uncertainty on the mass estimates in this regime.  

\begin{figure}
{\centering
	\includegraphics[width=0.95\columnwidth]{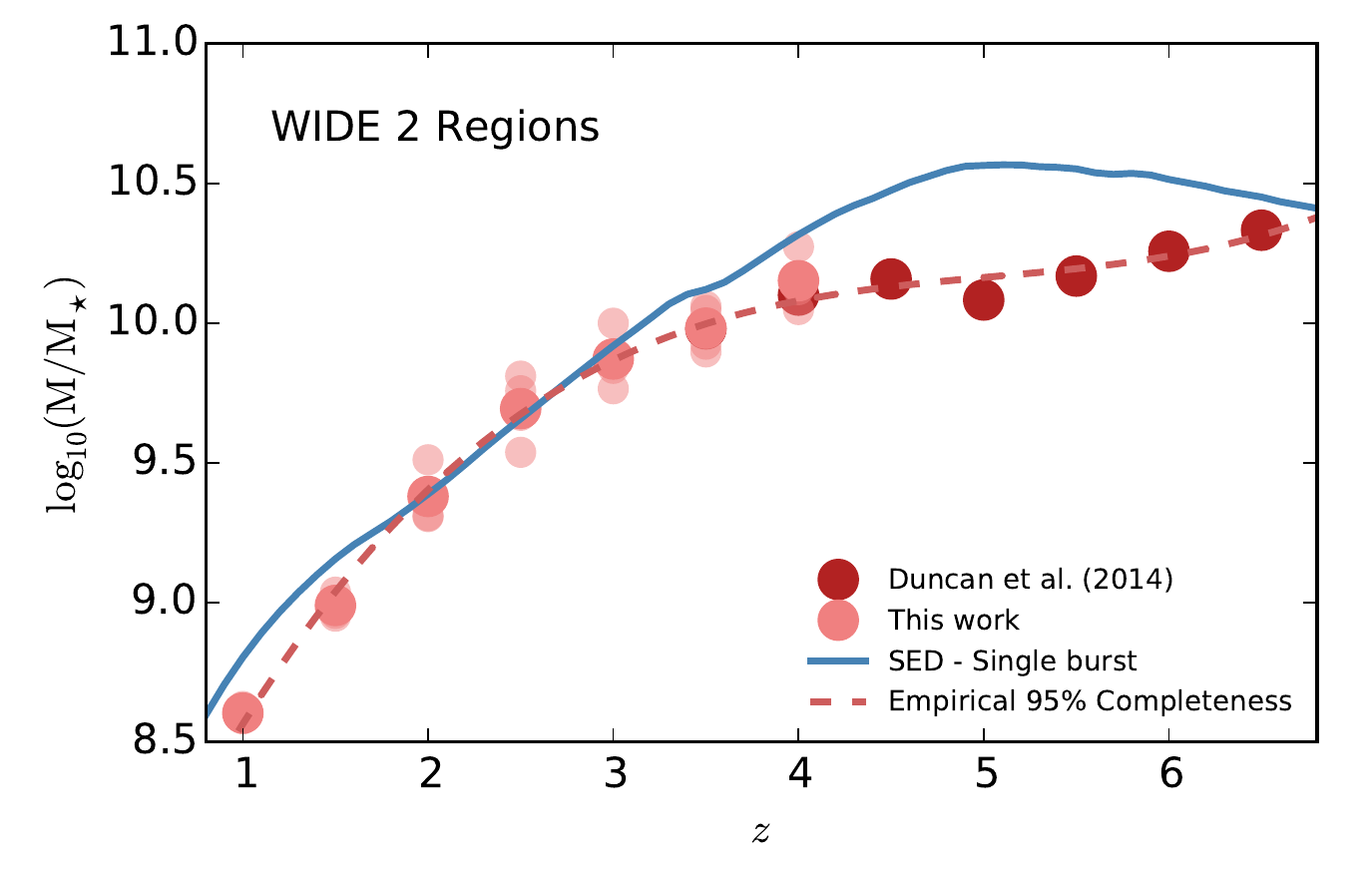}
	\includegraphics[width=0.95\columnwidth]{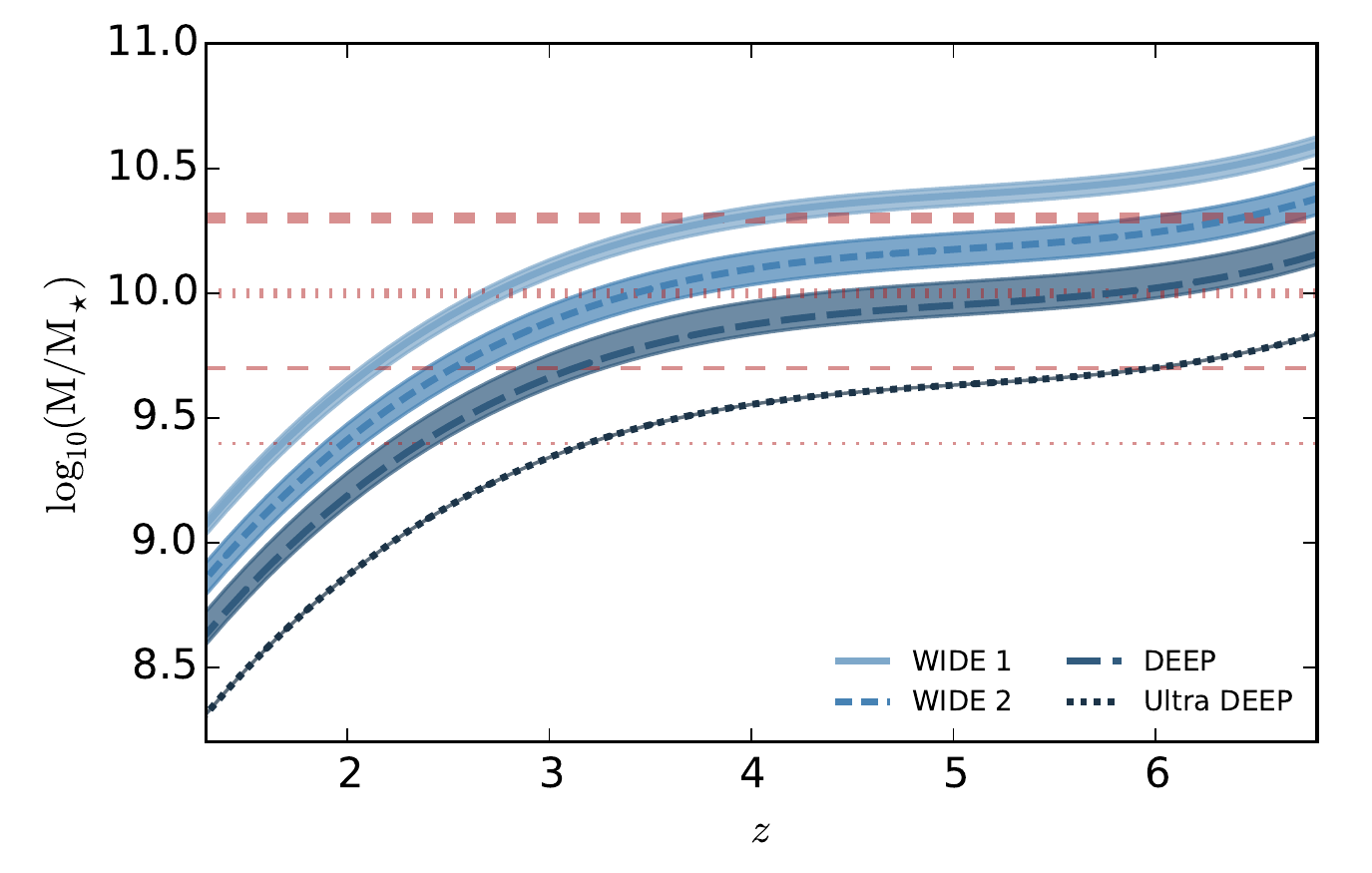}}
  \caption{\textit{Top:} Mass completeness limit corresponding to the flux limits of the WIDE 2 depth sub-fields in the CANDELS survey. Dark red circles correspond to the 95\% completeness limits at $z \geq 3.5$ derived from the stellar mass estimates of \citet{Duncan:2014gh}, lighter red circles show the equivalent estimates for the stellar mass estimates of this work for all five fields (smaller circles show estimates for individual fields). The continuous blue line shows the completeness limits corresponding to a maximally old (at a given redshift) single burst stellar population. The functional form (3rd order polynomial) fitted to the empirical mass completeness estimates is shown by the dashed red line. \textit{Bottom:} Estimated mass completeness limits for each of the sub-field depths: the functional form for the $95\%$ stellar mass-to-light limit has been scaled to the $80\%$ detection completeness limit for each sub-field (as determined in Section~\ref{merger-sec:completeness}). The shaded regions show the range of detection completeness limits covered by the CANDELS fields (Table~\ref{tab:completeness}) with the area-weighted average for each sub-field depth shown by the solid, dashed and dotted blue lines respectively. Relevant mass selection limits are shown as horizontal red dashed and dotted lines for illustrative purposes. }
  \label{merger-fig:mass_comp}
\end{figure}

The resulting mass completeness at $z > 1$ in bins with width $\Delta z = 0.5$ are shown in Fig.~\ref{merger-fig:mass_comp} assuming a flux-limit equal to the appropriate corresponding `WIDE 2'-depth $80\%$ detection completeness limit. 
Based on the binned empirical completeness limits, we then fit a simple polynomial function to the observed $\Mstar/\textit{L}$ redshift evolution. 
By doing so we can estimate the mass completeness as a continuous function of redshift. 

A common choice of template for estimating the strict $\Mstar/\textit{L}$ completeness is a maximally old single stellar population (continuous blue line in the top panel of Fig.~\ref{merger-fig:mass_comp}, assuming a formation redshift of $z = 12$ and sub-solar metallicity of $Z = 0.2 Z_{\odot}$). However, since the vast majority of galaxies above $z\sim3$ are expected to be actively star-forming, this assumption significantly overestimates the actual completeness mass at high-redshift (hence under-estimating the completeness). 

%While a dust extinction of $A_{V} = 2$ is higher than typically observed at high redshift \citep{2015A&A...574A..19S}, it is consistent with the dust extinction seen in sub-mm galaxies at $z > 2$ \citep{Targett:2013kg,Wiklind:2014gs,Smolcic:2015ke} and therefore represents a plausible limit for the highest mass-to-light ratios at high redshift. Instead of assuming a rising star-formation history and high fixed dust attenuation, we could have instead plausibly chosen a different star-formation history and allowed dust to vary with redshift in a manner which also matched the estimated empirical completeness.

The redshift-dependent mass limit, $\Mstar^{\rm{flux}}(z)$, is defined as
\begin{equation}\label{eq:fluxlim}
	\log_{10}(\Mstar^{\rm{flux}}(z)) = 	0.4\times( H_{\Mstar/\textit{L}}(z) - H^{\rm{lim}})
\end{equation}
where $H^{\rm{lim}}$ is the $H_{160}$ magnitude at the flux-completeness limit in the field or region of interest and $H_{\Mstar/\textit{L}}(z)$ is the $H_{160}$ magnitude at a given redshift of the fitted functional form normalized to 1 $\rm{M}_{\odot}$. 
In the bottom panel of Fig.~\ref{merger-fig:mass_comp} we show the redshift-dependent mass limit corresponding to each of the sub-field depths outlined in Section~\ref{merger-sec:completeness}. 
Also shown in this plot are lines corresponding to the stellar mass ranges we wish to probe for major mergers ($\mu > 1/4$) around galaxies with stellar mass of $9.7 < \log_{10} \Mstar \leq 10.3$ and $\log_{10} \Mstar \geq 10.3$ (hatched region).

\begin{figure}
{\centering
	\includegraphics[width=0.95\columnwidth]{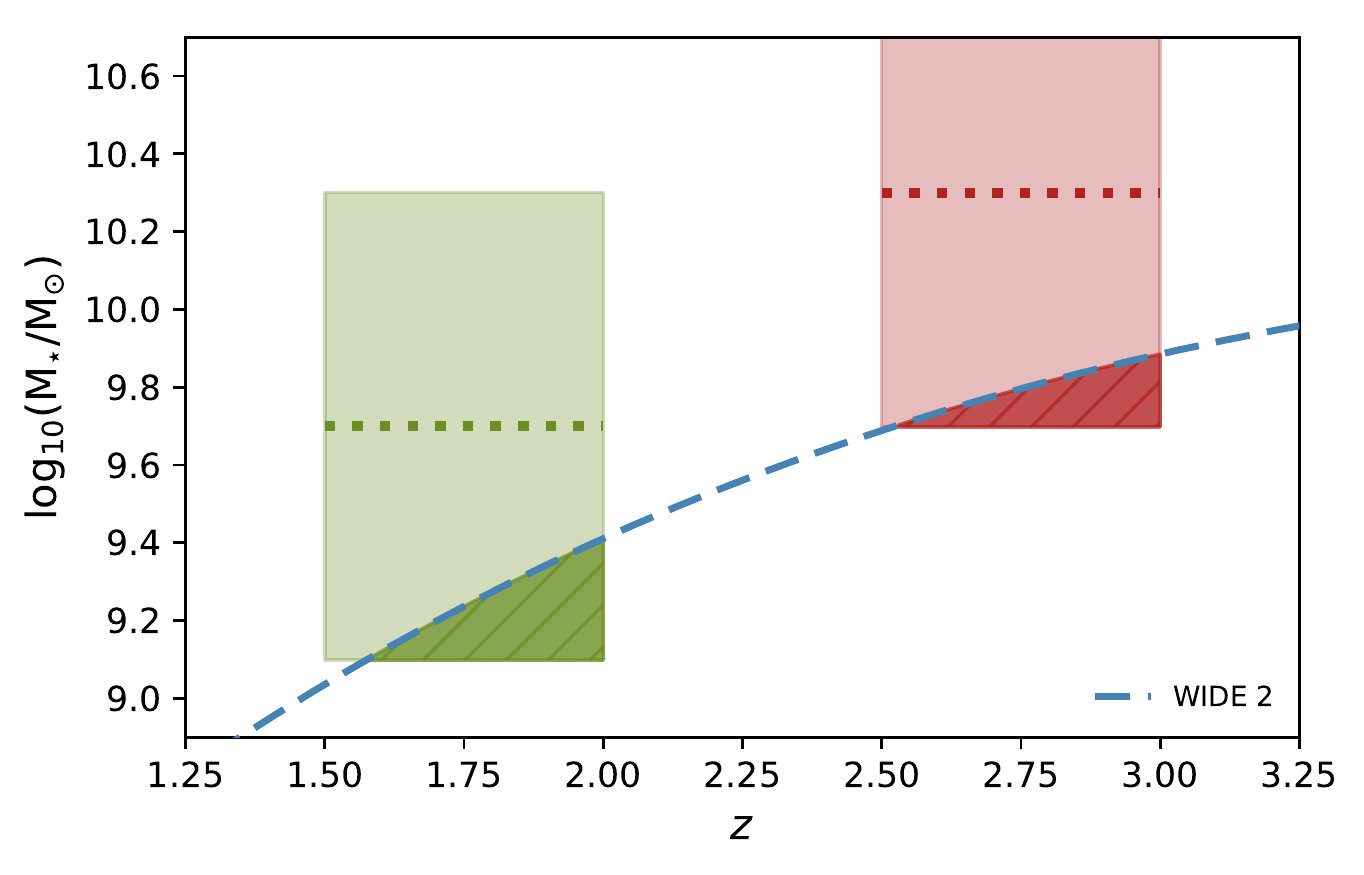}}
  \caption{Illustration of the parameter space where the statistical stellar-mass completeness correction is in effect. The example illustrates the relevant mass limits and selection ranges for a redshift bin of two different bins within the `WIDE 2' sub-fields: a primary mass selection of $9.7 < \log_{10}(\Mstar / \text{M}_{\odot}) < 10.3$ at $1.5 < z < 2$ (green) and a $\log_{10}(\Mstar / \text{M}_{\odot}) > 10.3$ selection at $2.5 < z < 3$ (red).}
  \label{merger-fig:mass_comp_eg}
\end{figure}

For a primary galaxy with a mass close to the redshift-dependent mass-limit imposed by the selection criteria $S(z)$, the mass range within which secondary pairs can be included may be reduced, i.e.  $\mu\Mstar^{1}(z) <  \Mstar^{\rm{lim}}(z) < \Mstar^{1}(z)$. 
In Fig.~\ref{merger-fig:mass_comp_eg} we illustrate this for a galaxy with $\log_{10} \Mstar \approx 10.3$ in the redshift range $2.5 < z \leq 3$ (red) and a $\log_{10} \Mstar \approx 9.7$ at $1.5 < z < 2$ (green). 
The darker shaded regions shows the area in the parameter space of $z$ vs $\Mstar$ where potential secondary galaxies with merger ratios $>1/4$ are excluded by the redshift-dependent mass-completeness cut.

To correct for the potential galaxy pairs that may be lost by the applied completeness limit, we make a statistical correction based on the stellar mass function at the redshift of interest - analogous to the luminosity function-based corrections first presented in \citet{Patton:2000kt}. 
The flux-limit weight, $w^{\text{flux}}_{2}(z)$, applied to every secondary galaxy found around each primary galaxy, is defined as 
\begin{equation}
	w^{\text{flux}}_{2}(z) = \frac{1} {W_{2}(z)},	
\end{equation}
where 
\begin{equation}
	W_{2}(z) = \frac{ \int_{\Mstar^{\text{lim}}(z)}^{\Mstar^{\text{1}}(z)} 
								\phi(\Mstar|z)d\Mstar} 
								{ \int_{\mu \Mstar^{1}(z)}^{\Mstar^{\text{1}}(z)}
								\phi(\Mstar|z)d\Mstar}
\end{equation}
and \( \phi(\Mstar|z) \) is the stellar mass function at the corresponding redshift. The redshift-dependent mass limit is \(\Mstar^{\text{lim}}(z) = max \{ \mu \Mstar^{1}(z), \Mstar^{\rm{flux}}(z) \} \), where \(\Mstar^{\rm{flux}}(z)\) is defined in Equation~\ref{eq:fluxlim} (dashed blue line in Fig.~\ref{merger-fig:mass_comp_eg}). By applying this weight to all pairs associated with a primary galaxy, we get the pair statistics corresponding to $\mu \Mstar^{1}(z) \leq \Mstar^{\text{2}}(z) \leq \Mstar^{\text{1}}(z)$ (the volume limited scenario, e.g. the total red or green shaded areas in Fig.~\ref{merger-fig:mass_comp_eg}). Note that because this correction is based on the statistically expected number density of galaxies as a function of mass, representative numbers of detected secondary galaxies above the completeness limit are still required. 

As in \citet{Patton:2000kt}, we also assign additional weights to the primary sample in order to minimize the error from primary galaxies that are closer to the flux limit (i.e. with redshift posteriors weighted to higher redshifts) as these galaxies will have fewer numbers of \emph{observed} pairs. The primary flux-weight, $w_{\text{flux}}^{1}(z)$ is defined as
\begin{equation}
	w^{\text{flux}}_{1}(z) =  W_{1}(z) = \frac{ \int_{\Mstar^{\text{lim}}(z)}^{\Mstar^{\rm{max}}} 
																	\phi(\Mstar|z)d\Mstar} { \int_{\Mstar^{\rm{min}}}^{\Mstar^{\rm{max}}}
																	\phi(\Mstar|z)d\Mstar}
\end{equation}
where $\Mstar^{\rm{min}}$ and $\Mstar^{\rm{max}}$ are the lower and upper limits of the mass range of interest for the primary galaxy sample, the redshift-dependent lower limit is defined as \( \Mstar^{\text{lim}}(z) = max \{ \Mstar^{\rm{min}}, \Mstar^{\rm{flux}}(z) \} \), and the remaining parameters are as outlined above. For volume-limited samples (where \(\Mstar^{\rm{flux}}(z) < \mu\Mstar^{1}(z) \) at all redshifts) both of the flux-limit weights are equal to unity.

The stellar mass functions (SMF) parameterizations as a function of redshift, \( \phi(\Mstar|z) \), are taken from \citet{Mortlock:2014et} at $z \leq 3$, \citet{Santini:2012jq} at $3 < z < 3.5$ and \citet{Duncan:2014gh} at $z \geq 3.5$. 
When selecting redshift bins in which to estimate the merger fraction, we ensure that the bins are chosen to match the bins in which the SMF are constrained (i.e. the SMF used to weight the merger fraction is the same across the bin).
Tests performed when applying the same methodology to wide-area datasets in \citet{mundy2017} indicate that results are robust to the choice of specific SMF and that results presented later in the paper would not be significantly affected if alternative SMF are assumed. 
Furthermore, we note that this correction assumes that the shape of the SMFs for satellite galaxies does not differ from those measured for the full population.
Observational constraints at low redshift indicate that such an assumption is valid \citep{2016MNRAS.459.2150W}, but direct constraints at higher redshift are not currently available.

\subsubsection{Image boundaries and excluded regions}\label{merger-sec:weights_area}
A second correction which must be taken into account is to the search area around primary galaxies that lie close to the boundaries of the survey region. Because of the fixed physical search distance, this correction is also a function of redshift, so it must be calculated for all redshifts within the range of interest.

In addition to the area lost at the survey boundaries, it is also necessary to correct for the potential search area lost due to the presence of large stars and other artefacts, around which no sources are included in the catalog (see Section~\ref{merger-sec:initial}). 

We have taken both of these effects into account when correcting for the search areas by creating a mask image based on the underlying photometry mosaics. Firstly, we define the image boundary based on the exposure map corresponding to the $H_{160}$ photometry used for object detection. Next, for every source excluded from the sample catalog based on its classification as a star or image artefact by our photometric or visual classification, the area corresponding to that object (from the photometry segmentation map) is set to zero in our mask image. Finally, areas of photometry which are flagged in the flag map (and excluded based on their corresponding catalog flags) are also set to zero.

To calculate the area around a primary galaxy that is excluded by these effects, we perform aperture `photometry' on the generated mask images.  Photometry is performed in annuli around each primary galaxy target, with inner and outer radii of $\theta_{\rm{min}}(z)$ and $\theta_{\rm{max}}(z)$ respectively. The area weight is then defined as
\begin{equation}
	w_{\rm{area}}(z) = \frac{1} {f_{	\rm{area}}(z)}
\end{equation}
where $f_{	\rm{area}}(z)$ is the sum of the normalized mask image within the annulus at a given redshift divided by the sum over the same area in an image with all values equal to unity. By measuring the area in this way we are able to automatically take into account the irregular survey shape and any small calculation errors from quantization of areas due to finite pixel size.

Despite the relatively small survey area explored in this study (and hence a higher proportion of galaxies likely to lie near the image edge), the effect of the area weight on the estimated pair fractions is very small. 
To quantify this, we calculate the pair averaged area weights, $\left \langle w_{\rm{area}} \right \rangle$, such that
\begin{equation}
	\left \langle w_{\rm{area}}^{i,j} \right \rangle = \frac{\int \text{PPF}^{i,j}(z) w_{\rm{area}}^{i}(z) \rm{d}z}{\int \text{PPF}^{i,j}(z) \rm{d}z},
\end{equation}
where $w_{\rm{area}}^{i}(z)$ is the redshift dependent area weight for a primary galaxy $i$, and $\text{PPF}^{i,j}(z)$ the corresponding pair-probability function for primary galaxy and a secondary galaxy $j$. Of the full sample of primary galaxies, less than $10\%$ have average area weights greater than 1.01 (where a primary galaxy has multiple pairs, we take the average of $\left \langle w_{\rm{area}}^{i,j} \right \rangle$ over all secondary galaxies). 
Furthermore, only $\approx 2\%$ of primary galaxies have average weights $\left \langle w_{\rm{area}}^{i,j} \right \rangle > 1.1$ and only $0.15\%$ have weights $>1.5$ (e.g. sources which lie very close to the edge of the survey field).
The effects of area weights on the final estimated merger fractions will therefore be minimal. Nevertheless, we include these corrections in all subsequent analysis.

\subsubsection{The Odds sampling rate}\label{merger-sec:weights_osr}
In the original method outlined in \citetalias{LopezSanjuan:2014uj}, and also applied in \citet{mundy2017}, an additional selection based on the photometric redshift quality, or odds $\mathcal{O}$ parameter. 
The original motivation for this additional selection criteria (and subsequent correction), as outlined partially in \citet{Molino:2014iz}, is that by enforcing the odds cut they are able to select a sample for which the posterior uncertainties are accurate.

Due to the extensive magnitude dependent photo-$z$ posterior calibration applied in this work and the fact that our resulting redshift posteriors are well calibrated at all magnitudes, we do not include this additional criteria.
Therefore, we do not apply the additional odds sampling rate weighting terms outlined in \citet{mundy2017}.

\subsubsection{The combined weights}
Taking both of the above effects into account, the pair weights for each secondary galaxy found around a galaxy primary are given by
\begin{equation}\label{eq:flux_weights_2}
w_{\text{2}}(z) = w_{1,\text{area}}(z) \times 
							w_{1,\text{flux}}(z) \times w_{2,\text{flux}}(z) %\times w^{\text{OSR}}_{1} \times w^{\text{OSR}}_{2}.
\end{equation}
The weights applied to every primary galaxy in the sample are then given by
\begin{equation}\label{eq:flux_weights_1}
w_{\text{1}} =  w_{1,\text{flux}}(z)  %\times w^{\text{OSR}}_{1}.
\end{equation}
These weights are then applied to the integrated pair-probability functions for each set of potential pairs to calculate the merger fraction.
The greatest contribution to the total weights primarily comes from the secondary galaxy completeness weights, $w_{2,\text{flux}}(z)$, with additional non-negligible contributions from the primary completeness. Furthermore, the largest additional uncertainty in the total weights results from the mass completeness weights. 

% In the implementation of the pair-count methodology used in this paper, it is not currently possible to fully incorporate the individual completeness weight errors in the overall merger fraction uncertainties. However, in future studies incorporating the full CANDELS datasets where Poisson and cosmic variance errors will be significantly reduced, propagating the SMF uncertainty into the final of increasing importance. For the results presented later in this paper, it is necessary to bear in mind that there may additional statistical errors in the measured pair-fraction up to $\sigma_{x}/x \lesssim 0.2$.

\subsection{Final integrated pair fractions}\label{merger-sec:pair_frac}
With the pair probability function and weights calculated for all potential galaxy pairs, the total integrated pairs fractions can then be calculated as follows. 
For each galaxy, $i$, in the primary sample, the number of associated pairs, $N_{\rm{pair}}^{i}$, within the redshift range $z_{\rm{min}} < z < z_{\rm{max}}$ is given by
\begin{equation}
N_{\text{pair}}^{i} = \sum_{j} \int_{z_{\rm{min}}}^{z_{\rm{max}}} w_{\text{2}}^{j}(z)\times \text{PPF}_{j}(z) dz	
\end{equation}
where $j$ indexes the number of potential close pairs found around the primary galaxy, $\text{PPF}_{j}(z)$ the corresponding pair-probability function (Equation~\ref{eq:PPF}) and $w_{\text{2},j}(z)$ its pair weight (Equation~\ref{eq:flux_weights_2}). The corresponding weighted primary galaxy contribution, $N_{1,i}$, within the redshift bin is
\begin{equation}
 N_{1,i} = \sum_{i}\int_{z_{\rm{min}}}^{z_{\rm{max}}} w_{1,i}(z) \times P_{i}(z) \times S_{1,i}(z){} dz	
\end{equation} 
where $S_{1,i}(z)$ is the selection function for the primary galaxies given in Equation~\ref{eq:pri_sel}, $P_{i}(z)$ its normalized redshift probability distribution and $w_{1,i}$ its weighting. In the case of a primary galaxy with stellar mass in the desired range with its redshift PDF contained entirely within the redshift range of interest, $N_{1,i} = w_{1,i}$, and hence always equal or greater than unity.

The estimated pair fraction $f_{\rm{P}}$ is defined as the number of pairs found for the target sample divided by the total number of galaxies in that sample. In the redshift range $z_{\rm{min}} < z < z_{\rm{max}}$,  $f_{\rm{P}}$ is then given by 
\begin{equation}
f_{\rm{P}} = \frac{\sum_{i} N_{\text{pair},i}}{\sum_{i} N_{1,i}}
\end{equation}
where $i$ is summed over all galaxies in the primary sample. For a field consisting of different sub-fields, this sum becomes 
\begin{equation}
f_{\rm{P}} = \frac{\sum_{k}\sum_{i} N_{\text{pair},k,i}}{\sum_{k}\sum_{i} N_{1,k,i}}
\end{equation}
where $k$ is indexed over the number sub-fields (e.g. 4: `Wide 1', `Wide 2', `Deep' and `Ultra Deep'). The mass completeness limit used throughout the calculations is set by the corresponding $H_{160}$ depth within each field.

\section{Results}\label{merger-sec:results}
 In this section we investigate the role of mergers in forming massive galaxies up to $z \approx 6$.  
 We first investigate and describe a purely observationally quantity, the pair fraction, using the full posterior pair-count analysis described in the previous section, within eight redshift bins from $z = 0.5$ to $z = 6.5$.  
 We carry this out within stellar mass cuts of $9.7 < \log_{10}(\Mstar / \text{M}_{\odot}) < 10.3$ and $\log_{10}(\Mstar / \text{M}_{\odot}) > 10.3$. We also perform the pair searches in annuli with projected separations of $5 \leq r_{\text{p}} \leq 30$. 
The minimum radius of 5 kpc is typically used in pair counting studies to prevent confusion of close sources due to the photometric or spectroscopic fibre resolution. 
Although the high-resolution HST photometry allows for reliable deblending at radii smaller than this \citep{2007PASP..119.1325L,Galametz:2013dd}, we adopt this radius for consistency with previous results.

Later in this section, we then calculate observational constraints placed on merger \emph{rates} for these galaxies, using physically motivated merger-time scales to explore both the merger rate per galaxy and the merger rate density over time since $z = 6$.
 
%Ken - I commented these below out of the paper, giving Evol. merger fraction 4.1
%\subsection{Tests of the estimated pair fraction}
%\subsubsection{Line-of-sight contamination}
%\subsubsection{Tests on mock galaxy catalogs}

\subsection{Evolution of the major pair fraction}\label{merger-sec:mergerfraction}

\subsubsection{Observed pair fractions in CANDELS}
\begin{figure*}
\centering
	\includegraphics[width=1.4\columnwidth]{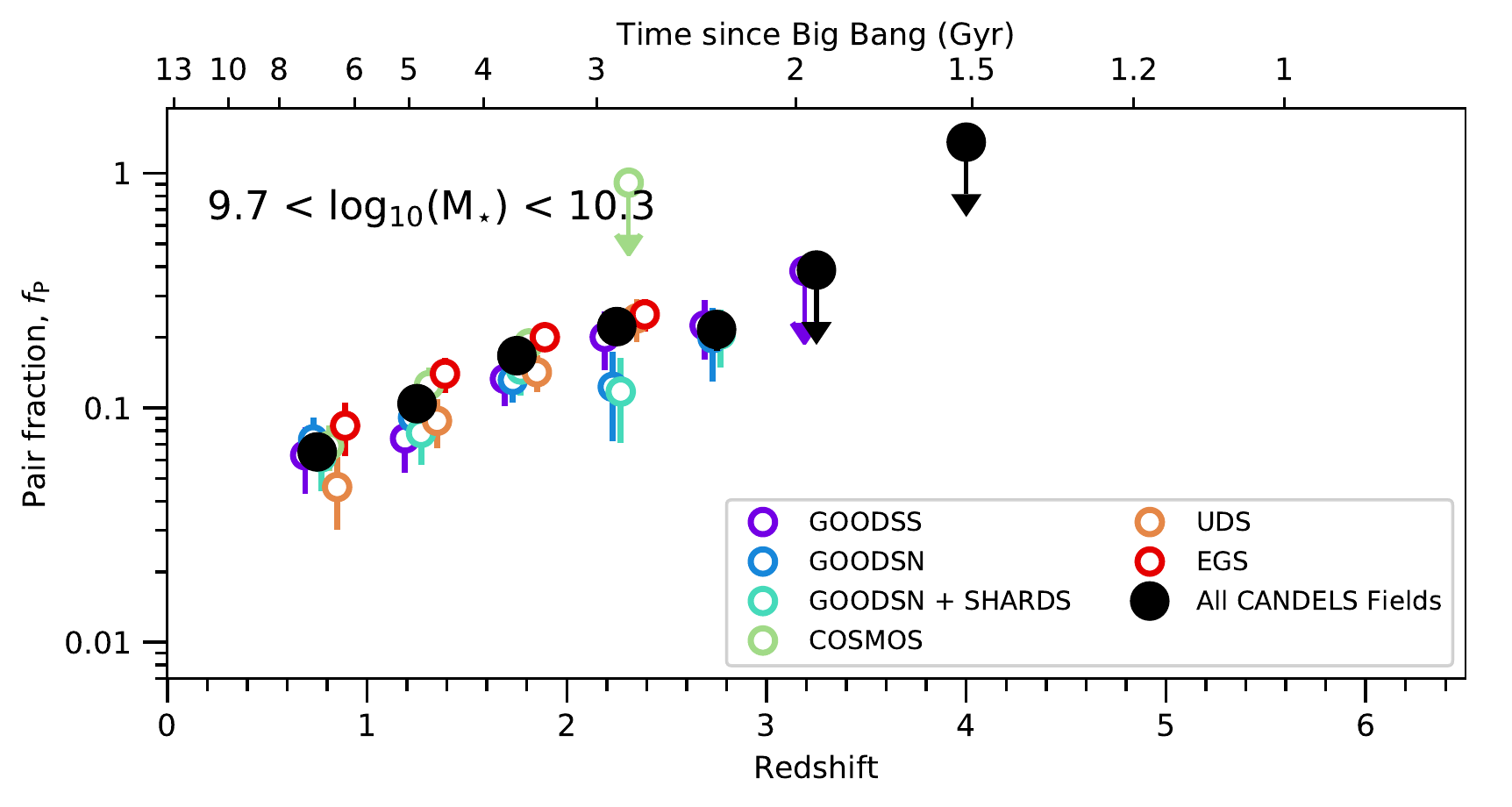}
	\includegraphics[width=1.4\columnwidth]{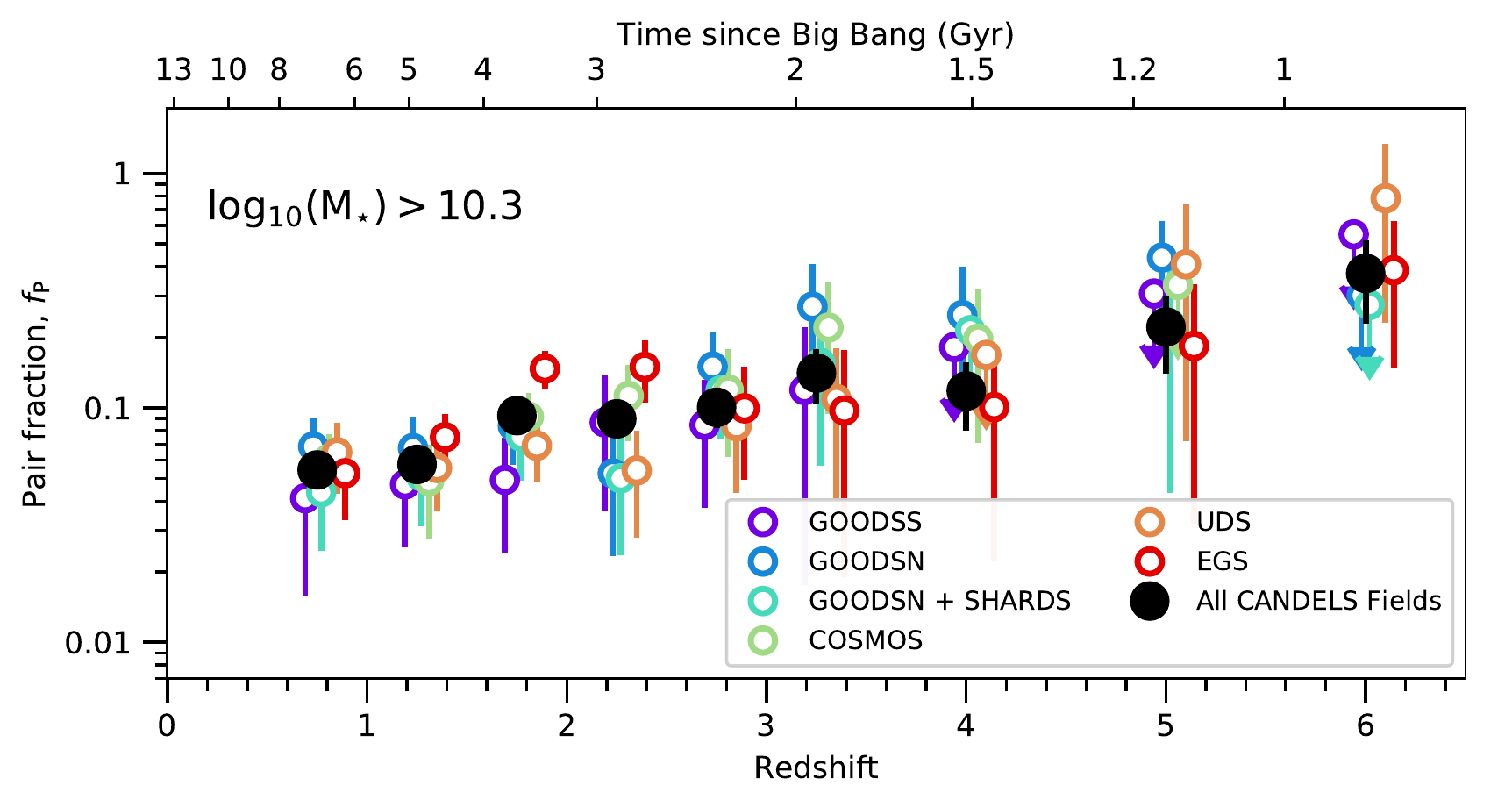}
 \caption{Estimated major merger fraction as a function of redshift for galaxies with stellar mass $9.7 < \log_{10}(\Mstar / \text{M}_{\odot}) < 10.3$ (top) and $\log_{10}(\Mstar / \text{M}_{\odot}) > 10.3$ (bottom). In each figure we show the estimated pair fraction for each individual CANDELS field alongside the combined estimate (larger black points). Results from each field are only plotted if the mass completeness requirements are satisfied.  Only our deepest data, from the Hubble Ultra Deep Field data in the GOODS-S, is shown in the upper panel at $z > 3$.}
  \label{merger-fig:merger_frac}
\end{figure*}

In this section we present measurements of the observed pair fraction, $f_{\text P}$ of massive galaxies from $z=0.5$ to $z \sim 6$ in the combined CANDELS multi-wavelength datasets.
Our results are shown in Fig.~\ref{merger-fig:merger_frac}, where we plot our derived pair fractions for each of the five fields as well the overall constraints provided by the combined measurements.
The measured values and their corresponding statistical errors are presented in Table~\ref{tab:fmerger}. 
The errors on our $f_\text{P}$ values are estimated using the common bootstrap technique of \citet{Efron:1979uf,EFRON:HJ2mD4hg}. The standard error, $\sigma_{f_{\text{P}}}$, is defined as
\begin{equation}\label{eq:bootstrap_err}
	\sigma_{f_{\rm{P}}} = 
	\sqrt{\frac{\sum_{i,N} \left (f_{\text{m},i} - \left \langle f_{\text{P}}  \right \rangle \right)^{2}}
	{(N-1)},
	} 
\end{equation}
where $f_{\text{P}}^{i}$ is the estimated merger fraction for a randomly drawn sample of galaxies (with replacement) from the initial sample (for $N$ independent realisations) and $\left \langle f_{\text{P}} \right \rangle =  \left( \sum_{i} f_{\text{P},i} \right) / N $. 

Only regions (i.e. `Wide 1', `Wide 2', `Deep' and `Ultra Deep') that are complete in stellar mass to the primary galaxy selection mass at the bin upper redshift limit are included in the estimate for a given field.
The same completeness cuts are applied when calculating the combined `All CANDELS' estimates, with only the contributing datapoints plotted in Fig.~\ref{merger-fig:merger_frac}.
When calculating the combined pair fraction estimates, we include only one measurement from GOODS North, specifically the estimates incorporating the SHARDS medium-band photometry.

As can be seen in Fig.~\ref{merger-fig:merger_frac}, there is a variance in the derived pair fraction across the five CANDELS fields. 
However, given the statistical uncertainties within each field, we find that the individual measurements are consistent across the wide range in redshifts.
In all fields, we find a systematic trend with redshift, such that the pair fraction increases towards higher redshifts for primary galaxies in both the $9.7 < \log_{10}(\Mstar / \text{M}_{\odot}) < 10.3$ and $\log_{10}(\Mstar / \text{M}_{\odot}) > 10.3$ mass selected samples.

In the lower stellar mass bin explored in this work, the fall in completeness for the shallower CANDELS fields is evident at higher redshifts, with constraints provided primarily by the Hubble Ultra-Deep Field region within the GOODS South field.
However, overall we find that the pair counts for the lower mass range show a similar increase in the pair fraction up to until $z \sim 3$.
Above this redshift the constraints are limited to measurements of the upper limit, i.e. finding no significant probability of pairs around the small number of galaxies that lie in the mass-complete sample \citep[where the upper limit therefore derives from the Poisson error upper limit on a count of zero; see][] {Gehrels:1986cx}.

\begin{deluxetable*}{cccccccc}
 \tablecaption{Major merger pair fractions in the CANDELS fields.}
 \tablehead{
\multicolumn{8}{c}{$9.7 < \log_{10}( \Mstar / \text{M}_{\odot}) < 10.3$} \\
\noalign{\smallskip}
$z$ & GS & GN & GN (SHARDS) & COSMOS & UDS & EGS & All 
}
\startdata
$0.5 \leq z < 1.0$ & $0.063 \pm 0.020$ & $0.073 \pm 0.018$ & $0.061 \pm 0.017$ & $0.069 \pm 0.015$ & $0.046 \pm 0.016$ & $0.084 \pm 0.022$ & $0.065 \pm 0.007$ \\
$1.0 \leq z < 1.5$ & $0.074 \pm 0.021$ & $0.091 \pm 0.023$ & $0.078 \pm 0.021$ & $0.125 \pm 0.023$ & $0.088 \pm 0.021$ & $0.140 \pm 0.024$ & $0.104 \pm 0.010$ \\
$1.5 \leq z < 2.0$ & $0.133 \pm 0.032$ & $0.132 \pm 0.026$ & $0.146 \pm 0.033$ & $0.188 \pm 0.030$ & $0.142 \pm 0.025$ & $0.200 \pm 0.027$ & $0.167 \pm 0.012$ \\
$2.0 \leq z < 2.5$ & $0.201 \pm 0.056$ & $0.123 \pm 0.051$ & $0.117 \pm 0.047$ & $< 0.913$ & $0.241 \pm 0.049$ & $0.250 \pm 0.039$ & $0.222 \pm 0.023$ \\
$2.5 \leq z < 3.0$ & $0.224 \pm 0.064$ & $0.198 \pm 0.069$ & $0.206 \pm 0.057$ &  &  &  & $0.216 \pm 0.041$ \\
$3.0 \leq z < 3.5$ & $< 0.384$ &  &  &  &  &  & $< 0.387$ \\
\hline
\multicolumn{8}{c}{$\log_{10}(\Mstar / \text{M}_{\odot}) > 10.3$} \\
$z$ & GS & GN & GN (SHARDS) & COSMOS & UDS & EGS & All \\
\hline
$0.5 \leq z < 1.0$ & $0.041 \pm 0.025$ & $0.068 \pm 0.023$ & $0.044 \pm 0.019$ & $0.061 \pm 0.016$ & $0.065 \pm 0.022$ & $0.053 \pm 0.019$ & $0.054 \pm 0.008$ \\
$1.0 \leq z < 1.5$ & $0.047 \pm 0.022$ & $0.067 \pm 0.025$ & $0.051 \pm 0.020$ & $0.049 \pm 0.021$ & $0.055 \pm 0.019$ & $0.075 \pm 0.019$ & $0.057 \pm 0.008$ \\
$1.5 \leq z < 2.0$ & $0.049 \pm 0.025$ & $0.084 \pm 0.027$ & $0.075 \pm 0.027$ & $0.092 \pm 0.024$ & $0.069 \pm 0.021$ & $0.147 \pm 0.028$ & $0.093 \pm 0.011$ \\
$2.0 \leq z < 2.5$ & $0.087 \pm 0.050$ & $0.053 \pm 0.029$ & $0.050 \pm 0.026$ & $0.112 \pm 0.040$ & $0.054 \pm 0.026$ & $0.150 \pm 0.045$ & $0.090 \pm 0.015$ \\
$2.5 \leq z < 3.0$ & $0.085 \pm 0.047$ & $0.150 \pm 0.060$ & $0.119 \pm 0.045$ & $0.120 \pm 0.058$ & $0.083 \pm 0.040$ & $0.099 \pm 0.050$ & $0.100 \pm 0.019$ \\
$3.0 \leq z < 3.5$ & $0.119 \pm 0.102$ & $0.270 \pm 0.140$ & $0.156 \pm 0.099$ & $0.220 \pm 0.126$ & $0.109 \pm 0.070$ & $0.097 \pm 0.078$ & $0.141 \pm 0.038$ \\
$3.5 \leq z < 4.5$ & $< 0.182$ & $0.249 \pm 0.153$ & $< 0.214$ & $0.197 \pm 0.126$ & $< 0.168$ & $0.101 \pm 0.078$ & $0.118 \pm 0.038$ \\
$4.5 \leq z < 5.5$ & $< 0.307$ & $0.437 \pm 0.190$ & $0.229 \pm 0.186$ & $< 0.334$ & $0.410 \pm 0.337$ & $0.184 \pm 0.152$ & $0.221 \pm 0.081$ \\
$5.5 \leq z < 6.5$ & $< 0.549$ & $< 0.301$ & $< 0.275$ &  & $0.783 \pm 0.552$ & $0.386 \pm 0.238$ & $0.374 \pm 0.146$ \\
\enddata
\tablecomments{Estimated pair fractions from PDF analysis, as plotted in Fig.~\ref{merger-fig:merger_frac}. Quoted errors include the bootstrapped errors calculated following Eq.~\ref{eq:bootstrap_err}. As discussed in the text, pair fractions presented only include regions (i.e. `Wide 1', `Wide 2', `Deep' and `Ultra Deep') in the estimate for a given field that are complete in stellar mass to the primary galaxy selection mass at the upper redshift limit for a given redshift bin.}
\label{tab:fmerger}
\end{deluxetable*}

\subsubsection{Comparison to literature}
A large number of previous studies have explored the redshift evolution of galaxy pair counts in mass or (absolute) magnitude selected samples \citep{LeFevre:2000iq,Conselice:2003jz,Kartaltepe:2007dv,Bluck:2009in,Bluck:2012dh,Bundy:2009jw,LopezSanjuan:2010cz,LopezSanjuan:2014uj,Man:2011jo, 2016ApJ...830...89M, 2014MNRAS.444..906F}.
However, these past studies employ a wide range of criteria in selecting close pairs (mass ranges, separation radius, line-of-sight selection/correction methods), making it difficult to direct comparisons with the observations presented in this work. 
The majority of merger rate studies typically focus on the most massive galaxies, i.e. $\log_{10}(\Mstar / \text{M}_{\odot}) > 11$. 
For studies at $z > 1$, such massive galaxies are above our typical flux and mass completeness limits and are bright enough for obtaining accurate spectroscopic redshift, they therefore represent the most robust samples studied to date \citep{Bluck:2009in,Man:2011jo}. 
However, given that these massive galaxies are increasingly rare at higher redshifts \citep{Ilbert:2013dq,Muzzin:2013bl,Mortlock:2014et,Duncan:2014gh}, the small field of view of the CANDELS fields does not probe a large enough volume to detect statistically significant samples of these galaxies.
We are therefore unable to compare our results with these previous works at the same mass limit $\log_{10}(\Mstar / \text{M}_{\odot}) > 11$, irrespective of any difference in pair selection radii.

\begin{figure*}
\centering
    \includegraphics[width=1.4\columnwidth]{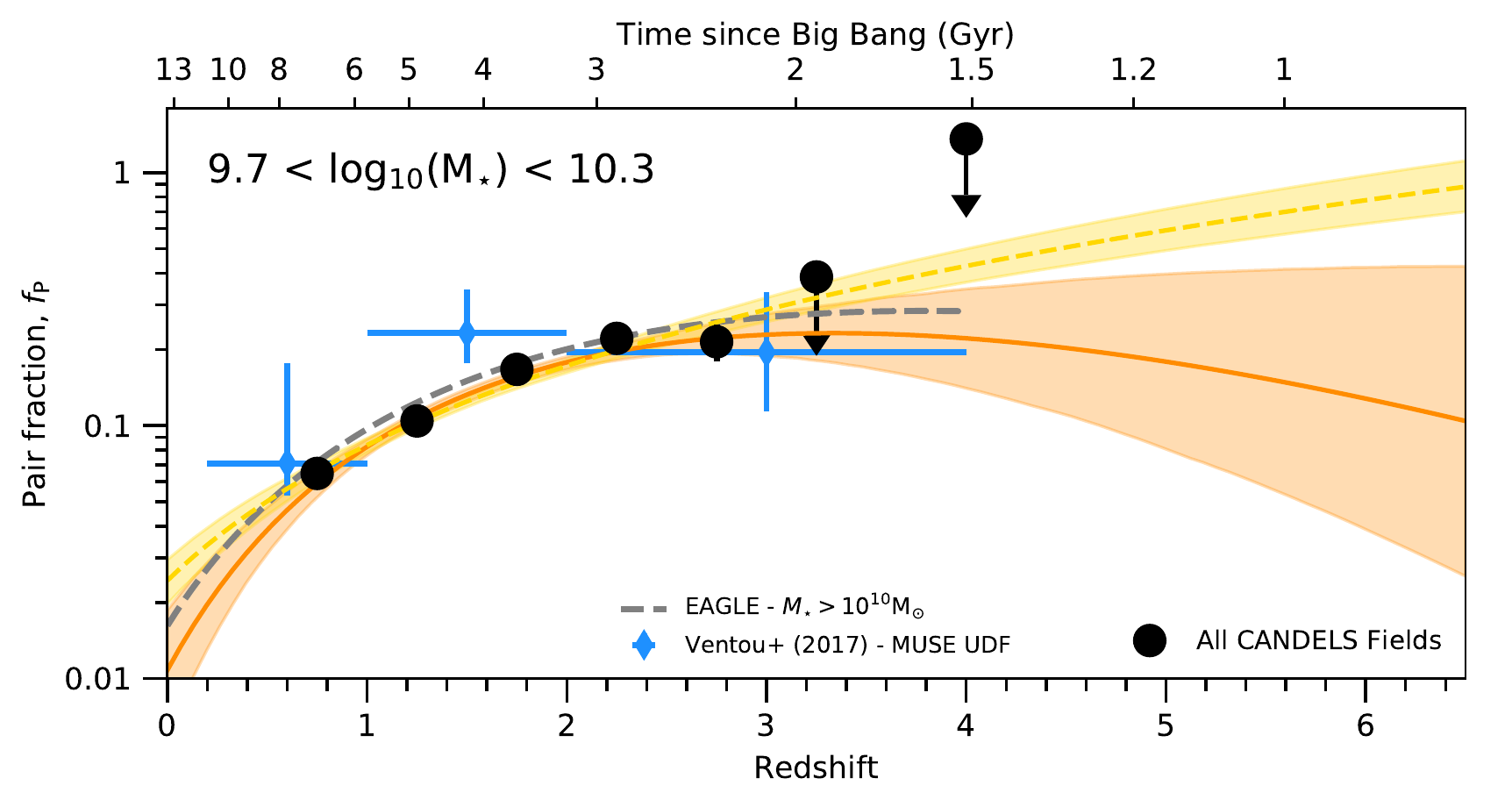}
	\includegraphics[width=1.4\columnwidth]{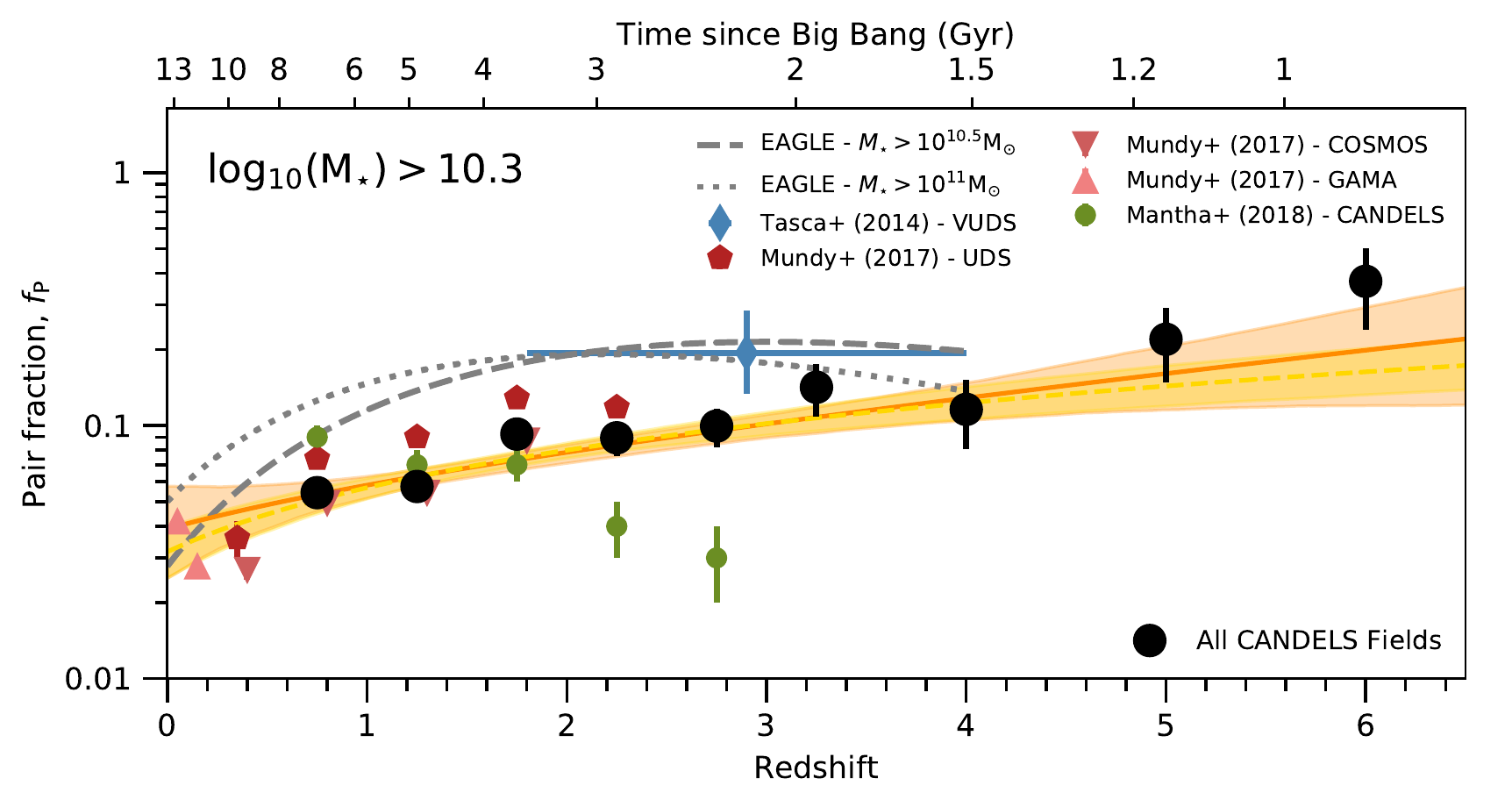}
  \caption{Estimated major merger fraction as a function of redshift for galaxies with stellar mass $9.7 < \log_{10}(\Mstar / \text{M}_{\odot}) < 10.3$ (top) and $\log_{10}(\Mstar / \text{M}_{\odot}) > 10.3$ (bottom). Also shown are the merger fractions from close pair statistics of \citet{Tasca:2014gz}, \citet{2017A&A...608A...9V}, \citet{mundy2017} and \citet{2018MNRAS.475.1549M}. Grey dotted and dashed lines indicate the function form for galaxy pair counts in the EAGLE simulation \citet{Qu:2016hd}. The gold and orange (power-law and power-law + exponential models respectively) lines and shaded regions show the median and 1-$\sigma$ range for our two model fits, based on 100 random draws from the final MCMC fitting chain.}
  \label{merger-fig:merger_frac2}
\end{figure*}

Nevertheless, a range of literature results that select galaxy pairs with comparable mass and pair separation criteria exist.
In Fig.~\ref{merger-fig:merger_frac2} we plot the combined CANDELS major merger pair count observations presented in this work alongside other published measurements that employ the same mass limits and projected separation cuts. 

From \citet{mundy2017} we plot the pair fractions for the three wide area optical surveys used in that work for a primary galaxy mass cut of $\log_{10}(\Mstar / \text{M}_{\odot}) > 10.3$ following the same method employed in this paper (priv. communication).
Additionally, we plot the recent results of \citet{2018MNRAS.475.1549M} who employ a different pair count methodology to the same underlying CANDELS photometric datasets.
To illustrate the latest results on spectroscopic pair counts at high redshift, in the upper panel of Fig.~\ref{merger-fig:merger_frac2} we plot the major pair fractions presented by \citet{2017A&A...608A...9V} for spectroscopically selected pairs with separation $<25$ kpc and primary galaxy stellar mass $\log_{10}(\Mstar / \text{M}_{\odot}) > 9.5$ (median masses from $9.9 \lesssim \log_{10}(\Mstar / \text{M}_{\odot}) \lesssim 10.3$).
In the lower panel of Fig.~\ref{merger-fig:merger_frac2} we plot the pair fraction over the redshift $1.9 < z < 4$ as presented by \citet{Tasca:2014gz}, with pairs also defined by $<25$ kpc separation and a median primary galaxy mass of $\log_{10}(\Mstar / \text{M}_{\odot})= 10.3$.
Both sets of spectroscopic measurements are in good agreement with the higher pair fractions measured in this work.
\footnote{We note that while naming convention varies between studies \citep[e.g. `companion fraction';][]{2018MNRAS.475.1549M}, all literature values plotted correspond to the same observational quantity: the number of galaxy pairs divided by the number of primary galaxies within the sample.}

Finally, we also plot the parameterized pair fraction evolution calculated for the EAGLE \citep{Schaye:2014gk} hydrodynamical simulation presented by \citet{Qu:2016hd}.
Although the mass limits and merger ratio selections presented in \citet{Qu:2016hd} match closely the ranges explored in this work, we note that the pair separation criteria employed are dependent on the half-stellar mass of each primary galaxy and are therefore mass and redshift dependent (typically between 10 and 30 kpc  for the redshift and mass range presented here).
We therefore caution against over-interpretation of any comparison between the simulation results and those presented in this work.

In addition to the literature comparison, in Fig.~\ref{merger-fig:merger_frac2} we also plot our best-fit parameterization of the observations presented in this work.
The redshift evolution of the galaxy pair fraction has been previously parametrized in a number of ways, but primarily as a power-law with respect to $(1+z)$ such that the observed pair fraction goes as

\begin{equation}\label{eq:power}
f_{\rm P}(z) = f_{0} \times (1+z)^{\rm m}.  
\end{equation}

However, other studies have found that the pair, and thus inferred merger, fraction shows evidence of a decline at redshifts higher than around $z \sim 1.5$ to $z\sim2.5$ \citep[e.g. ][]{Conselice:2008de,2016ApJ...830...89M,2018MNRAS.475.1549M}.  
To test whether there is any statistical evidence for a turn-over in the pair fraction at high redshift we therefore fit both the power-law form and a two-component model of a power-law form and an exponential:

\begin{equation}\label{eq:power+exp}
f_{\rm P}(z) = f_{0} \times (1+z)^{\rm m} \times \exp(\tau (1+z)).
\end{equation}

\noindent We fit these two models to the observational results in both mass ranges using a likelihood-based regression optimised through Markov chain Monte Carlo fitting \citep{ForemanMackey:2013io} and incorporating an additional intrinsic scatter term, $s$, within the uncertainties such that $\sigma^{2}_{\textup{tot}} = \sigma_{f_{P}}^{2} + s^{2}f_{\rm P}(z)^{2}$.
In all fits we use a permissive prior that is flat in linear space with very broad boundary conditions for the shape parameters and a flat log prior for the intrinsic scatter, $s$.
The resulting median values and marginalized 1-$\sigma$ uncertainties for both sets of parameterizations are presented in Table~\ref{tab:fitvalues} alongside the Bayesian Information Criterion (BIC) for each fit.

\begin{deluxetable*}{cccccc}
\tablecaption{Evolution of the major merger pair fraction}
\tablehead{
   Mass Bin &  $f_{0}$ & $m$ & $\tau$ & $s$ & BIC }
\startdata
  \multicolumn{6}{c}{Power-law} \\
 $9.7 < \log_{10} \textup{M}_{\star} < 10.3$ & $0.024^{+0.005}_{-0.004}$ &  $1.775^{+0.205}_{-0.196}$ & - & $0.009^{+0.100}_{-0.009}$ & -121.5 \\
  $\log_{10} \textup{M}_{\star} > 10.3$ & $0.032^{+0.009}_{-0.007}$ &  $0.844^{+0.216}_{-0.235}$ & -  & $0.002^{+0.036}_{-0.002}$ & -218.1\\
  & & & & \\
  \multicolumn{6}{c}{Power-law + Exponential} \\
$9.7 < \log_{10} \textup{M}_{\star} < 10.3$ & $0.030^{+0.009}_{-0.007}$ &  $4.431^{+1.721}_{-1.590}$ & $-1.028^{+0.621}_{-0.672}$ &  $0.010^{+0.094}_{-0.010}$  & -120.2 \\
 $\log_{10} \textup{M}_{\star} > 10.3$ & $0.033^{+0.008}_{-0.007}$ &  $0.439^{+1.085}_{-0.939}$ & $0.131^{+0.291}_{-0.363}$ & $0.001^{+0.024}_{-0.001}$ & -214.7 \\
 \noalign{\smallskip}
 \enddata
 \tablecomments{Median and marginalized 1-$\sigma$ uncertainties for the fits to the combined pair counts of this work (Table~\ref{tab:fmerger}) for the power-law and power-law plus exponential functional forms in Eq.\ref{eq:power} and Eq.\ref{eq:power+exp} respectively. Fits assume a prior that is flat in linear space for the shape parameters and a flat log prior for the intrinsic scatter, $s$, with very broad boundary conditions. Also shown are the corresponding Bayesian Information Criterion (BIC) parameters for each fit.}
 \label{tab:fitvalues}
\end{deluxetable*}

Based on the BIC, we find that there is no strong statistical evidence ($\Delta\textup{BIC} > 10$) for a power-law plus exponential form for the evolution of the pair fraction in either mass bin.
Rather, we find that the two models are formally indistinguishable ($0 < \Delta\textup{BIC} < 4$) given our statistical uncertainties.
This result is in contrast to the conclusions drawn by \citet{2018MNRAS.475.1549M} from pair count measurements based on the same underlying datasets.
We attribute this difference primarily to the incorporation of flux-limit corrections that account for pairs which are un-observed due to selection effects \citep[as is also done in][]{mundy2017}.

We note that in choosing to fit the power-law distribution to binned data, we are potentially subject to biases in the best-fitting power-law slope \citep{2004EPJB...41..255G,2007EPJB...58..167B}.
Quantitative comparison of the best-fitting slopes should therefore be made with this caveat in mind.
However, our key conclusions regarding the statistical evidence for or against a redshift turnover are robust to this problem.

\subsubsection{The effects of photometric redshift precision on measured pair-counts}\label{merger-sec:pz_precision}
In Fig.~\ref{merger-fig:merger_frac} we present pair fraction measurements for the CANDELS GOODS North field using two separate photo-$z$ estimates, both with and without the inclusion of the SHARDS medium-band photometry \citep{2013ApJ...762...46P}.
As illustrated in Fig.~\ref{merger-fig:specz_comp1}, the photo-$z$ estimates incorporating SHARDS are $\sim5\times$ more precise at $z \lesssim 1.5$ than those without. 
We are therefore able to explore the effect of redshift precision on the results obtained by our pair-count methodology given the same galaxy sample.

Across all redshift bins, we find that the observed pair fractions between both GOODS North estimates are in agreement within the statistical uncertainties.
However, the GOODS North + SHARDS pair fractions are systematically lower by $\approx 30\%$ on average at these redshifts - comparable to the scatter observed between different CANDELS fields.

To further investigate the effect of redshift uncertainty and the reliability of our pair-count method, we perform an additional test to investigate the potential for residual contamination of the observed pair-counts by chance line-of-sight projections.
Previous attempts to estimate pair-counts using photo-$z$s have estimated the number of true galaxy pairs by subtracting a statistical estimate of the number of random line-of-sight pairs from the observed pair counts.
This correction is typically done using Monte Carlo simulations where the source positions have been randomized across the field, \citep[e.g.][]{Kartaltepe:2007dv,2018MNRAS.475.1549M}. 
In \citet{2018MNRAS.475.1549M}, the \emph{chance} pairs at separations of $<30$ kpc were found to contribute between $\sim75$ to 85$\%$ of the observed pairs for a stellar mass cut of $\log_{10} \textup{M}_{\star} > 10.3$.

A key advantage of our method is that it does not treat the projected pairs as binary, i.e. contributing either 0 or 1 to the pair count.
Rather, the probabilistic pair-count accounts for the fact that even though the 1-$\sigma$ photo-$z$ uncertainties of two galaxies may overlap, the integrated possibility of the two galaxies being at the same redshift will be less than one.\footnote{Conversely, two galaxies separated in redshift by more than 1-$\sigma$ will still have a non-zero possibility of being at the same redshift.}
If the method is performing as designed, chance projected pairs that are unassociated should therefore not contribute significantly to the pair count.

However, as illustrated by the comparison with spectroscopic pairs in Section~\ref{merger-sec:specz_pairs}, there may still remain some contamination at low-redshift from chance projections due to imperfect or outlier photo-$z$s. 
Due to the inhomogeneity in depth across many of the CANDELS fields, creation of fully releastic random catalogs that account for the variation in depth (and hence relative source-counts) are non trivial.
We therefore perform our test on EGS as it is the most homogeneous field with more than 80\% of its area having almost identical $H_{160}$ limiting magnitudes and the remaining area having very similar depths.  These results can be generalized accross all of our CANDELS fields. 

To estimate the residual contamination from un-associated projected pairs, we produce 10 catalogs where the source positions have been re-drawn randomly from within the $H_{160}$ observation footprint and run the full pair-count analysis for the $\log_{10} \textup{M}_{\star} > 10.3$ stellar mass cut.
The background contamination is then estimated based on the median pair-count over the 10 random catalogs.
Averaged over all redshift bins, we find that the random pairs can account for 29\% of the observed pairs in this field -- directly comparable to the difference we see for the high-precision SHARDS sample compared to the broadband only measurements.
This fraction also represents a conservative upper limit due to increased signal from the larger scale clustering at a given redshift over the field (while positions were randomized, the redshift distributions still represent those of the small survey area).
Regardless, the maximum size of this effect is not large.

When fitting the power-law and power-law plus exponential models to the EGS field data points alone, we find that our conclusions on the redshift evolution of the pair fraction are unchanged.
The best-fitting power law for the EGS pair-fractions before subtracting the contamination is 
$$f(z) = 0.045^{+0.019}_{-0.014} \times (1+z)^{0.762^{+0.328}_{-0.359}}.$$ 
While after subtracting the contamination for chance pairs we find
$$f(z) =0.043^{+0.026}_{-0.017} \times (1+z)^{0.488^{+0.474}_{-0.545}}.$$
The power-law only parametrisation remains the best fit after subtraction of the random pairs, but formally the two models are still statistically indistinguishable ($\Delta\textup{BIC} = 2.7$).
As this effect is not large enough to affect any of the conclusions presented in the following section and has not been applied to previous  \citep{mundy2017}, we do not apply the correction to the full pair fraction results.
In Section~\ref{merger-sec:discussion}, we discuss further how this systematic might effect the conclusions on the merger history of massive galaxies.

\subsection{Minor merger pair fractions}\label{merger-sec:minor}

Minor mergers, with mass ratios between 10:1 and 4:1, are predicted in some galaxy formation models as one of the dominant ways in which mass is added to massive galaxies. 
However, almost no direct observational information is available to determine the role of minor mergers \citep[some studies such as][observationally infer their importance]{Ownsworth:2014gt}.  
This quantity was previously examined in more massive galaxies by \citet{Bluck:2012dh} for the GOODS NICMOS Survey, and more recently by \citet{2016ApJ...830...89M} and \citet{mundy2017}.
The depth of CANDELS data used in this work means we can investigate the pair fraction for galaxies in our sample down to mass ratios as low as 20:1 or lower. 
While we are not able to measure these ratios out to our highest redshifts of $z \sim 6$ due to the mass completeness limits, we can investigate the evolution of these minor pairs over the epoch of peak galaxy formation ($z \lesssim 3$). 

\begin{figure*}
{\centering
	\includegraphics[width=1\textwidth]{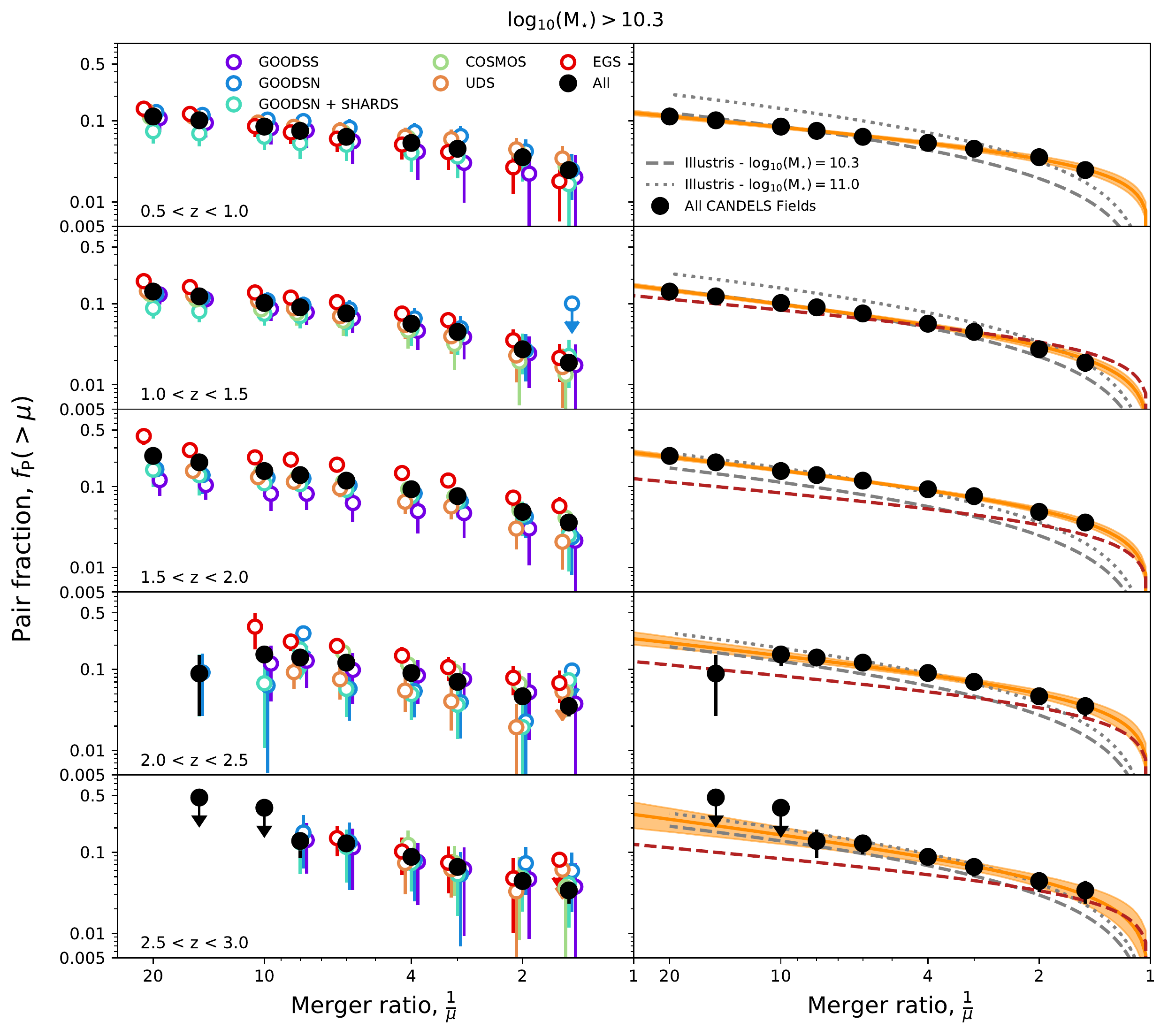}}
  \caption{Measured pair fractions as a function of merger ratio, $1/\mu$, for three different redshift bins. Left-hand panels show the pair fractions for each individual field alongside the combined estimate. In the right-hand column we present the combined measurement for all five CANDELS fields (black points) and the best-fit parameterizations (Equation~\ref{eq:mu_func}) and corresponding 1-$\sigma$ uncertainties. The best-fit curve from $0.5 \leq z < 1.0$ is plotted again for comparison in the higher redshift bins (dashed red line).}
  \label{merger-fig:merger_frac_mu}
\end{figure*}

In Fig.~\ref{merger-fig:merger_frac_mu} we show the measured \emph{cumulative} pair fraction for five different redshift ranges between $0.5 \leq z < 3$. 
We plot these pair fractions $f_{\rm P}$ as a function of the mass ratio $\mu =$ M$_{\star,\rm pri}$/M$_{\star,\rm sec}$ where `pri' and `sec' denote the stellar mass of the more and less massive galaxy involved in the merger, respectively.  
As expected, we find that the cumulative merger fraction smoothly increases with mass ratio. 
To parametrize the pair fractions as a function of mass ratio $\mu$, we fit the following functional form for each redshift bin:

\begin{equation}\label{eq:mu_func}
f_{\rm P}(>\mu) = A \times \left(\frac{1}{\mu}  -1 \right)^{B}.
\end{equation}

\noindent Table~\ref{tab:minor_func} shows the corresponding parameter fits for each of the redshift bins.  
As can be seen through these fits, there is no significant change in the slope of this relation between merger mass ratio and the resulting pair fraction.
The shallow slope we find for the \emph{cumulative} pair fractions indicates that at larger mass ratio differences (smaller $\mu$), the observed pair fraction decreases for greater mass ratios (more minor mergers).
This result qualitatively confirms the findings of \citet{2016ApJ...830...89M} and \citet{mundy2017} for more massive samples of galaxies.

\begin{deluxetable}{ccc}
  \tablecaption{Merger ratio dependence of pair fractions}
  \tablehead{
   Redshift  & $\log_{10}(A)$ & $B$
   }
\startdata
$0.5 \leq z < 1.0$ & $-1.472^{+0.037}_{-0.040}$ & $0.413^{+0.042}_{-0.041}$ \\
$1.0 \leq z < 1.5$ & $-1.522^{+0.037}_{-0.039}$ & $0.540^{+0.040}_{-0.039}$ \\
$1.5 \leq z < 2.0$ & $-1.291^{+0.032}_{-0.033}$ & $0.515^{+0.041}_{-0.040}$\\
$2.0 \leq z < 2.5$ & $-1.299^{+0.046}_{-0.051}$ & $0.491^{+0.078}_{-0.076}$\\
$2.5 \leq z < 3.0$ & $-1.346^{+0.065}_{-0.078}$ & $0.582^{+0.160}_{-0.147}$\\
  \enddata
  \tablecomments{Best-fit parameters for the functional form fitted to the cumulative pair fraction as a function of merger ratio (see equation 37.)}
  \label{tab:minor_func}
\end{deluxetable}

Similarly, within this range we also do not see a significant decline in the values for the normalization ($A$), such that the observed history of galaxy pairs over this redshift range from $0.5 < z < 3$ is fairly constant, as seen previously in the redshift evolution of the pair fraction for major mergers (Fig.~\ref{merger-fig:merger_frac2}).  
This suggests that minor mergers are following the major mergers in terms of their commonality at these redshifts.

\subsection{Evolution of galaxy merger rates}\label{merger-sec:mergerrate}
The major or minor merger pair fraction is a purely observational quantity and not a fundamental parameter for deriving evolution (such as the star-formation rate).  
Furthermore, comparisons of pair fractions between different redshift bins and methodologies can be difficult, as different methods of finding mergers having different time-scale sensitivities.  
This is analogous to measuring the star-formation rate using e.g., UV fluxes, H$\alpha$ fluxes or FIR fluxes.  
Each flux is a representation of some aspect of the star-formation rate, but each one is sensitive only to certain types of stars and over certain time-scales.  
Thus the conversion between flux and star-formation rate for these different fluxes has to be done differently for each method.  
Likewise, a similar situation exists when examining pair and merger fractions measured using different mass/luminosity criteria, different separations and when using pairs or structure/morphology.

A more fundamental property of interest is therefore the merger rate, either the average time between mergers per galaxy ($\mathcal{R}$) or the overall merger rate, specifically the merger rate density measured in units of co-moving Mpc$^{3}$ (denoted $\Gamma$ in this work).  

\subsubsection{Major merger rates}
Conversion of the observed pair fraction to a merger rate per galaxy is typically defined as

\begin{equation}\label{eq:old_rate_conv}
	\mathcal{R} (>\Mstar, z) = \frac{f_{\text{P}}(>\Mstar, z) \times C_{\rm merg}}{\tau_{\text{m}}(z)}
\end{equation}

\noindent where $f_{\text{P}}(>\Mstar, z)$ is the pair fraction at redshift $z$ and masses greater than $\Mstar$ (Section~\ref{merger-sec:mergerfraction}), $C_{\rm merg}$ is the average fraction of those pairs that will eventually merge into a single galaxy and $\tau_{\text{m}}(z)$ the corresponding merger timescale at a given redshift.

The merger timescale can be derived either empirically \citep{Conselice:2009bi} or through simulations \citep{Kitzbichler:2008gi,Lotz:2010ie,Lotz:2010hf,2017MNRAS.468..207S}, with different morphology or pair criteria having different timescales within the merger process.   
Simulations using N-body models of this merger process have measured the time-scales for mergers of galaxies with different masses, mass ratios, and other merger properties \citet{Lotz:2010ie}.  
Typically these have been found by e.g., \citet{Conselice:2009bi,Lotz:2010ie} to be around $\tau_{\rm{m}} = 0.3-0.7$ Gyr for pairs with projected separation of $\leq 20$ and $\leq 30$ kpc, respectively. 
These values are based on the average timescales for those separations and similar (baryonic) mass ratios of 1:3. 

The additional factor, $C_{\rm merg}$, is necessary because two galaxies that appear as a pair only have some probability to merge over a given time-scale.  
The orbital parameters of some galaxy pairs can result in a very long dynamical friction time-scale, resulting in merger timescale longer than the Hubble time.  
From simulations, this value computed over all possible merging scenarios is typically $C_{\rm merg} = 0.6$ \citep{Conselice:2014ct} but this value will also depend on the specific mass and redshift dependent.

In this work we estimate the merger rates using the redshift dependent merger observability timescale of \citet{2017MNRAS.468..207S}, such that
\begin{equation}
	\mathcal{R} (>\Mstar, z) = \frac{f_{\text{P}}(>\Mstar, z)}{\tau_{\text{P}}(z)}.
\end{equation}

\noindent The redshift dependent merger observability timescale, $\tau_{\text{P}}(z)$, is calculated by modelling the timescale required to reconcile the intrinsic merger rates of galaxies in the Illustris simulation \citep{Vogelsberger:2014gw,Genel:2014dh} with the estimated pair counts of galaxies from the simulation. 
This evolving time-scale incorporates the effects accounted for by $C_{\rm merg}$ in Eq.~\ref{eq:old_rate_conv}, and is defined as:
\begin{equation}\label{eq:merger_timescale}
\tau_{\text{P}}(z) = 2.4 \times (1+z)^{-2} \textup{Gyr}.
\end{equation}

\noindent We note that the pair criteria employed by \citet{2017MNRAS.468..207S} differ from those in this work, with a primary galaxy mass range of $10.5 < \log_{10}(\Mstar / \text{M}_{\odot}) < 11$ and pair separation radii of 10 to 50 kpc.
The overall normalization of the timescales therefore represents a significant systematic uncertainty, particularly in the case of the $9.7 < \log_{10}(\Mstar / \text{M}_{\odot}) < 10.3$ sample.
Despite these systematic uncertainties, our assumed observability timescales presented by \citet{2017MNRAS.468..207S} represent the best currently available and the most plausible avenue for inferring merger rates from observed pair counts.
In addition to these systematic uncertainties, we also highlight that there is likely significant scatter in the merging timescales on a pair-by-pair basis \citep[see Fig.~6 of][]{2017MNRAS.468..207S}.

\begin{figure}
{\centering
	\includegraphics[width=\columnwidth]{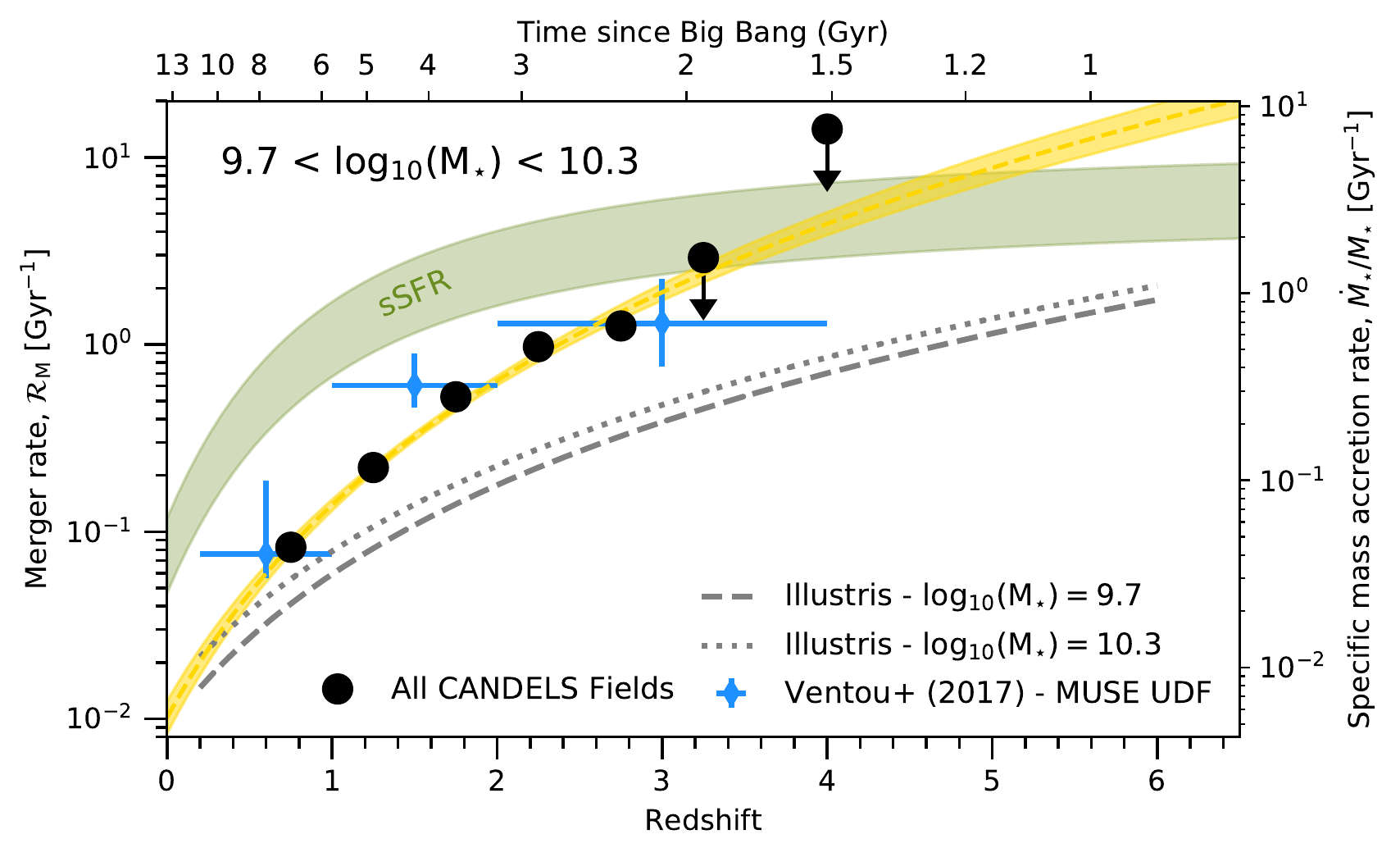}
	\includegraphics[width=\columnwidth]{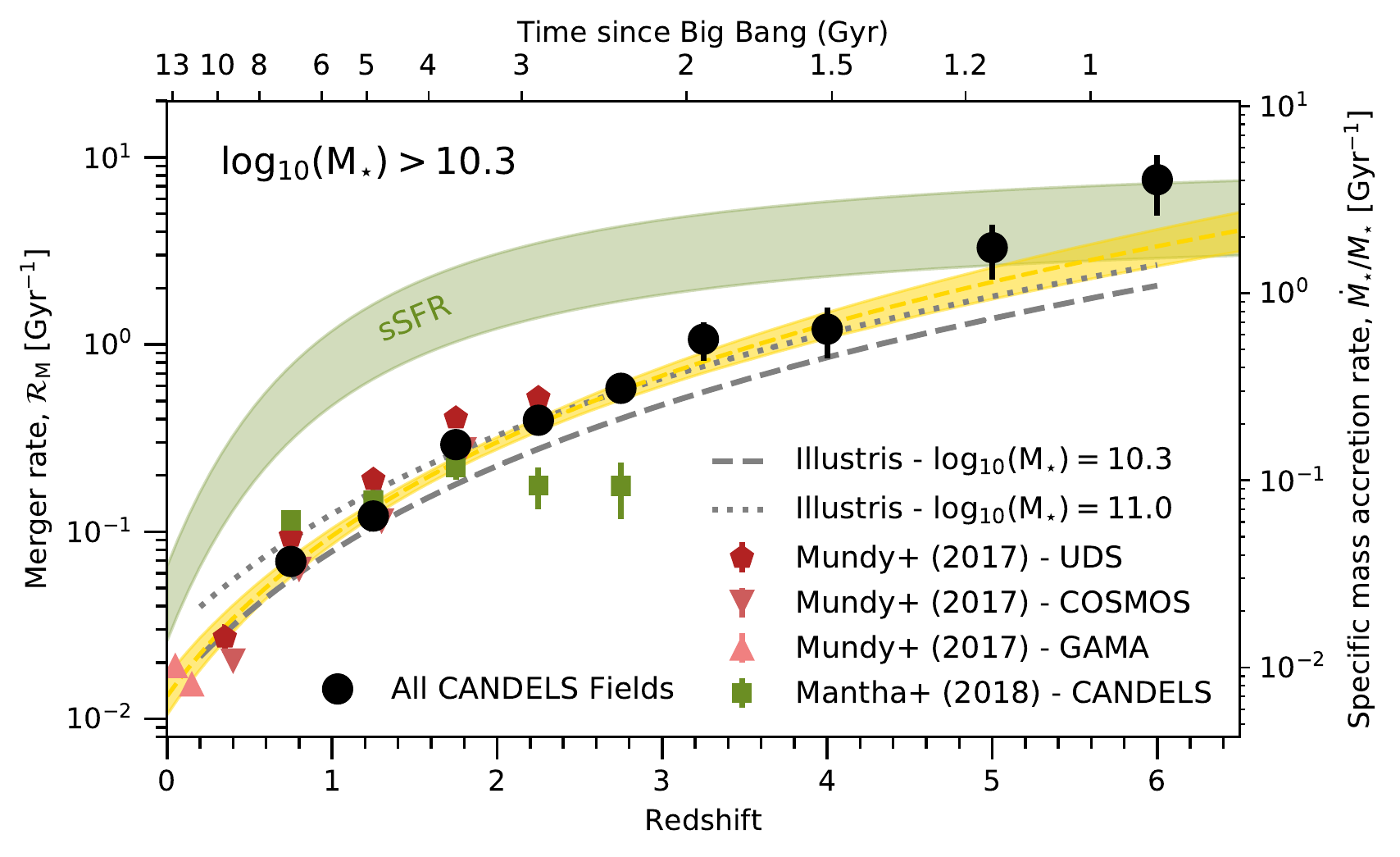}}
  \caption{Estimated major merger rate per galaxy as a function of redshift for galaxies with stellar mass $9.7 < \log_{10}(\Mstar / \text{M}_{\odot}) < 10.3$ (top) and $\log_{10}(\Mstar / \text{M}_{\odot}) > 10.3$ (bottom) assuming the redshift dependent merger timescales of \citet{Snyder:2016ta}. Also shown are the merger rates based on the close-pair statistics of \citet{2017A&A...608A...9V}, \citet{mundy2017} and \citet{2018MNRAS.475.1549M}, assuming the same redshift dependent timescale. The gold line and shaded region in each figure show the best-fitting power-law model from Figure~\ref{merger-fig:merger_frac} converted to merger rates using our assumed merger timescales (Equation~\ref{eq:merger_timescale}). The right-hand scale illustrates the inferred specific mass accretion rate through major mergers based on the observed merger rate (see text). For reference, we also show the observed specific star-formation rates for similar mass galaxies as a function of redshift \citep[green shaded region; ][]{2014ApJS..214...15S}.}
  \label{merger-fig:merger_rate_per_galaxy}
\end{figure}

With these caveats in mind, in Fig.~\ref{merger-fig:merger_rate_per_galaxy} we present the merger rate per galaxy as a function of redshift implied by the observed pair counts in this work. 
We find an increase in the merger rate over all redshifts such that the highest merger rates are found for galaxies at the highest redshifts where we can probe.
In Fig.~\ref{merger-fig:merger_rate_per_galaxy} we also plot the best-fit power law and power-law + exponential parameterizations from Table~\ref{tab:fitvalues} convolved with the observability timescale.
The pair count results of \citet{mundy2017}, \citet{2018MNRAS.475.1549M} and \citet{2017A&A...608A...9V} converted using the same merger timescale are also shown. 

In the higher mass bin we find that there is excellent agreement with the merger rates measured in the Illustris hydrodynamical simulation by \citet{RodriguezGomez:2015hw}.
However, at $9.7 < \log_{10}(\Mstar / \text{M}_{\odot}) < 10.3$, the pair counts measured for the CANDELS fields imply merger rates that are significantly larger than those presented in Illustris.
\citet{Snyder:2016ta}.

\begin{deluxetable}{cc}
  \tablecaption{Merger rate per galaxy, $\mathcal{R}$.}
  \tablehead{
     Redshift & Merger rate per galaxy, $\mathcal{R}$ [Gyr$^{-1}$]
     }
\startdata
    \multicolumn{2}{c}{$9.7 < \log_{10}(\Mstar / \text{M}_{\odot}) < 10.3$} \\ 
    \noalign{\smallskip}
    \hline
$0.5 \leq z < 1.0$ & $0.08 \pm 0.01$ \\
$1.0 \leq z < 1.5$ & $0.22 \pm 0.02$ \\
$1.5 \leq z < 2.0$ & $0.53 \pm 0.04$ \\
$2.0 \leq z < 2.5$ & $0.97 \pm 0.09$ \\
$2.5 \leq z < 3.0$ & $1.26 \pm 0.21$ \\
$3.0 \leq z < 3.5$ & $< 2.91$ \\
$3.5 \leq z < 4.5$ & $< 14.15$ \\
 & \\
 \hline
   \multicolumn{2}{c}{$\log_{10}(\Mstar / \text{M}_{\odot}) > 10.3$} \\ \noalign{\smallskip}
   \hline 
$0.5 \leq z < 1.0$ & $0.07 \pm 0.01$ \\
$1.0 \leq z < 1.5$ & $0.12 \pm 0.02$ \\
$1.5 \leq z < 2.0$ & $0.29 \pm 0.03$ \\
$2.0 \leq z < 2.5$ & $0.39 \pm 0.06$ \\
$2.5 \leq z < 3.0$ & $0.58 \pm 0.10$ \\
$3.0 \leq z < 3.5$ & $1.07 \pm 0.25$ \\
$3.5 \leq z < 4.5$ & $1.21 \pm 0.37$ \\
$4.5 \leq z < 5.5$ & $3.29 \pm 1.07$ \\
$5.5 \leq z < 6.5$ & $7.59 \pm 2.69$ \\
  \enddata
  \tablecomments{Based on the merger fractions presented in Table~\ref{tab:fmerger}. As discussed in the text, conversion from pair fractions to merger rates assumes a redshift dependent merger timescale, Eq.~\ref{eq:merger_timescale}, from \citet{Snyder:2016ta}.}
  \label{tab:mergerpergal}
\end{deluxetable}

Although more informative than the merger fraction alone, the merger rate per galaxy is an average over all galaxies at a given mass and redshifts.  
We are also interested in knowing what the true merger rate is - that is how many merger are occurring per unit time per unit volume as a function of redshift. 
Similarly to previous studies, we define the comoving merger rate density, $\Gamma$, as:
\begin{equation}\label{eq:merger_rate}
	\Gamma(>\Mstar, z) = f_{\text{p}}(>\Mstar, z)n_{\text{c}}(\Mstar, z)\tau_{\text{P}}(z)^{-1},
\end{equation}

\noindent where $f_{\text{p}}(>\Mstar, z)$ is, as before, the mass and redshift-dependent galaxy pair fraction, $n_{\text{c}}(>\Mstar, z)$ the comoving number density for galaxies with stellar mass $>\Mstar$ and $\tau_{\text{P}}(z)$ the redshift dependent merger observability timescale. 
The comoving number densities for galaxies with $9.7 < \log_{10}(\Mstar / \text{M}_{\odot}) < 10.3$ and $\log_{10}(\Mstar / \text{M}_{\odot}) \geq 10.3$ are estimated from the same stellar mass function parameterizations used for the mass completeness weights: \citet{Mortlock:2014et} at $z \leq 3$, \citet{Santini:2012jq} at $3 < z < 3.5$, and \citet{Duncan:2014gh} at $z \geq 3.5$. 

Errors on the number densities are estimated by perturbing the Schechter function parameters based on their quoted errors and recalculating the integrated number density. 
This step is then repeated $10^{4}$ times and the lower and upper 1-$\sigma$ errors are taken as the \nth{16} and \nth{84} percentiles.

\begin{figure}
{\centering
	\includegraphics[width=\columnwidth]{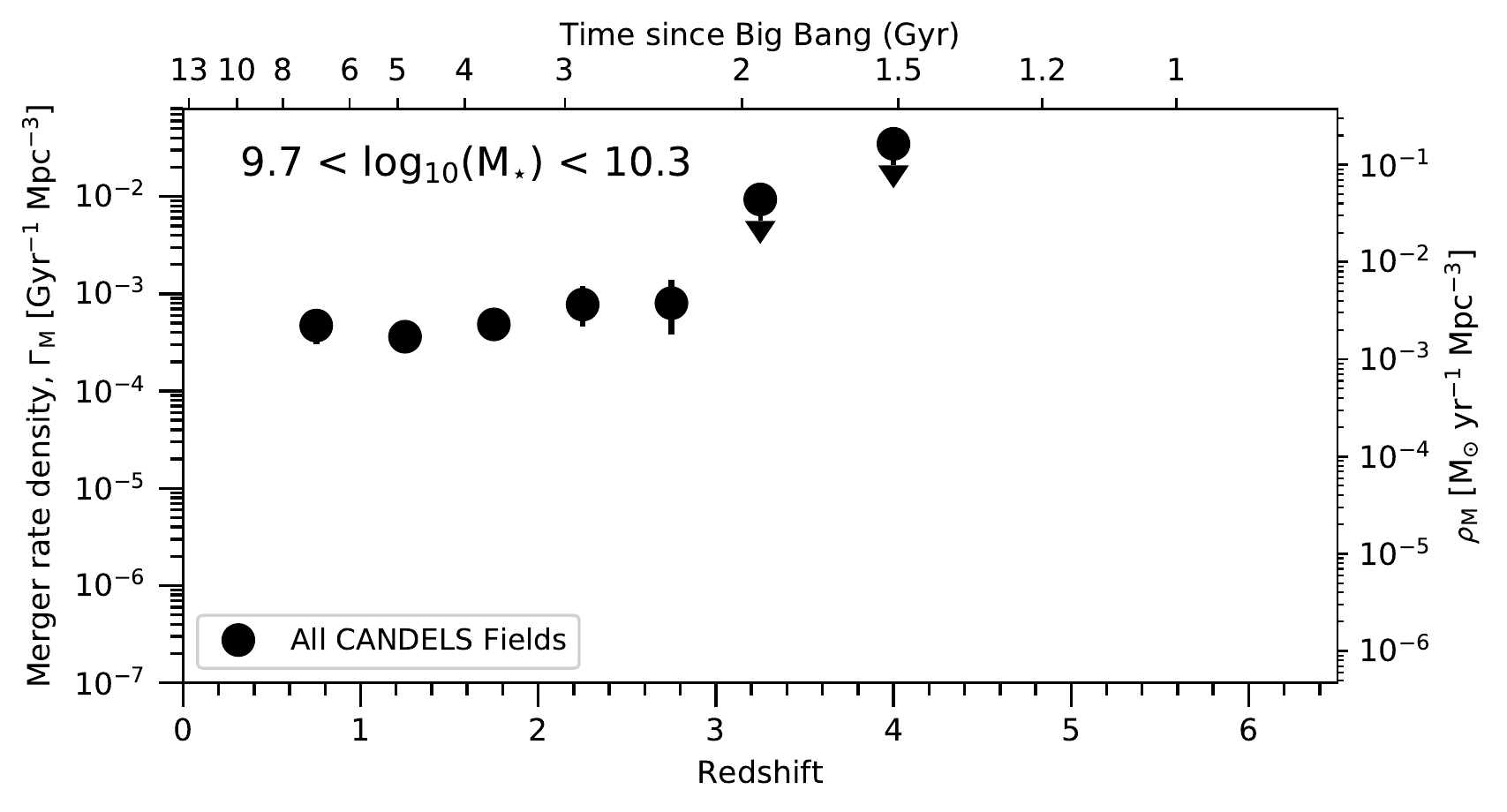}
	\includegraphics[width=\columnwidth]{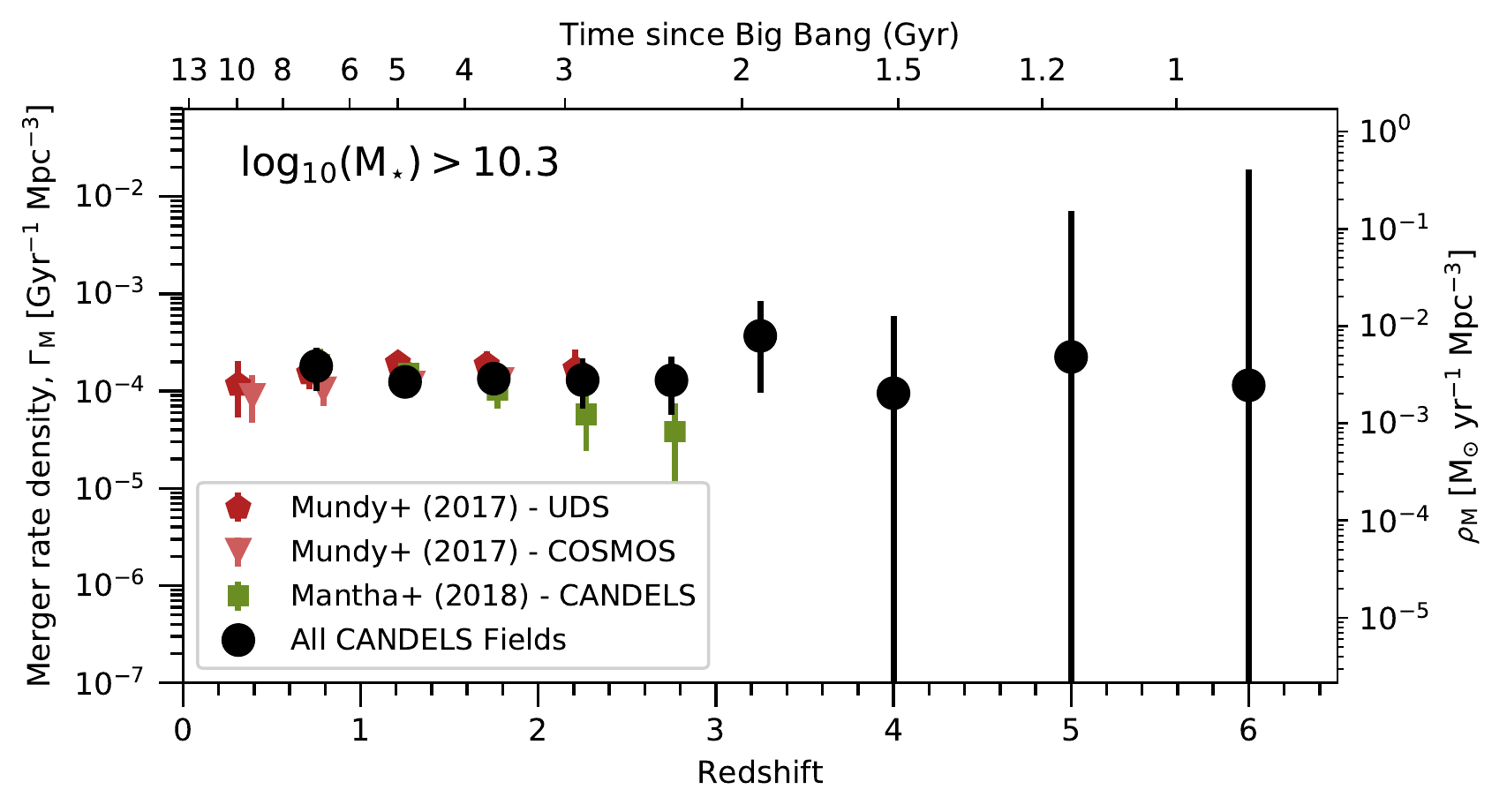}}
  \caption{Estimated comoving major merger rate as a function of redshift for galaxies with stellar mass $9.7 < \log_{10}(\Mstar / \text{M}_{\odot}) < 10.3$ (top) and $\log_{10}(\Mstar / \text{M}_{\odot}) > 10.3$ (bottom) assuming the redshift dependent merger timescales of \citet{2017MNRAS.468..207S}. Also shown are the merger rates based on the close-pair statistics of \citet{mundy2017} and \citet{2018MNRAS.475.1549M}, assuming the same redshift dependent timescale. The right-hand scale illustrates the inferred mass accretion rate density from major mergers based on the observed merger rate (see text).}
  \label{merger-fig:merger_rate}
\end{figure}

In Fig.~\ref{merger-fig:merger_rate} we show the resulting merger rates calculated following Equation~\ref{eq:merger_rate}. 
We also compare in Fig.~\ref{merger-fig:merger_rate} our results with those from Mundy et al. (2017) and \citet{2018MNRAS.475.1549M}.  

Here we see that the volume merger rates for the mass-selected samples are relatively constant with redshift, albeit with significant uncertainties in the highest redshift bins (the statistical uncertainties are dominated by the poor constraints on the high mass end of the stellar mass function). 
Given the additional statistical uncertainties in the cumulative number densities, the results for $\Gamma$ are in significantly less tension than for $\mathcal{R}$. 

\begin{deluxetable}{cc}
  \tablecaption{Comoving merger rate density, $\Gamma$.} 
  \tablehead{Redshift  & Merger rate density, $\Gamma$\tablenotemark{a}}
\startdata
     \multicolumn{2}{c}{$9.7 < \log_{10}(\Mstar / \text{M}_{\odot}) < 10.3$} \\ 
     \noalign{\smallskip}
     \hline
$0.5 \leq z < 1.0$ & $4.72^{+2.06}_{-1.69}$ \\
$1.0 \leq z < 1.5$ & $3.61^{+0.93}_{-0.84}$ \\
$1.5 \leq z < 2.0$ & $4.85^{+1.64}_{-1.31}$ \\
$2.0 \leq z < 2.5$ & $7.76^{+4.35}_{-3.17}$ \\
$2.5 \leq z < 3.0$ & $7.99^{+5.67}_{-4.12}$ \\
$3.0 \leq z < 3.5$ & $< 94.83$ \\
$3.5 \leq z < 4.5$ & $< 345.28$ \\
\hline
   \multicolumn{2}{c}{$\log_{10}(\Mstar / \text{M}_{\odot}) > 10.3$} \\ \noalign{\smallskip}
   \hline
$0.5 \leq z < 1.0$ & $1.80^{+0.99}_{-0.79}$ \\
$1.0 \leq z < 1.5$ & $1.24^{+0.37}_{-0.34}$ \\
$1.5 \leq z < 2.0$ & $1.35^{+0.52}_{-0.44}$ \\
$2.0 \leq z < 2.5$ & $1.29^{+0.82}_{-0.62}$ \\
$2.5 \leq z < 3.0$ & $1.28^{+1.00}_{-0.70}$ \\
$3.0 \leq z < 3.5$ & $3.65^{+4.70}_{-2.69}$ \\
$3.5 \leq z < 4.5$ & $0.92^{+4.60}_{-1.08}$ \\
$4.5 \leq z < 5.5$ & $2.28^{+64.87}_{-2.99}$ \\
$5.5 \leq z < 6.5$ & $1.01^{+171.86}_{-1.37}$ \\
  \enddata
\tablenotetext{a}{Merger rate density in $10^{-4}~\text{Gyr}^{-1}~\text{Mpc}^{-3}$.}
\tablecomments{Comoving number densities for the mass selected samples are calculated from the corresponding stellar mass functions as described in the text.}
  \label{tab:merger_rate_dens}
\end{deluxetable}

\subsubsection{Minor merger rates}
Despite the strong observational constraints on the evolution of the pair fraction as a function of merger ratio, it is not currently possible to derive strong conclusions on the actual minor merger \emph{rates} and their corresponding mass growth.
As illustrated in the previous section, assumptions on the merger timescales used to convert pair fractions to merger rates have significant effects on the estimated merger rates.
Detailed simulations of the merger timescale as a function of mass, mass-ratio and redshift (whether physical or observability) are not currently available.

Simulations of isolated mergers at low-redshift indicate that the timescales of minor mergers could be longer than those of major mergers \citep{Lotz:2010ie,Lotz:2010hf}, for mass ratios of 1:9 the average timescale increases only by $\approx 50\%$. 
The effect of longer timescales would be to decrease the predicted merger rate for minor mergers compared to that for major mergers, reducing their importance as a channel for galaxy growth. 
However, these simulations do not take into account the broader effects of projection effects and redshift uncertainties explored by \citet{2017MNRAS.468..207S} that may dominate the merger timescales for pair counts at high redshift.
The observational results presented here illustrate that when such simulations are available, it will be possible to place detailed constraints on the complete merger histories of galaxies out to these high redshifts.

% One quantity we can calculate with these merger rates for both the minor and major mergers is how much mass is added to galaxies due to these mergers and the specific merger mass accretion rate.  
% These are in some ways the ultimate equations and quantities for determining the role of different types of mergers on the formation of galaxies.  
% We include a discussion of this in Section~5.

\subsection{Merger Mass Accretion Rates}
The rapid rise in merger rates per galaxy observed in Fig.~\ref{merger-fig:merger_rate_per_galaxy} mirrors that observed in the specific star-formation rate evolution of galaxies over this period, \citep[e.g.][and references therein]{Stark:2013ix,Schenker:2013ep,Duncan:2014gh,2014ApJS..214...15S}.
Growth through major mergers may therefore still represent a significant role in the formation of the earliest galaxies at $z > 3$, a fundamental prediction of hierarchical structure formation.
Based on simple assumptions for the average mass accreted per major merger, in this section we present estimates of the stellar mass growth corresponding to the merger rates presented in Section~\ref{merger-sec:mergerrate}.
Due to the large systematic uncertainty inherent in the conversion from pair fractions to merger rates, interpretation of results from more complex modelling approaches would still be dominated by the same systematic limitations.

\subsubsection{Specific mass accretion rates}
Analogous to the specific star-formation rate, the specific mass accretion rate can simply defined as $\dot{M}/M  = \mathcal{R}(z)\bar{\mu}$, where $\bar{\mu}$ is the median mass ratio.
By integrating the distribution of pair fraction as a function of merger ratio presented at $2.5 < z < 3$ in Fig.~\ref{merger-fig:merger_frac_mu}, we calculate an average `major merger' mass ratio of $\bar{\mu} = 0.53$.
Based on the lack of observed evolution in pair fraction as a function of mass ratio (Fig.~\ref{merger-fig:merger_frac_mu}) we make the assumption of $\bar{\mu} = 0.53$ at all redshifts.
In Fig.~\ref{merger-fig:merger_rate_per_galaxy} we plot the resulting specific mass accretion rate for our sample in the right-hand twin axis.
We find values for the specific mass accretion rate which vary between $0.07$ and $\sim7$ Gyr$^{-1}$ for the major mergers in our sample.   

Also plotted in Fig.~\ref{merger-fig:merger_rate_per_galaxy} for reference is the median specific star-formation rate (sSFR, plus intrinsic scatter) for star-forming galaxies out to $z\sim6$, as described by the functional form presented in \citet{2014ApJS..214...15S}.
During the period of peak galaxy formation ($1 < z < 3$), star-formation in massive galaxies is clearly the dominant form of mass growth.
However, modulo the large systematic uncertainties in both estimates, the sSFR and specific merger mass accretion rate begin to converge at $z > 3$.  
This implies that at the highest redshifts, the amount of mass added to galaxies through major mergers may be directly comparable to that added by in-situ star-formation.

However, it is also the case that some of the star formation we see is being produced in the merging events associated with these galaxies.  
We cannot separate at this point the merger contributed to the non-merger triggered star formation, but suffice it to say, a significant fraction of the mass in these galaxies is being added in some form by the merger process.

\subsubsection{Mass accretion rate density}
A second important observational property is the integrated mass accretion rate density from major mergers, $\rho_{\textup{M}}$.
As above, we make a simple assumption that the average mass added per merger event is equal to $\bar{\mu}\times \bar{M_{\star}}$, where $\bar{\mu} = 0.53$ based on the $2.5 < z < 3$ bin and $\bar{M_{\star}}$ is calculated from the stellar mass function in this same bin.
The resulting mass accretion rate density estimates are illustrated by the twin axis in Fig.~\ref{merger-fig:merger_rate}.

We note that while there is variation in $\bar{M_{\star}}$ between redshift bins, typically $\pm 0.05~\textup{dex}$, it is smaller than the large systematic uncertainties in the merger timescales used to derive $\Gamma$.
We therefore present only this fiducial conversion in Fig.~\ref{merger-fig:merger_rate} for ease of interpretation.

Interpreting the estimated $\rho_{\textup{M}}$ presented in Fig.~\ref{merger-fig:merger_rate}, we find that the merger rate density is fairly constant, and this extends down to the lowest redshifts when we include results from \citet{mundy2017}.
We find no clear peak in the integrated merger rate, at least for galaxies with masses M$_{*} > 9.7$.   
This is in stark contrast to the cosmic star-formation rate density which peaks at $z \sim 2$ \citep{Madau:2014gt}.

The difference in merger rate and merger rate density redshift evolution can be reconciled by the fact that, while the number of mergers per galaxy is going down at lower redshifts, the number of galaxies above that mass limit is increasing and these two effects average each other out -- at least for the mass range probed in this work.

\section{Discussion - The evolution of galaxy mergers at $0.5 < z < 6$}\label{merger-sec:discussion}

It is fairly well established that there appears to be a disagreement between the observed merger history and models, particularly at high redshifts \citep[e.g.][]{Bertone:2009jc,Jogee:2009iz}.
Recent studies have attempted to alleviate this discrepancy with the idea that it results from observational studies selecting galaxies (and their merger ratios) by stellar mass, while model predictions have often been based on \emph{baryonic} mass \citep{2016ApJ...830...89M}.
The significant rise in the gas-fraction of galaxies at higher redshift would mean that pairs of merging galaxies with stellar mass ratio of $\mu << 1/4$ could have a baryonic mass ratio that would classify it as a major merger $\mu > 1/4$ - therefore increasing the observed number of major mergers.
Two recent observational studies \citep{2016ApJ...830...89M,2018MNRAS.475.1549M} have supported this picture, finding significantly greater numbers of major merger pairs at $z > 1$ based on flux ratios when compared with stellar mass ratios.

However, simulations that explore mergers as a function of \emph{stellar} mass \citep{RodriguezGomez:2015hw} can also significantly over-predict major merger rates at high redshift with respect to those presented in observational studies \citep{2016ApJ...830...89M,mundy2017}.
It is important to remember that the conversion of observed pair fractions to a merger rate requires the assumption of a corresponding merger timescale.
This merger time-scale is critical but difficult to measure, and in the past has been taken to be a constant through cosmic time.
 
\citet{Snyder:2016ta} revealed through forward modelling of galaxy pair-counts in simulations that the merger time-scale for galaxy pairs declines as $\sim (1+z)^{2}$.
When we use these new evolving time-scales to estimate the merger rate from pair fractions, the
`observed' merger rate is found to increase with redshift at a rate that is more comparable to those predicted by hydrodynamical simulations that previous work would suggest \citep{RodriguezGomez:2015hw}.  
Because the merger time-scale is shorter at higher redshifts, this means that although we see a gentle rise in pair fraction with redshift, there are significantly more mergers actually occurring due to the fact that the time-scale for these mergers to occur is much faster at higher redshifts.    
This suggests that mergers are a more common process, by a factor of $>10$ at $z = 6$ compared with $z = 1$.  
The reason we do not see as many mergers ongoing is clearly because the time-scales for them to occur is much quicker than it is at lower redshifts.

Based on the results presented in Section~\ref{merger-sec:mergerrate}, we conclude that the assumed timescale is the origin of the discrepancy between the observations and simulation results for the merger history (and not necessarily the use of stellar mass selections).   
What we generally find is that, while we agree well with the predicted merger rates at higher masses, our observations now imply a higher merger rate than predicted for galaxies with $9.7 < \log_{10}(\Mstar / \text{M}_{\odot}) < 10.3$.
Some of this discrepancy may be accounted for by the expected mass-dependency of merger timescales.
\citet{Kitzbichler:2008gi} find a merger timescale in $N$-body simulations that varies as $\propto M_{\star}^{-0.3}$, yielding expected timescales for $9.7 < \log_{10}(\Mstar / \text{M}_{\odot}) < 10.3$ that are $\approx40\%$ longer than for $\log_{10}(\Mstar / \text{M}_{\odot}) > 10.3$, and hence merger rates that are lower by the same amount.
Further investigation is required to establish whether any remaining offset is physical or a result of additional mass-dependence in the merger observability timescales.

For minor mergers ($\mu < \frac{1}{4}$) we find that difference between the observations and theory gets larger at the lowest mass ratio of mergers. 
At face value, our observations suggest that minor mergers may not be as common or as important in the galaxy formation process than what is predicted in the Illustris simulation \citep{RodriguezGomez:2015hw}.
However, given the simplistic prescription used in this work convert from pair count to merger rates (and vice-versa), the source of this discrepancy may also lie in this critical assumption.
Only a small mass-ratio dependence in the merger observability timescales would be required to alleviate the observed tension.

Finally, given that the redshift evolution of pair fractions for massive galaxies ($\log_{10}(\Mstar / \text{M}_{\odot}) \geq 10.3$) observed in this work is in good agreement with other recent studies \citep{2016ApJ...830...89M, mundy2017}, our key conclusion on the rapid rise in merger rates is not necessarily unique to our observed pair fractions.
Furthermore, although the recent work by \citet{2018MNRAS.475.1549M} using the same CANDELS dataset find different evolution in the observed galaxy pair fraction, when incorporating the evolving merger timescales of \citet{Snyder:2016ta} the authors draw similar conclusions to those presented in this study. 
However, the higher observed pair fractions at $2 < z < 3$ in this work (see Fig.\ref{merger-fig:merger_frac}) and the extension to higher redshift mean that this is the first instance where the observed merger rates per galaxy are shown to rise at a rate that so closely matches those of simulations out to the very earliest epoch of galaxy formation.

Despite these advances, there still remain key uncertainties in estimating galaxy merger rates that future studies can address.
From the additional tests performed in this study (see Sections~\ref{merger-sec:specz_pairs} and \ref{merger-sec:pz_precision}), we know there are still systematic uncertainties in the pair fractions obtained from photometric redshifts on the order of $\sim30\%$.
However, in the final inferred merger rates, these uncertainties are dominated by the larger uncertainty in the merger timescales (or observability timescale).
With larger simulation volumes and the improved number statistics these allow, extensions to the forward modelling of \citet{Snyder:2016ta} would enable estimates of merger observability timescales (and the scatter therein) as a function of mass, merger ratio and redshift to much greater precision.
This increased understanding of merger timescales, more than any increase in redshift precision or reliability, is key to placing meaningful observational constraints on the assembly history of massive galaxies.

\section{Summary}\label{merger-sec:summary}
% We present a systematic analysis of the pair fraction and resulting merger rates for galaxies up to $z=6$ as measured in all five of the CANDELS fields.
% This is the first analysis of this type to such early times in the universe, and uses the deepest data over a relatively large area where this type of analysis can be preformed.  
% The results of this will have implications for a host of other areas of galaxy formation and astrophysics, including star-formation triggering, black hole growth and AGN activity, galaxy assembly as well as the number of super-massive black hole mergers in the early universe.  
% This last issue is an important one for gravitational wave detections through future missions such as LISA.

Using the full CANDELS data set we present a study of galaxy major mergers up to $z = 6$, and minor mergers up to $z = 3$, for massive galaxies with $\log_{10}(\Mstar / \text{M}_{\odot}) \geq 10.3$ and $9.7 < \log_{10}(\Mstar / \text{M}_{\odot}) < 10.3$. 
This is the first analysis at such early times in the universe, and uses the deepest data over a relatively large area where this type of analysis can be preformed.  
The results of this study have implications for a host of other areas of galaxy formation and astrophysics, including star-formation triggering, black hole growth and AGN activity, galaxy assembly as well as the number of super-massive black hole mergers in the early universe.  
This last issue is an important one for gravitational wave detections through future missions such as LISA (Conselice \& Duncan 2019, in prep).

As part of our analysis we have made new stellar mass and photometric redshift measurements for galaxies in all five CANDELS fields, including the full photometric redshift posteriors and stellar masses estimate at all likely redshift steps. 
The summary of our findings are:

\begin{enumerate}
\item For both $\log_{10}(\Mstar / \text{M}_{\odot}) \geq 10.3$ and $9.7 < \log_{10}(\Mstar / \text{M}_{\odot}) ,\leq 10.3$, the fraction of galaxies in major pairs (mass ratios of $0.25 \leq \mu \leq 1$) increases monotonically as a function of redshift out to $z\sim6$.

\item We furthermore find that the merger rate increases up to the highest redshifts explored ($z \sim 6$). This is due to the fact that we use new scaling laws from simulations which show that the merger observability time-scale declines at higher redshifts at $\sim (1+z)^{2}$ \citep{2017MNRAS.468..207S}. This differs significantly from previous work whereby the merger rate appeared to decline at higher redshifts.  

\item Based on our observed merger rates, we infer that at $z > 3$, major mergers may play an increasingly important role in the mass growth of star-forming galaxies - significantly more so than at the peak of galaxy formation.

\item While the cumulative pair fraction increases for more minor mergers down to a mass ratio of 1:20 for galaxies at $1 < z < 3$, the relative number of minor mergers is lower than predicted by simulations. Between these redshifts we also do not find a significant change in the fraction of galaxies merging at any merger mass ratio we probe, suggesting that the merger history for both minor and major mergers mimic each other at these epochs.
\end{enumerate}

Overall, our conclusions are that observational constraints of mergers in massive galaxies are now consistent with hierarchical models of galaxy formation.
At the highest redshifts, mass growth from major mergers may be comparable to or even higher than in-situ star-formation.
To probe at even higher redshifts, or lower mass galaxies at $z < 6$, will require deeper surveys with the James Webb Space Telescope. 
Alternatively wide-area surveys at comparable depths to CANDELS will probe volumes sufficient to provide samples of even more massive galaxies that are large enough to perform similar analyses \citep[e.g. the Euclid Deep fields;][]{2011arXiv1110.3193L}. 
In addition to providing vital new observational constraints on galaxy formation, our results can be used to predict the number of likely events gravitational wave detectors such as LISA will find due to merger of super-massive black holes that exist at the centres of these galaxies. Thus, overall our results lead to a suite of implications that we will explore in future papers.

\section*{Acknowledgements}
The research leading to these results has received funding from the European Union Seventh Framework Programme FP7/2007-2013/ under grant agreement number 607254. This publication reflects only the author's view and the European Union is not responsible for any use that may be made of the information contained therein. KJD acknowledges support from the ERC Advanced Investigator programme NewClusters 321271. 
We would also like to acknowledge funding from the Science and Technology Facilities Council (STFC) and the Leverhulme Trust.
PGP-G wishes to acknowledge
support from Spanish Government MINECO Grant AYA2015-63650-P.
KBM acknowledges support from the HST archival research grant HST-AR-15040.
This work is based on observations taken by the CANDELS Multi-Cycle Treasury Program with the NASA/ESA HST, which is operated by the Association of Universities for Research in Astronomy, Inc., under NASA contract NAS5-26555.
The VUDS spectroscopic data is in this work is based on data obtained with the European Southern Observatory Very Large Telescope, Paranal, Chile, under Large Program 185.A-0791, and made available by the VUDS team at the CESAM data center, Laboratoire d'Astrophysique de Marseille, France.

\footnotesize{
    \bibliographystyle{mn2e}
    \bibliography{bibtex_lib}
}

\appendix
\section{Consistent completeness simulations for all CANDELS fields}\label{app:completeness}
Our methodology for determining the detection completeness follows a procedure similar to those presented in the CANDELS release papers \citep{Galametz:2013dd,Guo:2013ig}, adding a representative range of mock galaxies to the $H_{160}$ detection images and attempting to recover them using the same photometry procedures used to produce the science catalogs.

For the morphological distribution of our input mock sources, we assume an empirical distribution that is dependent on apparent magnitude.  Firstly, we divide the galaxies with parametric morphology measurement of \citet{vanderWel:2012eu} into bins of apparent $H_{160}$ magnitude. Next, for every mock galaxy with a given assigned magnitude, we assign a morphology (effective radius, Sersic index and ellipticity) by randomly sampling a morphology drawn from the corresponding magnitude bin of the real galaxy sample. To maximise the final number statistics of detected sources at faint magnitudes, we assume a power-law magnitude distribution which results in $\approx 10\times$ more input sources at the faint magnitude limit ($H_{160} = 30$) as at the bright limit ($H_{160} = 22$).

One critical assumption to note is that we assume the morphological distribution of sources below the magnitude limit of the \citet{vanderWel:2012eu} sample morphologies is similar to those just brighter than the limit. While in true physical terms this assumption is not likely to be valid for the key properties such as size, for the image resolution of HST any further evolution in size would have minimal effect.
Additionally, the observed distribution of morphologies in our faintest bin is very similar to the completeness corrected morphology distribution for Lyman break galaxies observed by \citet{2004ApJ...600L.107F} and therefore likely represent a valid assumption.

\begin{figure}
\centering
\includegraphics[width=0.5\textwidth]{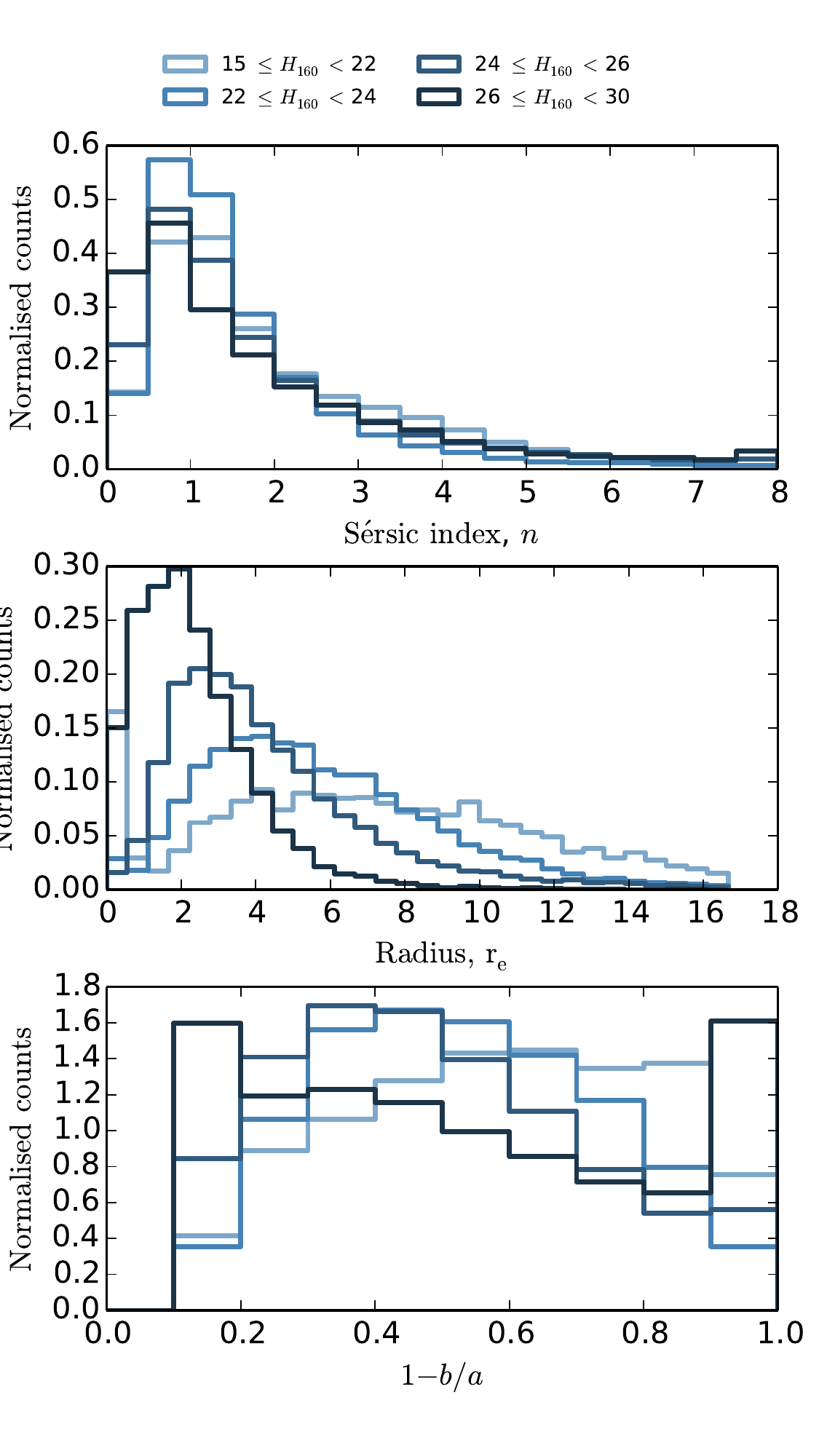}
  \caption{Size and morphology distributions of the mock sources inserted into the CANDELS $H_{160}$ images for completeness simulations. The measured distributions are taken from the catalog presented in \citet{vanderWel:2012eu}. Top: Surface brightness distrubution parametrized by Sersic indices. Middle: Circularised effective radii (kpc). Bottom: Measured ellipticity.}
  \label{merger-fig:MorphologyDists}
\end{figure}

Once the morphologies of the mock sources has been assigned, we then insert the mock galaxy images into the respective $H_{160}$ image for each field, 3000 sources at a time, and then process the images through the same \textsc{SExtractor} process as used to produce the original CANDELS photometry catalogs. This process was repeated 75 times for each field, yielding an average number of mock sources detections per field of 70,000 (typically 100 to 3000 \textit{detected} galaxies in the magnitude bins corresponding to 50\% completeness). 

\begin{figure*}
\centering
\includegraphics[width=0.45\textwidth]{plots/GS_completeness.pdf}
\includegraphics[width=0.45\textwidth]{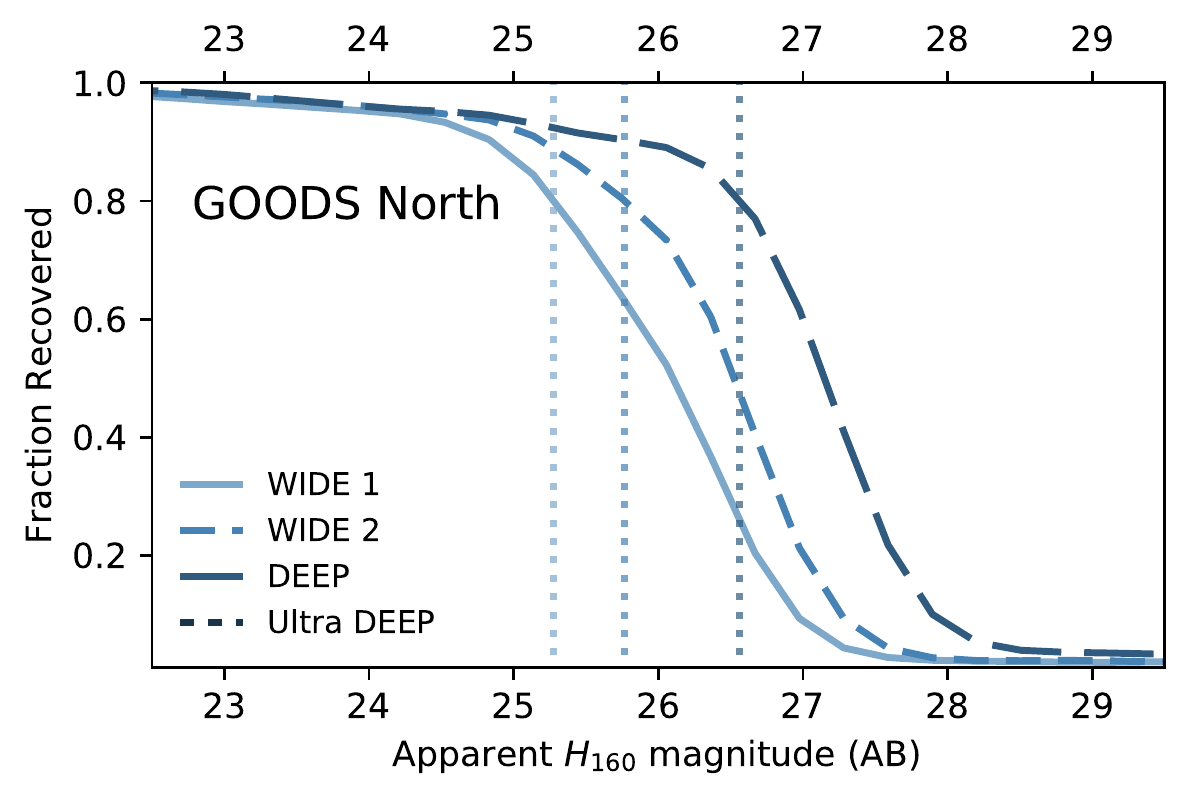}  
\includegraphics[width=0.45\textwidth]{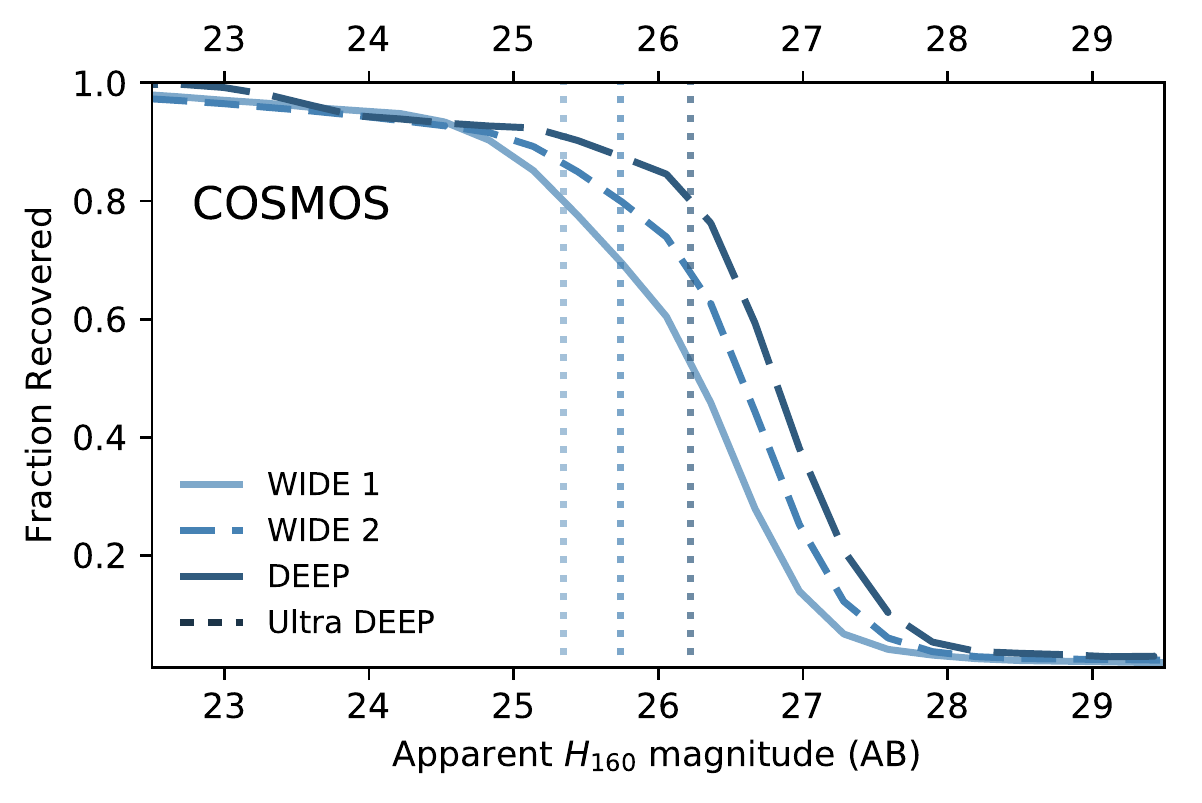}  
\includegraphics[width=0.45\textwidth]{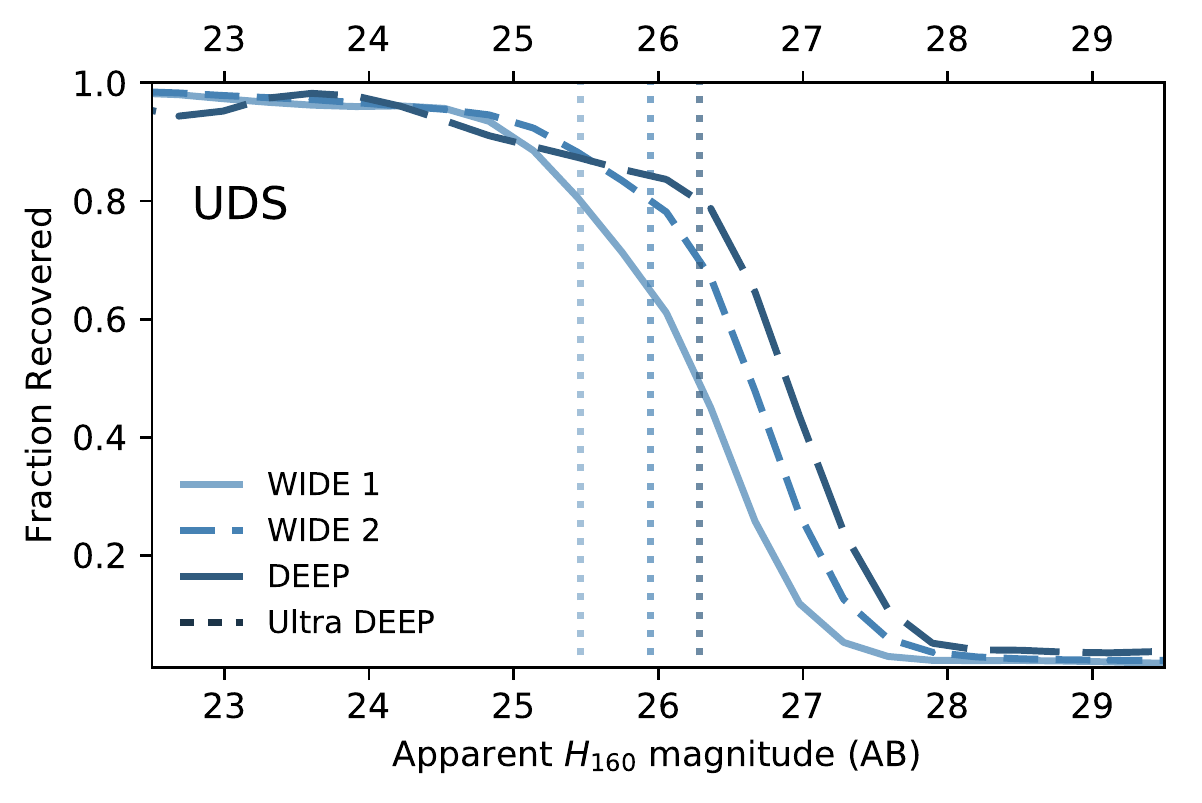}  
\includegraphics[width=0.45\textwidth]{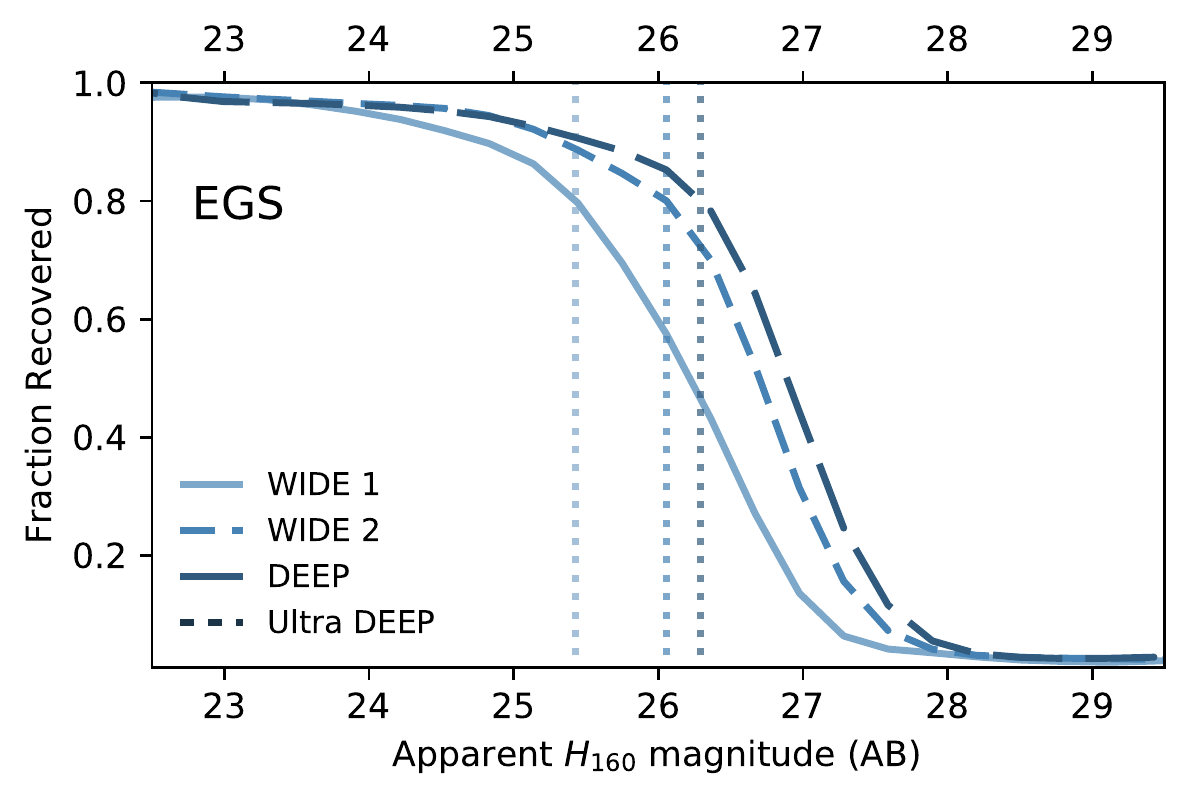}
  \caption{Source recovery fraction as a function of apparent $H_{160}$ magnitude for each of the CANDELS fields used in this analysis. The vertical dashed lines show the magnitude at which the recovery fraction equals 80\% for each sub-field.}
  \label{merger-fig:all_completeness}
\end{figure*}

We note that specific care was taken to ensure that the correct combination of image release, software version and extraction parameters were used for each CANDELS field. With no additional mock sources were added to the $H_{160}$ science image, we confirm that we obtain the exact number of galaxy detections as presented in the official CANDELS releases (we refer the reader to the respective release papers for precise numbers).

\section{Stellar mass consistency checks}

Although there is clearly visible scatter, the majority of mass estimates are in very good agreement with the team estimate with no significant bias and relatively small scatter. Furthermore, the scatter is most significant at masses of $\log_{10}(\Mstar / \text{M}_{\odot}) < 9$, well below the range probed in this analysis. Note that the redshift assumed for the stellar mass fits differs between the results of this paper and the team redshift, much of the scatter is therefore a result of small differences in redshift and due to issues in the mass estimate. When assuming identical redshifts (i.e. the best available redshift from the CANDELS photo-z releases) the scatter and biases are reduced even further.

\begin{figure}
\centering
	\includegraphics[width=0.55\textwidth]{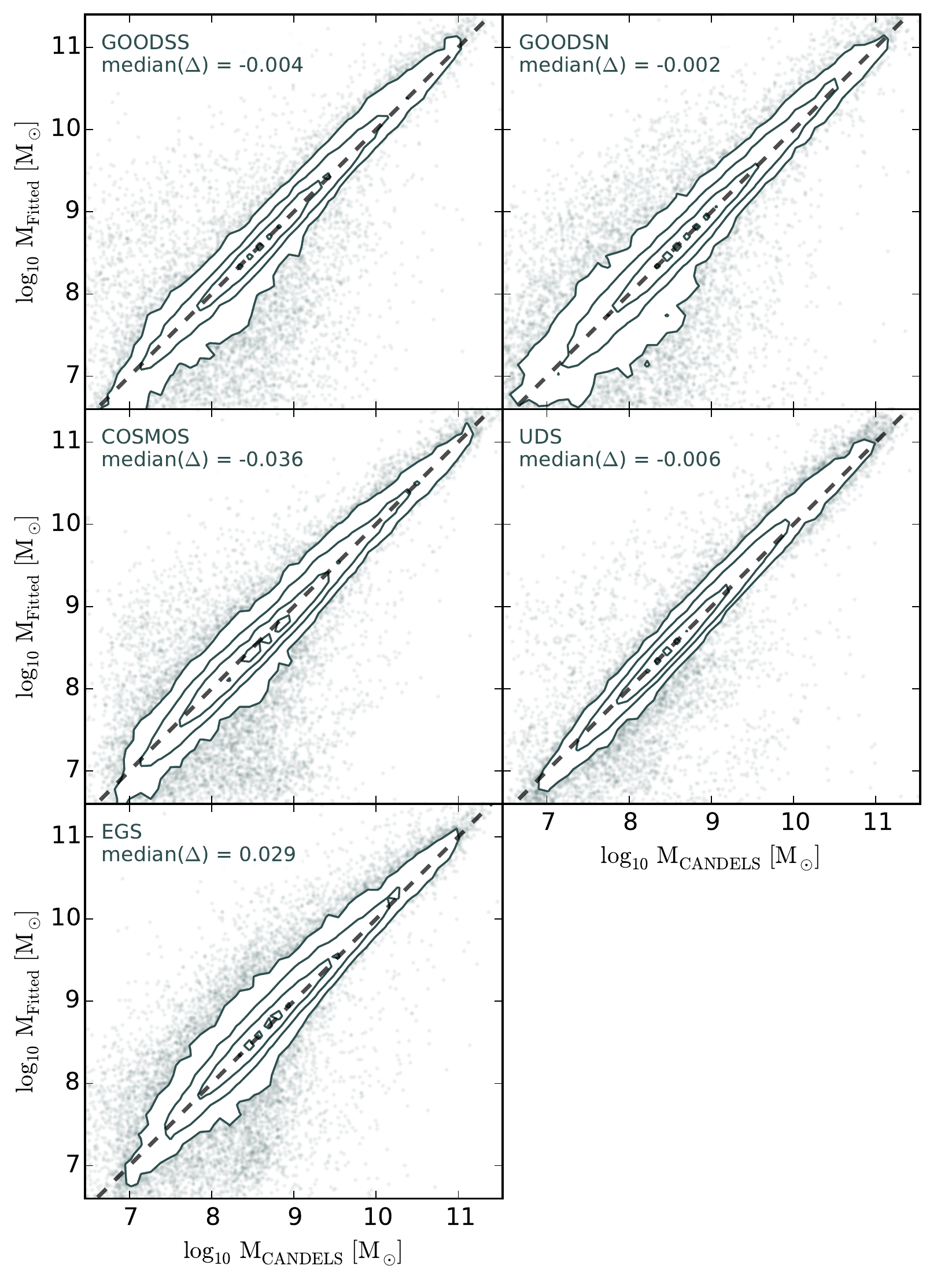}
  \caption{Comparison of stellar mass estimates from this work (minimum-$\chi^{2}$) with those of the median CANDELS team mass estimates for each of the full CANDELS photometry catalog. The median offset in each field is indicated in each panel.}
  \label{merger-fig:mass_candels}
\end{figure}
\label{lastpage}

\end{document}